\documentclass[a4paper,fleqn,usenatbib]{mnras}

\usepackage{mathptmx}

\usepackage[T1]{fontenc}
\usepackage{ae,aecompl}

\usepackage{graphicx,subfig}
\usepackage{epsfig}
\usepackage{color}
\usepackage{xcolor}
\usepackage{verbatim}
\usepackage{amssymb}
\usepackage{lscape}
\usepackage{float}
\usepackage{amsmath}
\usepackage{longtable}
\usepackage{supertabular}
\usepackage{journal_names}
\usepackage{rotating}
\usepackage{pdflscape}
\usepackage{dcolumn}
\usepackage{caption}
\usepackage{multicol}
\usepackage{soul}

\newcommand*{\thead}[1]{\multicolumn{1}{c}{ #1}}

\def\la{\mathrel{\hbox{\rlap{\hbox{\lower4pt\hbox{$\sim$}}}\hbox{$<$}}}}
\def\ga{\mathrel{\hbox{\rlap{\hbox{\lower4pt\hbox{$\sim$}}}\hbox{$>$}}}}

\def\arcmin{\hbox{$^\prime$}}
\def\arcsec{\hbox{$^{\prime\prime}$}}

\def\fdg{\hbox{$.\!\!^\circ$}}

\newcommand{\dg}{^{\circ}}

\newcommand{\MSUN}{${\rm M}_\odot$}

\newcommand{\kms}{{\,km\,s$^{-1}$}}

\newcommand{\HI}{\mbox{\normalsize H\thinspace\footnotesize I}}

\def\apj    {{ApJ }}
\def\apjl   {{ApJL }}
\def\apjs   {{ApJS }}
\def\aj     {{AJ }}

\def\mnras  {{MNRAS }}
\def\nat    {{Nature }}

\def\pasa   {{PASA }}

\title[The WSRT ZoA Perseus-Pisces Filament wide-field H{\Large I} imaging survey I ]{The WSRT ZoA Perseus-Pisces Filament wide-field H{\Large I} imaging survey I.  H{\Large I} catalogue and atlas.}

\author[M. Ramatsoku et al.]{M. Ramatsoku$^{1,2,3}$\thanks{E-mail: mpati@astro.rug.nl}, M.A.W Verheijen$^{1}$, R.C. Kraan-Korteweg$^{2}$, G.I.G. J\'ozsa$^{4}$, 
\newauthor
A.C. Schr\"{o}der$^{5}$, T.H. Jarrett$^{2}$, E.C Elson$^{2}$, W. van Driel$^{6}$, W.J.G. de Blok$^{3,2,1}$,
\newauthor
P.A. Henning$^{7}$.\ \\
$^{1}$Kapteyn Astronomical Institute, University of Groningen, Landleven 12, 9747 AV Groningen, The Netherlands\\
$^{2}$Astrophysics, Cosmology and Gravity Centre (ACGC), Department of Astronomy, University of Cape Town, Private Bag X3,\\ 
Rondebosch 7701, South Africa\\
$^{3}$ASTRON, Netherlands Institute for Radio Astronomy, Postbus 2, 7990 AA Dwingeloo, The Netherlands\\
$^{4}$SKA South Africa, Radio Astronomy Research Group, 3rd Floor, The Park, Park Road, Pinelands 7405, South Africa\\
$^{5}$South African Astronomical Observatory (SAAO), PO Box 9, 7935 Observatory, Cape Town, South Africa\\
$^{6}$GEPI, Observatoire de Paris, CNRS, Universit\'e Paris Diderot, 5 place Jules Janssen, 92190 Meudon, France\\
$^{7}$Department of Physics and Astronomy, University of New Mexico, 1919 Lomas Blvd. NE, MSC07 4220, Albuquerque NM 87131-0001, USA}

\date{Accepted 2016 April 21. Received 2016 April 21; in original form 2015 November 24}

\pubyear{2016}

\begin{document}
\label{firstpage}
\pagerange{\pageref{firstpage}--\pageref{lastpage}}
\maketitle

\begin{abstract}
We present results of a blind 21cm \HI-line imaging survey of a galaxy overdensity located behind the Milky Way at $\ell,b$ $\approx$ 160$\dg$, 0.5$\dg$. The overdensity corresponds to a Zone-of-Avoidance crossing of the Perseus-Pisces Supercluster filament. Although it is known that this filament contains an X-ray galaxy cluster (3C\,129) hosting two strong radio galaxies, little is known about galaxies associated with this potentially rich cluster because of the high Galactic dust extinction. We mapped a sky area of $\sim$9.6~sq.deg using the Westerbork Synthesis Radio Telescope in a hexagonal mosaic of 35 pointings observed for 12 hours each, in the  radial velocity range $cz = 2400 - 16600$ \kms. The survey has a sensitivity of 0.36 mJy/beam rms at a velocity resolution of 16.5 \kms. We detected 211 galaxies, 62\% of which have a near-infrared counterpart in the UKIDSS Galactic Plane Survey. We present a catalogue of the \HI\ properties and an \HI\ atlas containing total intensity maps, position-velocity diagrams, global \HI\ profiles and UKIDSS counterpart images. For the resolved galaxies we also present \HI\ velocity fields and radial \HI\ surface density profiles. A brief analysis of the structures outlined by these galaxies finds that 87 of them lie at the distance of the Perseus-Pisces Supercluster ($cz \sim 4000 - 8000$ \kms) and seem to form part of the 3C\,129 cluster. Further 72 detections trace an overdensity at a velocity of $cz \approx$ 10000 \kms\ and seem to coincide with a structure predicted from mass density reconstructions in the first 2MASS Redshift Survey.
\end{abstract}



\section[]{Introduction}\label{intro}
Major steps forward have been achieved in mapping the large-scale distribution of galaxies forming the Cosmic Web, thanks to dedicated wide-field (on-sky), deep (in terms of redshift) galaxy redshift surveys such as the 2dF Galaxy Redshift Survey (2dFGRS; \citealp{Colless2001}), the 2MASS Redshift Survey (2MRS, \citealp{Huchra2012}), the Sloan Digital Sky Survey (SDSS; \citealp{Eisenstein2011}) and the 6dF Galaxy Survey (6dFGS; \citealp{Jones2004}). The largest areal coverage is provided by the 2MRS with 44000 galaxies. It is based on the brightest objects in the near-infrared 2MASS Extended Source Catalogue (2MASX; \citealp{Jarrett2000}) and has a relatively low median velocity of $cz_{\rm med} = 9000$ \kms. A version slightly deeper in velocity ($cz_{\rm med} = 15000$ \kms) is achieved through a combination of the 2MRS, SDSS and 6dFGS catalogues into the 2M$++$ galaxy compilation by \citet{Lavaux2011}, however, it is not homogenous in sky coverage. \citet{Bilicki2014} have therefore created a much deeper catalogue ($cz_{\rm med} \approx 60000$ \kms) based on photometric redshifts, (the 2MASS Photometric Redshift catalog; 2MPZ).

What all these 'all-sky' surveys have in common is the lack of information on galaxies hidden behind our own Milky Way due to dust extinction and confusion by high stellar densities. This is what creates the so called Zone of Avoidance (ZoA; for reviews see \citealp{kraan00} and \citealp{kraan05}). We still have a highly incomplete census of the galaxy distribution in the local Universe along the $360\dg$-circle of  the inner ZoA, i.e., for Galactic latitudes $|b| \le 5\dg$, with a broader ZoA around the Galactic Bulge ($|b| \la 10\dg$). 

Uncovering galaxies in the ZoA is important for studying the kinematics of the Local Group (LG) and large-scale structures (LSS) in the local Universe. This includes understanding the origin of the dipole observed in the Cosmic Microwave Background (CMB; \citealp{Fixsen2011}). Earlier studies focused on the significance of the Great Attractor (GA) at $45 - 50$ Mpc and the Perseus-Pisces Supercluster (PPS) at 57-114 Mpc as likely contributors (\citealp{Dressler1987}, \citealp{RowanRobinson1990}, \citealp{Hudson1993}). More recent  studies, however, suggest contributions to the dipole from more distant LSS  at $\sim$200 Mpc like the Shapley Supercluster (\citealp{Branchini1996}, \citealp{Plionis1998}, \citealp{Kocevski2006}, \citealp{Lavaux2010}, \citealp{Lavaux2011}). It has been proposed that incomplete mapping of the ZoA is to a certain degree responsible for these uncertainties (\citealp{erdogdu2006}, \citealp{Loeb2008}).

Observing the 21 cm line emission of neutral hydrogen (\HI) in gas-rich galaxies is an effective method of mapping the most obscured regions behind the Galactic Plane (GP). This is due to the transparency of foreground interstellar dust at this wavelength and the Doppler shift of the emission line. To take advantage of this, considerable efforts have been devoted to \HI\ surveys to find galaxies in the ZoA and measure their redshifts. 

Earlier \HI\ surveys of the ZoA include the 91-m Green Bank radio telescope blind survey out to $\sim$7200 \kms\ \citep{Kerr1987} and the more systematic Dwingeloo Obscured Galaxy Survey (DOGS; \citealp{kraan1994}), with an rms of 40\,mJy at a channel width of 4 \kms\ out to 4000 \kms. The Parkes Multi-beam Receiver has been used to systematically map the ZoA in the southern hemisphere (HIZOA-S; Staveley-Smith et al. 2015) with an extension to the north (HIZOA-N; \citealp{Donley2005}) for the most obscured part of the ZoA ($|b| < 5\dg$). With an rms of $\sim$6 mJy and channel width of 13.2 \kms\ out to $\sim$12000 \kms\ these surveys together cover  the ZoA from  $\ell = 196\dg$ across the Galactic Bulge to $\ell = 52\dg$. Another recent and more sensitive survey is the Arecibo L-Band Feed Array Zone of Avoidance Survey (ALFA ZoA; \citealp{Henning2010}, \citealp{McIntyre2015}) with an rms  of 1\,mJy at 9 \kms\ resolution which is mapping the ZoA accessible to the Arecibo telescope ($30\dg \la \ell \la 75\dg$ and $175\dg \la \ell \la 207\dg$). All of these surveys have contributed to uncovering gas rich galaxies and structures they belong to in the ZoA. However, they have so far left a major part of the northern ZoA mostly unexplored ($80\dg \la \ell \la 180\dg$).

For this reason a systematic \HI\ follow-up pointed survey of  2MASX bright galaxy candidates in the ZoA without previous redshift information ($\sim$1200 observed to date) was started with the 100m-class Nan\c{c}ay Radio Telescope (NRT)\footnote{http://www.nrt.obspm.fr/} to an rms of $\sim$3 mJy with a velocity resolution of 18 \kms\ out to $v \leq 12000$ \kms\ (\citealp{vandriel09}, \citealp{Ramatsoku2014}, Kraan-Korteweg et al, in prep).

Among features uncovered by the 250 NRT detections was a filament crossing the  Galactic plane at  $\ell  \approx 160\dg$ with a recession velocity range of $4000 -  7000$ \kms\ (see Fig.~1 in Ramatsoku et al. 2014). Its position and  velocity suggest it to be part of the ZoA crossing of the PPS, which had been hypothesised to exist previously \citep{Focardi1984}. This structure was earlier mapped with \HI-observations of optically identified galaxies (e.g. \citealp{Chamaraux1990}) and tentative indicators for this filament were found, but it was sparsely sampled. This filament encompasses an X-ray galaxy cluster, 3C\,129 at $\ell, b \approx 160.52\dg, 0.28\dg$ (\citealp{Ebling2000}, \citealp{Leahy2000}). The X-ray cluster hosts two bright radio galaxies, a head-tail source (3C\,129) and a double-lobed giant elliptical radio galaxy (3C\,129.1), at $cz = 6236$ and $6655$ \kms, respectively \citep{Spinrad1975}. The presence of such radio sources usually implies a rich cluster environment. The cluster has a total X-ray luminosity of $L_{\rm X} =  1.89 \times 10^{44}$~h$_{50}^{-2}$erg~s$^{-1}$ as listed in the ''Clusters In the ZoA'' survey (CIZA; \citealp{Ebling2000}) from the ROentgen SATellite (ROSAT\footnote{http://www.xray.mpe.mpg.de/cgi-bin/rosat/rosat-survey}; \citealp{Truemper1982}). While not the most luminous ROSAT X-ray cluster listed, it should be noted that it might be more massive than suggested by its X-ray luminosity since the intervening high Galactic gas column density ($N_{\rm H} \approx 5.76 \times 10^{21}$~cm$^{-2}$; \citealp{Stark1992})  may well have reduced the X-ray flux in the ROSAT $0.2 - 2.4$ keV band by more than 30\% \citep{Leahy2000}.

Due to the high extinction layer of the Milky Way, most of the galaxies belonging to this cluster had not been observed before. Hence we decided to conduct a deep blind \HI\ imaging survey with the Westerbork Synthesis Radio Telescope\footnote{http://www.astron.nl/radio-observatory/} (WSRT) of a mosaic covering a 9.6~sq.deg sky area which comprises the cluster and the surrounding filament within which the NRT survey showed it to be embedded.

Our aims are to firstly investigate in detail the structure associated with the 3C\,129 galaxy cluster. The \HI\ data and accompanying near-infrared photometry will give distance estimates using the Tully-Fisher relation. These will be used to determine the cluster's relevance to flow fields around it and the larger Perseus-Pisces Complex, notably evident in the whole-sky 2MASS depictions of the large scale structure that appears to pierce the cluster as it crosses from the southern to northern Galactic hemispheres \citep{Jarrett2004}. Moreover, these data will aid  the 2MASS Tully-Fisher survey (2MTF; \citealp{Masters2008}) efforts, by complementing the  inner $b \approx |5\dg|$ ZoA regions excluded from optical spectroscopy. Secondly, taking advantage of the wide areal coverage of our \HI\ imaging survey of the cluster and its environments, we will be able to conduct an examination of environmental effects on the \HI\ properties of these galaxies.

A further aspect of this project is to provide interferometric data cubes to assess data handling algorithms such as pipeline reduction, calibration, source finding and characterisation. Having a good handle on these algorithms will prove invaluable in the planning and preparation of the forthcoming \HI\ surveys to be carried out with AperTIF at the WSRT (\citealp{Verheijen2008}, \citealp{Oosterloo2010}) and SKA  precursor instrument \HI\ surveys to be conducted with MeerKAT and ASKAP (\citealp{DeBlok2012}, \citealp{Duffy2012}, respectively). It is for this reason that our velocity search range (2400 $-$ 16600~\kms), extends far beyond the mean velocity of the 3C\,129 cluster.

In this first of a series of papers based on the WSRT imaging survey described above, we present the \HI\ data of the galaxies unveiled in the ZoA, both as a catalogue containing \HI\  parameters as well as an atlas of \HI\ distributions and kinematics. We also include a discussion of the large-scale structures traced by the detected galaxies. Subsequent papers will focus in detail on the other aforementioned science goals.

This paper is organised as follows: in Sect.~2 we describe the observing strategy and WSRT data reduction. Section 3 provides a description of our source finding procedure. Methods used to determine the \HI\ properties of detected galaxies are described in Sect.~4. The resulting \HI\ atlas products are described in Sect.~5, while the full atlas is presented in supplementary material online ($http://www.ast.uct.ac.za/$$\sim$mpati/). Also in Sect.~5, the completeness of the survey and counterparts search are discussed. In Sect.~6 we discuss the measured \HI\ parameters and their characteristics. Lastly, structures revealed by this survey are presented in Sect.~7.

We assume a $\Lambda$ cold dark matter cosmology with $\Omega_{\rm M} = 0.3, \Lambda_{\Omega} = 0.7$ and a Hubble constant H$_{0}$ = 70 \kms\ Mpc$^{-1}$ throughout this paper.

\section{Observations and Data Reduction}\label{sec2}
\begin{figure} 
  \centering
    \includegraphics[width = 92mm, height = 83mm]{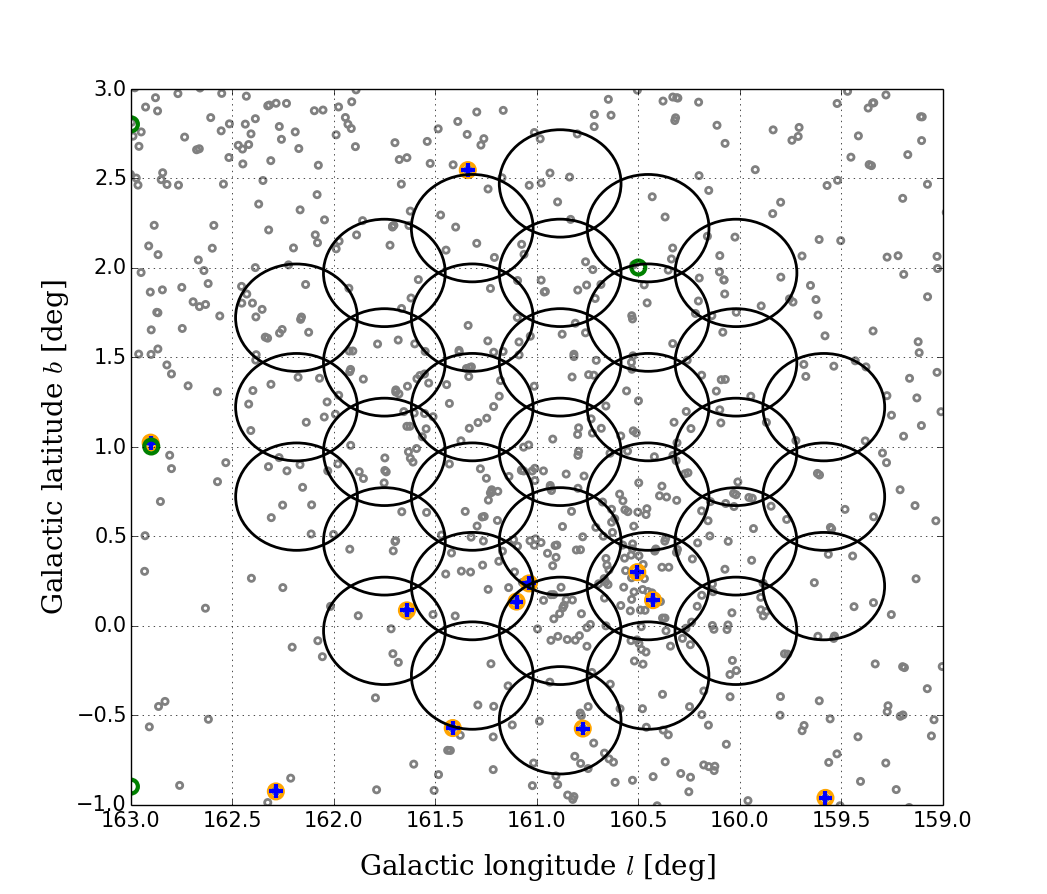}
   \caption{The projected sky distribution of the observed hexagonal mosaic. The circles indicate the HPBW of the individual WSRT pointings. The small grey circles show the 2MASX near-infrared extended sources within this region, highlighting the 3C129 cluster centred at $\ell, b \approx 160.52\dg, 0.28\dg$. The orange open circles and blue crosses mark 7 galaxies within the observed area with redshifts listed in NED and HyperLEDA, respectively. Galaxies with \textsc{Hi} detections from the NRT survey are shown in green open circles.}\label{wsrtmos}
\end{figure}

Observations were carried out using the WSRT between August and November 2012. We imaged a mosaic comprising 35 fields arranged in a hexagonal grid covering 9.6~sq.deg, each separated by $\Delta = 0\fdg5$. This setup is similar to that of a single AperTIF phased array feed pointing, minus two fields. In Fig.~\ref{wsrtmos}, we present the layout of the mosaic with positions of galaxies from the literature shown in this field. We chose this particular mosaic grid to be able to cover the largest possible field of view with an acceptable noise variation of 20\% between pointing centres. The total sky area covered is large enough to map the network between the PPS filament and the 3C129 cluster.

Each field was observed for a total on-source integration time of $2 \times 6$ hours, in between half-hour long pre- and post-calibrations. We obtained data in eight IF-bands per pointing, each 10 MHz wide, with 256 channels in dual polarisation mode, overlapping by 16\% in frequency. This configuration covered a total effective bandwidth of 67 MHz with 1717 channels that are 39 kHz (8.25 \kms) wide at $z$ = 0. We therefore cover a velocity range of $cz = 2400 - 16600$ \kms\ corresponding to a volume depth of about 214 Mpc. Therefore we map our main target with acceptable velocity resolution while still being able to survey the more distant universe. The total volume surveyed is 26300 Mpc$^3$.

The survey configuration allows the detection of galaxies with \HI\ masses greater than M$_{\text{HI}}$ = 3.0 $\times$ 10$^{8}$M$_{\odot}$ at the median distance of the PPS ($cz \approx$ 6000 \kms), assuming a line width of 150 \kms\ at the $6\sigma$ noise level such that we could attain acceptable statistics to map the cluster.

The observational parameters for the 35 pointings are summarised in Table~\ref{obs_summary}. Column 1 gives the pointing number. Columns 2 \& 3 and columns 4 \& 5 are the central equatorial and Galactic coordinates of the pointings, respectively. The synthesised beam sizes and position angles are given in columns 6 and 7, respectively. In column 8 the average rms noise for each pointing is given. 
Column 9 gives the flux densities of the brightest continuum sources in each pointing.

\begin{table*}
\small
\caption {Observing parameters of the WSRT PP\,ZoA survey.}\label{obs_summary}
\begin{tabular}{lcccrcrcll}
\hline

Field       & R.A (J2000)               & Dec (J2000)            &  $\ell$                    &  \textit{b}                     &Synth. beam& PA                         & rms noise                & Cont.  sources                    & \\
 
            &       hh:mm:ss.ss              &        dd:dm:ds.ss              &$\dg$      & $\dg$      & $\arcsec \times \arcsec$& $\dg$  & mJy/beam                &  mJy                                       & \\
(1)         &  (2)                   &(3)                     &(4)                                      &   (5)                                    &(6)                                      &(7)            & (8)                             & (9)                                          &\\
\hline

 01$^{a}$			&  			04:49:34.85 &       45:02:32.89   &       160.452 & 0.220    &21.2 $\times$ 15.2 & 2.0 &  0.37 &  4001, 935      &      \\
 02$^{b}$			&  			04:47:25.35 &       44:43:13.83   &       160.452 & $-$0.280 &22.9 $\times$ 16.1 & 1.3 &  0.41 &  4001, 935      &  \\       
 03$^{c}$			& 		      04:46:54.93 &       45:12:44.14   &       160.019 & $-$0.030 &21.8 $\times$ 16.3 & 4.1 &  0.35 &  4001, 935      &   \\      
 04			&				  04:49:05.37 &       45:32:05.54   &       160.019 & 0.470    &23.2 $\times$ 16.4 & 1.6 &  0.38 &  4001, 935, 703 &                                \\
 05			&  				04:51:45.83 &       45:21:42.96   &       160.452 & 0.720    &23.4 $\times$ 16.4 & 4.7 &  0.41 &  4001, 935      &                                \\
06			&		04:52:13.84        		&		44:52:07.62        		&	160.885	&	0.470	&		23.8	$\times$	16.3		&		0.0		&			0.36			&			4001, 935			&	 	\\			
07			&		04:50:03.82        		&		44:32:59.77        		&	160.885	&	$-0.030$	&		23.3	$\times$	16.3		&		1.1		&			0.41			&			4001, 935, 414  	&	 			\\			
08			&		04:54:25.33        		&		45:11:06.33        		&	160.885	&	0.970	&		23.4	$\times$	16.6		&		1.1		&			0.38			&							&	 \\			
09			&		04:53:58.31        		&		45:40:43.68        		&	160.452	&	1.220	&		23.2	$\times$	16.6		&		1.2		&			0.38			&							&	 \\			
10			&		04:54:51.88        		&		44:41:28.59        		&	161.318	&	0.720	&		24.3	$\times$	16.7		&		1.7		&			0.39			&			426				&	 \\			
11			&		04:57:03.82        		&		45:00:15.90        		&	161.318	&	1.220	&		24.3	$\times$	16.7		&		1.6		&			0.40			&			703, 414 				&	 \\			
12			&		04:56:38.28        		&		45:29:55.52        		&	160.885	&	1.470	&		23.4	$\times$	16.3		&	  -0.1		&			0.38			&			426				&	 \\			
13			&		04:59:17.21        		&		45:18:53.53        		&	161.318	&	1.720	&		23.5	$\times$	16.3		&		0.1		&			0.38			&			426				&	 \\			
14			&		04:52:41.37        		&		44:22:31.99        		&	161.318	&	0.220	&		22.8	$\times$	17.3		&	  12.6		&			0.38			&			414				&	 \\			
15$^{d}$			&		04:51:17.33        		&		45:51:17.74        		&	160.019	&	0.970	&		22.4	$\times$	17.3		&	  14.2		&			0.40			&			703				&	 \\			
16			&		04:58:52.69        		&		45:48:34.80        		&	160.885	&	1.970	&		23.3	$\times$	16.7		&		1.0		&			0.41			&			703				&	 \\			
17			&		04:59:41.29        		&		44:49:11.95        		&	161.751	&	1.470	&		23.7	$\times$	16.7		&		1.7		&			0.38			&			426				&	 \\			
18			&		04:57:28.93        		&		44:30:36.07        		&	161.751	&	0.970	&		23.9	$\times$	16.6		&		2.0		&			0.38			&			343				&	 \\			
19			&		04:56:12.28        		&		45:59:34.63        		&	160.452	&	1.720	&		23.2	$\times$	16.7		&		1.6		&			0.39			&			414				&	 \\			
20			&		04:55:17.98        		&		44:11:50.75        		&	161.751	&	0.470	&		24.3	$\times$	16.6		&		1.1		&			0.40			&			703				&	 \\			
21			&		04:50:32.30        		&		44:03:26.40        		&	161.318	&	$-0.280$	&		24.1	$\times$	16.6		&		2.1		&			0.40			&			426, 343			&	 	 \\			
22$^{e}$			&		04:48:35.38        		&		46:01:37.44        		&	159.586	&	0.720	&		21.7	$\times$	17.0		&		5.5		&			0.45			&							&	 \\			
23			&		04:53:30.82        		&		46:10:20.35               	&	160.019	&	1.470	&		21.5	$\times$	16.3		&		4.3		&			0.42			&							&	 \\			
24			&		04:47:55.26        		&		44:13:43.16        		&	160.885	&	$-0.530$	&		24.3	$\times$	16.6		&		1.3		&			0.37			&			426				&	 \\			
25			&		04:46:23.99        		&		45:42:13.80        		&	159.586	&	0.220	&		23.5	$\times$	16.6		&		1.2		&			0.37			&			703, 414			&	 	\\			
26			&		04:58:27.75        		&		46:18:15.45        		&	160.452	&	2.220	&		23.6	$\times$	17.1		&		3.6		&			0.40			&			426				&	 \\			
27			&		05:01:32.02        		&		45:37:21.12        		&	161.318	&	2.220	&		23.9	$\times$	17.0		&		3.1		&			0.39			&			426				&	 \\			
28			&		05:01:55.06        		&		45:07:38.02        		&	161.751	&	1.970	&		23.7	$\times$	16.7		&		1.9		&			0.38			&			414				&	 \\			
29			&		05:00:04.97        		&		44:19:30.33        		&	162.184	&	1.220	&		24.0	$\times$	16.6		&		1.7		&			0.37			&			703				&	 \\			
30			&		04:57:53.61        		&		44:00:56.31               	&	162.184	&	0.720	&		24.2	$\times$	16.6		&		0.9		&			0.40			&			703				&	 \\			
31			&		04:53:08.43               	&		43:52:56.36               	&	161.751	&	$-0.030$	&		24.2	$\times$	16.6		&		1.0		&			0.40			&			426				&	 \\			
32			&		04:50:48.33        		&		46:20:51.66        		&	159.586	&	1.220	&		23.1	$\times$	16.7		&		1.2		&			0.39			&			343				&	 \\			
33			&		04:55:45.82        		&		46:29:12.97        		&	160.019	&	1.970	&		23.0	$\times$	16.7		&		0.7		&			0.38			&			414				&	 \\			
34			&		05:01:08.57        		&		46:07:03.79        		&	160.885	&	2.470	&		23.3	$\times$	16.7		&		1.2		&			0.38			&			703				&	 \\			
35			&		05:02:17.70        		&		44:37:54.77        		&	162.184	&	1.720	&		24.3	$\times$	16.8		&		0.8		&			0.37			&			426, 343				&	 \\
	  \hline
\end{tabular}
\begin{flushleft}
\small{$^{a}$3C\,129 \& 3C\,129.1 (centred)}\\
\small{$^{b,c}$3C\,129 \& 3C\,129.1 (off-centred)}\\
\small{$^{d,e}$RFI on all short baselines}
\end{flushleft}
\end{table*}

\subsection{Data Processing}
The uv-data were flagged, calibrated and Fourier transformed using the NRAO Astronomical Imaging Processing System package\footnote{http://www.aips.nrao.edu/index.shtml} (AIPS; \citealp{vanMoorsel1996}, \citealp{Greisen1990}).

\subsubsection{RFI flagging, Amplitude and Phase Calibration}
Telescope-based gain and phase corrections were determined by observing the standard calibrators 3C\,48 (J0137+331) with a flux density of 15.56 Jy and 3C\,147 (J0542+498) with a flux density of 21.58 Jy. After loading the uv-data into AIPS, the system temperature (T$_{sys}$) was checked to assess the behaviour of antennae throughout the observation, followed by applying the time-dependent T$_{sys}$ corrections. In all cases solar interference was excised from the observations by deleting visibilities from the short 9A, 9B and AB baselines for time ranges during which the Sun was above the horizon. This measure is justified by the expected small sizes of our target sources. Additionally, Radio Frequency Interference (RFI) that affected flux and phase calibrators were flagged using the \textit{clip} task which flags RFI by searching for amplitudes that are out of range. For the calibrators 3C\,48 and  3C\,147 we clipped amplitudes that were above 23 Jy and 29 Jy, respectively, after which we conducted an automated frequency dependent RFI flagging using \textit{rflag}. In total, about 5\% of all the uv-data was flagged. Gain and phase solutions were then determined and transferred to the visibility data of the science targets. The same calibrators were used for bandpass corrections. Firstly, rough bandpass (BP) solutions were determined (task \textit{bpass}) and examined. This was followed by further RFI flagging, if any, and the production of new BP solutions. The process was repeated, each time improving on the BP solutions. Once satisfied with the calibration results, we applied these to the uv-data of the science target. No self-calibration was performed.

\subsubsection{Imaging Data Cubes}
The calibrated uv-datasets were Fourier transformed into image cubes using the \textit{imagr} task. For each pointing, eight data cubes were made as per the number of IF bands. The size of each image was 512 $\times$ 512 pixels with a pixel size of $6\arcsec \times 6\arcsec$ and 256 channels. They were made using a \textit{robust} weighting scheme set to 0.4, this being the best trade-off between the WSRT beam size and rms noise. Dirty beam image cubes corresponding to the antenna patterns were made for CLEANing purposes. They were 1024 $\times$ 1024 pixels in size,  twice that of the image cubes. Since this is a blind survey, with the location of the \HI\ emission not known beforehand, no CLEANing and continuum subtraction were done before or during the imaging process. 

\subsubsection{Continuum Subtraction}
The removal of continuum sources as well as their instrumental responses from the data cubes was performed in the image plane for each of the eight IF bands in each cube using the Groningen Image Processing SYstem (GIPSY; \citealp{vanderhulst1992}). As an initial step the strongest continuum sources were masked and CLEANed. The initial CLEANing was conducted to mitigate the variability of their sidelobes from channel to channel, caused by the frequency dependent RFI flagging. This was followed by fitting a 0th-order polynomial baseline to the spectrum at each pixel including possible \HI\ emission. After this initial baseline fit and subtraction, the rms noise per channel of each continuum-subtracted channel map was inspected. Data points above and below a chosen rms noise level were clipped and excluded from successive baseline fits. This procedure was repeated, each time increasing the baseline fit order until the  3rd order was reached, whilst gradually decreasing the rms noise clip levels down to $\pm 2 \sigma$ so as to exclude the faint  \HI\ emission from the baseline fitting. The final 35 image cubes were produced by combining the eight IF bands for each pointing, excluding the first 5 and last 36 channels in each cube due to increased noise levels after the bandpass correction.

\begin{figure}
  \centering
    \includegraphics[width = 80mm, height = 50mm]{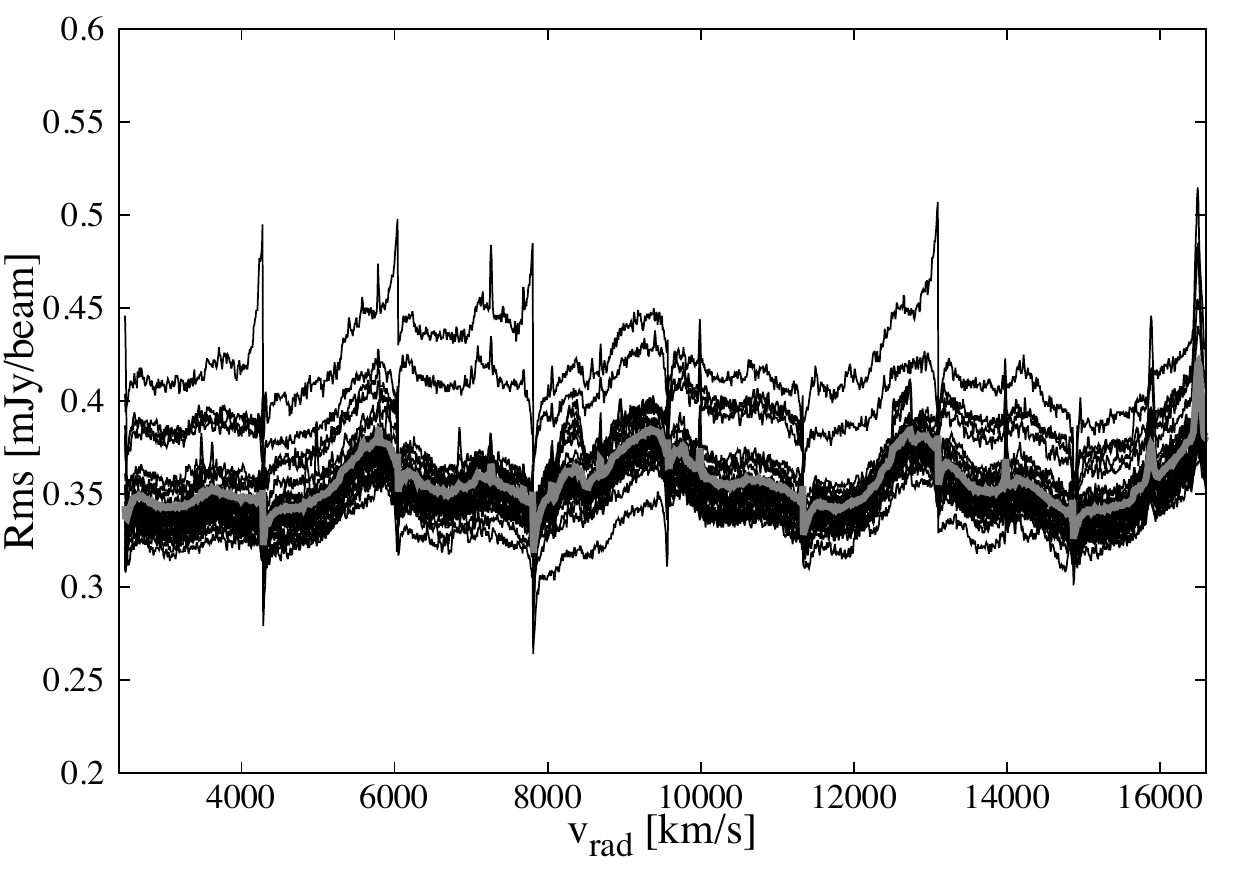}
   \caption{The rms noise as a function of velocity in the 35 cubes after continuum subtraction and RFI flagging. The average noise level in each individual cube is mostly uniform at 0.36 mJy/beam over the whole observed velocity range. The average rms noise of all the individual cubes is illustrated by the thick grey line. Variations in the rms noise are frequency-dependent and caused by the bandpass roll-off in the overlapping regions of the eight IF bands, RFI residuals and the 17 MHz standing wave ripple. The increased rms noise in two of the fields is caused by severe RFI, which resulted in complete flagging of all their short baselines.}\label{rmscubes}
\end{figure}

Figure~\ref{rmscubes} shows the rms noise as a function of velocity after continuum subtraction and Hanning smoothing in velocity. We achieved an average rms noise of 0.36 mJy/beam per channel, for a resolution of 16.5 \kms, with a frequency dependent variation of less than 20\% on average over the observed velocity range. This variation is due to RFI residuals as well as the bandpass roll-off at the edge of the IF bands and the 17 MHz standing wave ripple. Additionally 2 of the 35 pointings had severe residual RFI on all short baselines (i.e., less than 144m) which resulted in those baselines being completely flagged, hence increasing the noise in those cubes. Flagging all short baselines in these pointings decreased our inner UV coverage, which reduced the size of the largest structures to which our observations are sensitive. For instance in these fields we are sensitive to structures up to 7\arcmin\ in size, which is about 4$\times$ smaller than what we would be sensitive to without RFI on the short baselines.

\newpage
\subsubsection{CLEANing and Mosaicking}
To remove sidelobes of the synthesised beam, each of the 35 cubes were first CLEANed blindly to the $5\sigma$ noise level to remove the sidelobes from the brightest sources. The clean components  were then restored with a Gaussian beam of  $23\arcsec \times 16\arcsec$ FWHM (PA = $0\dg$). We then conducted a source finding procedure, as described in Sect.~\ref{sec3}. The procedure resulted in defining those areas within the cube that contain only the identified \HI\ signal in each channel. These areas were then used as masks to perform a new CLEAN image deconvolution down to the $0.33\sigma$ noise level. The final CLEANed cubes were combined into a mosaic to improve the sensitivity of the survey within the overlapping regions of the pointings. Mosaicking was conducted using the \textit{flatn} task in AIPS which combines image cubes produced by \textit{imagr}. Overlapping regions of images were averaged with a weighting scheme that accounts for the primary beam attenuation and the number of flagged visibilities.

\begin{figure}
  \centering
    \includegraphics[width = 85mm, height = 75mm]{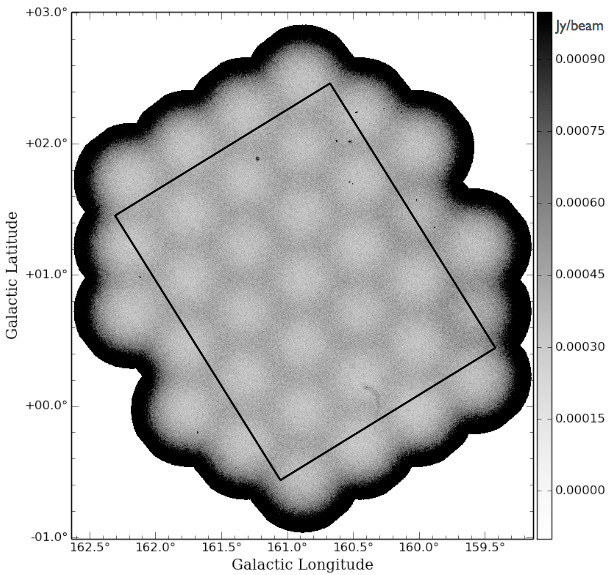}
   \caption{An illustration of the spatial noise variation over 50 channels (nos. 411 to 460, or 6903 $-$ 7307 \kms\ in radial velocity) out of the 1717 channels in our full, primary-beam corrected WSRT survey mosaic. Note the increase in the noise near the edges due to the applied correction for primary-beam attenuation. The black box outlines the inner region within which the spatial noise distributions, shown in Fig.~\ref{rms_histogram_spatial}, were calculated. The rms noise level values in Jy/beam are shown by the greyscale bar.}\label{final_mosaic}
\end{figure}

\begin{figure} 
  \centering
    \includegraphics[width = 85mm, height = 55mm]{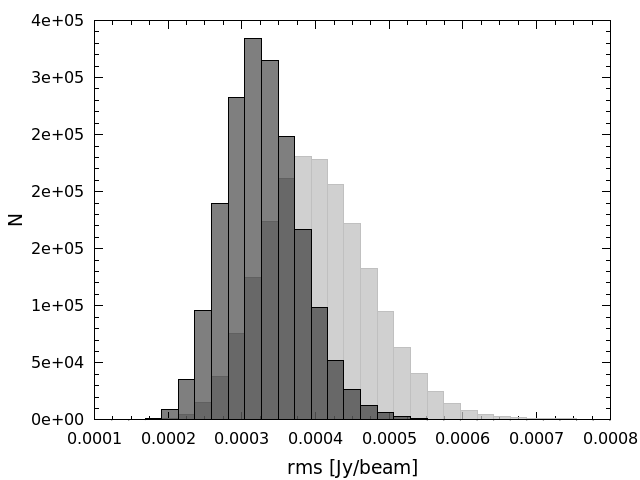}
   \caption{An Illustration of the effect of the primary beam correction on the rms noise distribution. Histograms of the rms noise distribution measured over 50 channels within the inner regions (outlined by the black box in Fig.~\ref{final_mosaic}) of our full WSRT survey mosaic. The light grey distribution is measured from the primary-beam corrected mosaic and the dark grey from a mosaic without the primary-beam corrections.}\label{rms_histogram_spatial}
\end{figure}

An example of typical spatial noise variations in a mosaicked channel map, measured for each pixel over a range of 50 channels (nos. 411 to 460, i.e., the velocity range of 6903 to 7307 \kms) in the full primary-beam corrected mosaic is shown in Fig.~\ref{final_mosaic}. In these channels the average rms noise is $\sim$0.4 mJy~beam$^{-1}$ in the central region (black box in Fig.~\ref{final_mosaic}). It increases to 0.8~mJy~beam$^{-1}$ toward the edges due to the applied correction for primary-beam attenuation. Fig.~\ref{rms_histogram_spatial} shows histograms of the spatial noise distribution for the primary-beam corrected mosaic (light grey) and the un-corrected mosaic (dark grey) in these channels as measured inside the area outlined by the black box. As expected, after applying the primary-beam correction, the peak of the distribution shifts to higher noise values while the distribution becomes broader.

\newpage
\section{Source Finding}\label{sec3}
\begin{figure}  
\centering    
\includegraphics[width = 85mm,height = 50mm]{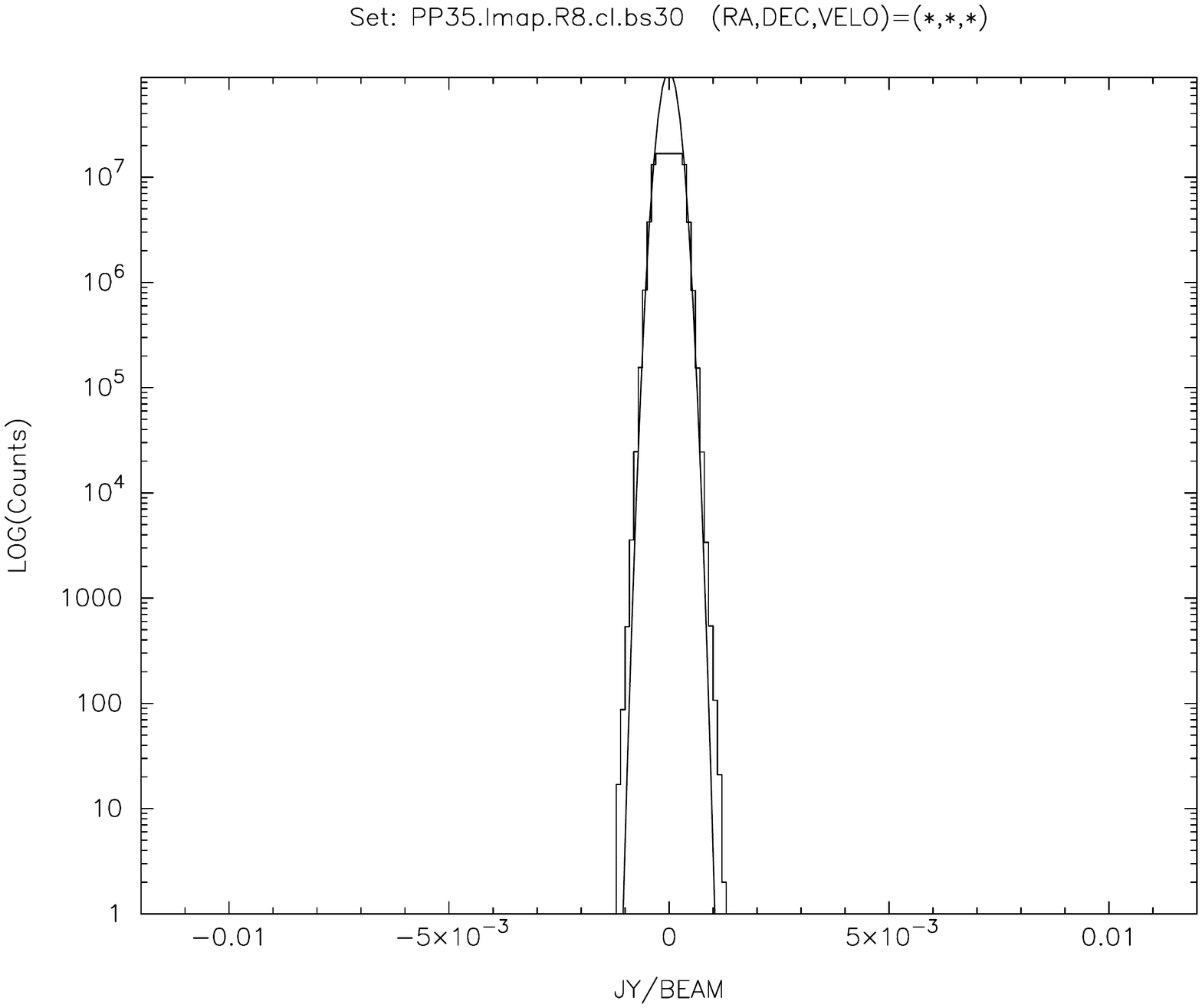}
\includegraphics[width = 85mm,height = 50mm]{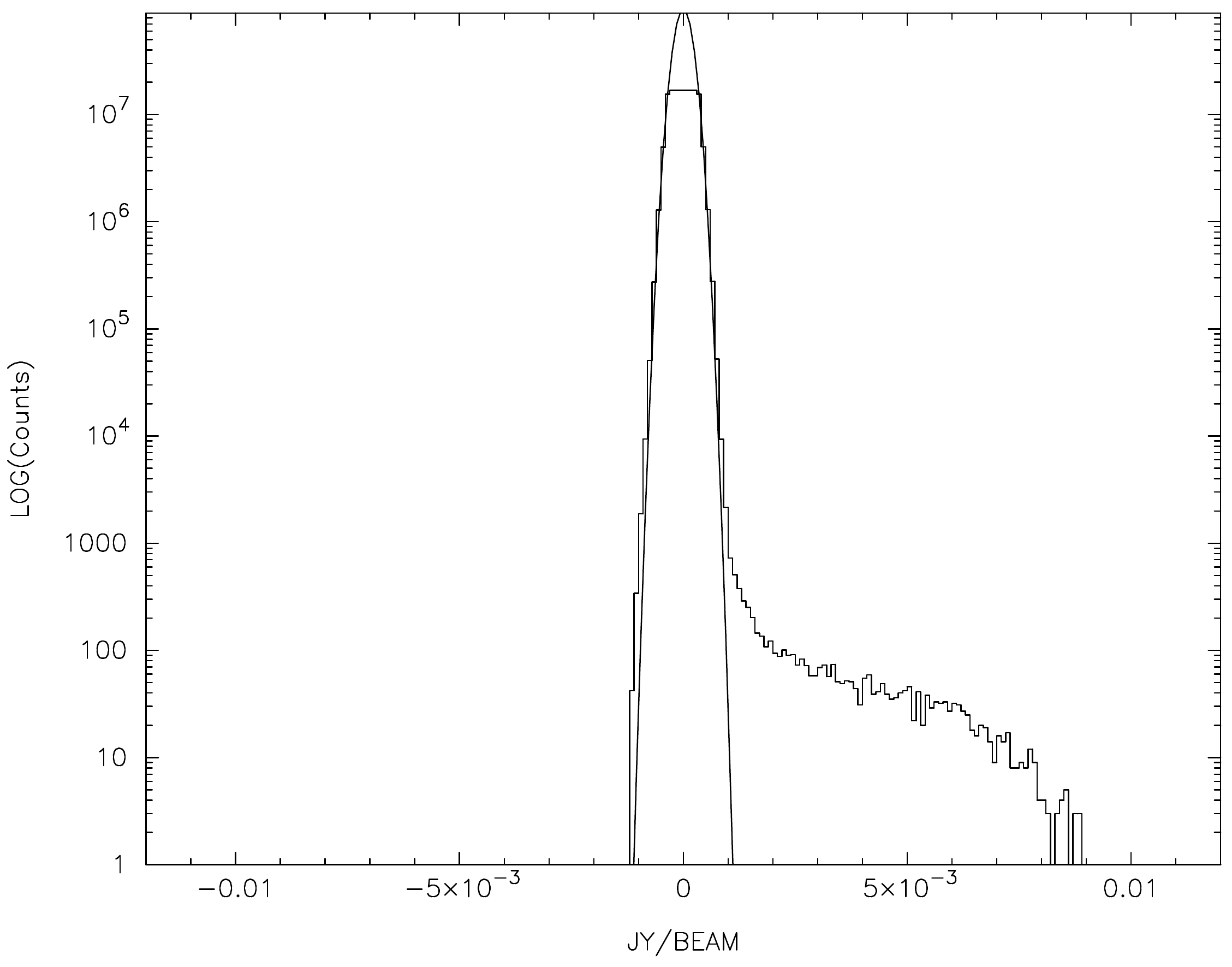}
\includegraphics[width = 85mm,height = 50mm]{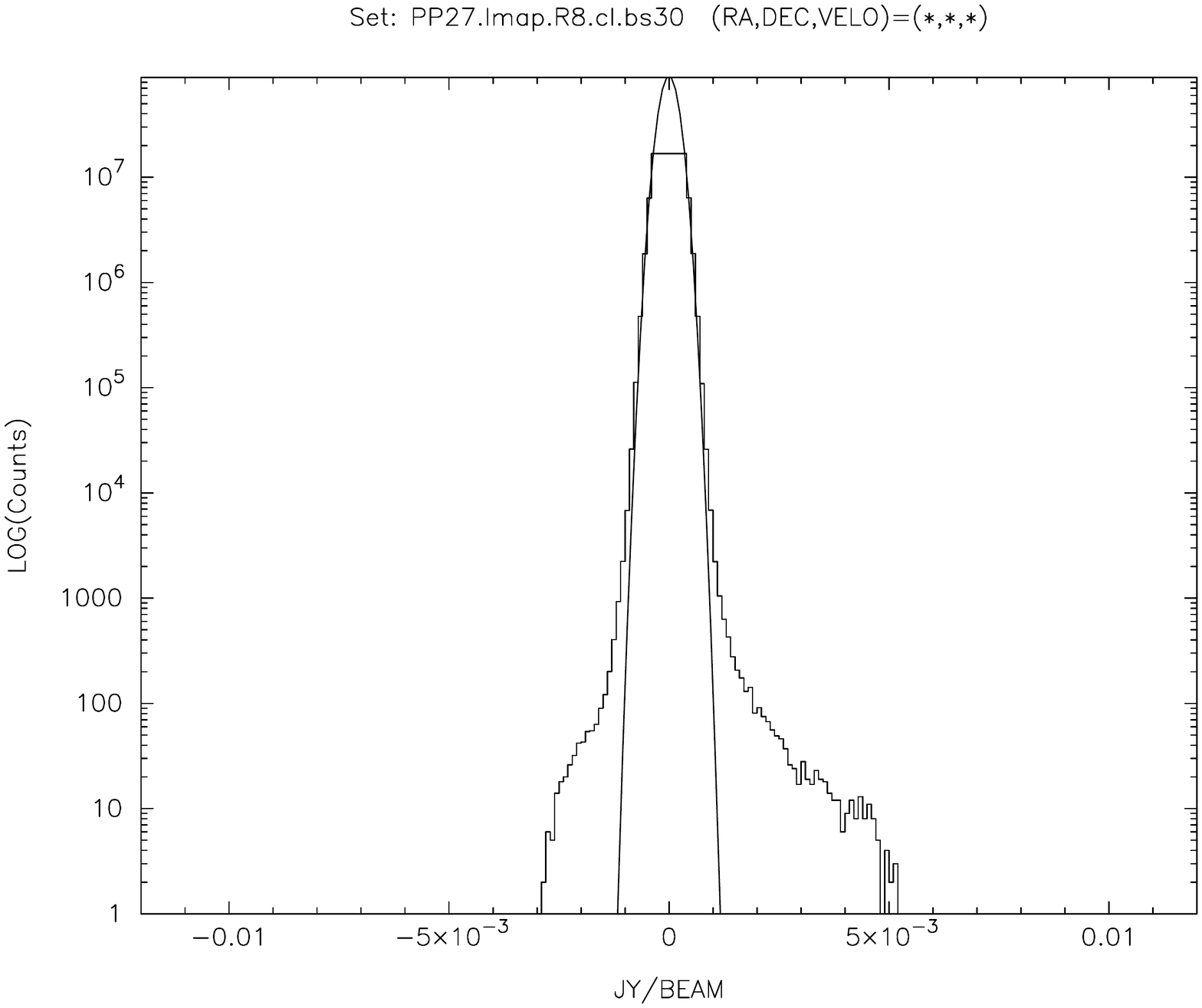}
\includegraphics[width = 85mm,height = 50mm]{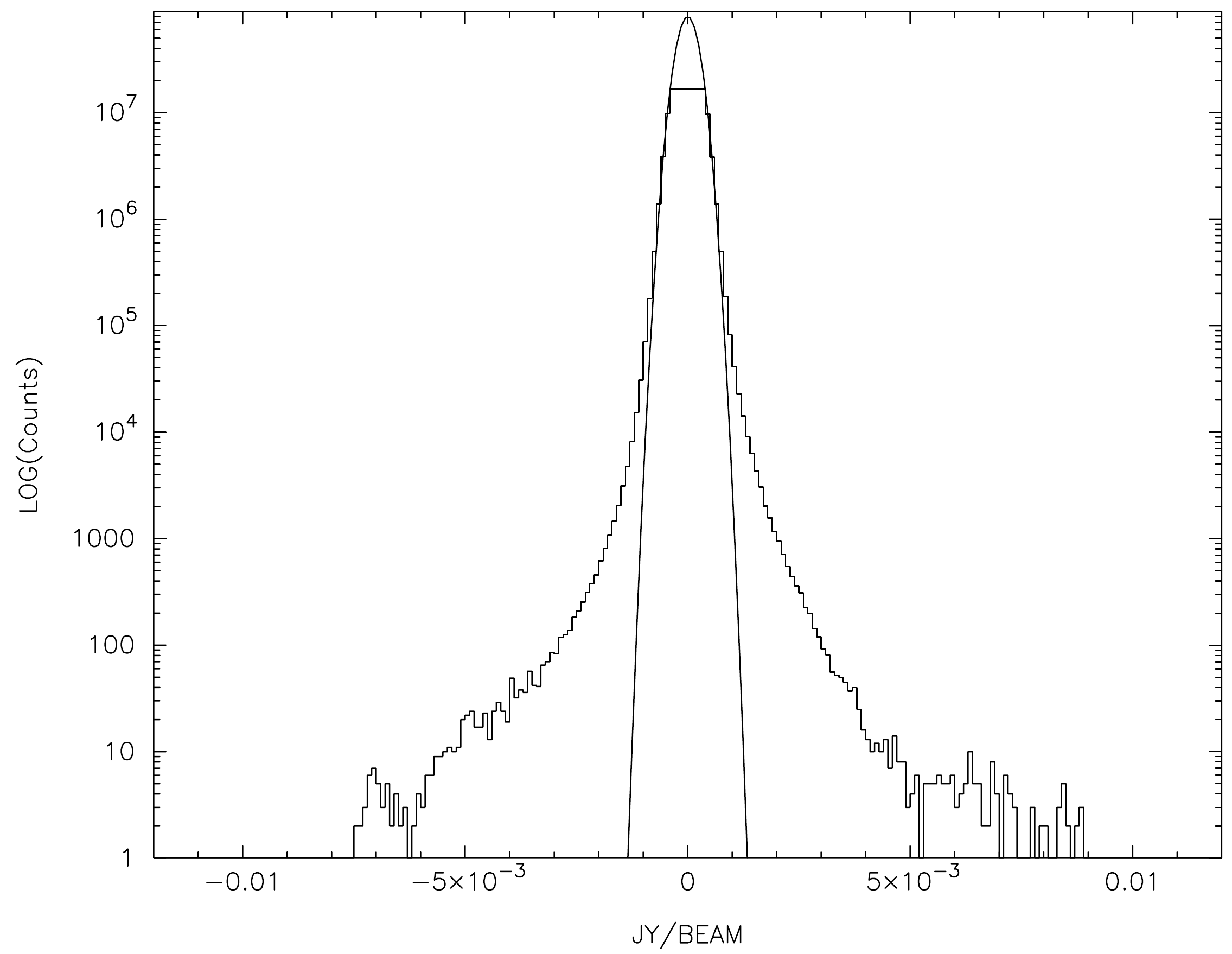}
\caption{Statistics of pixel values in individual cubes with a Gaussian curve overlaid, illustrating the four distinct characteristic cases identified. The top panel represents  cases where there are neither imaging artefacts nor \textsc{Hi} detections. In the second panel there are no imaging artefacts but many \textsc{Hi} detections. The third panel illustrates cases in which there are imaging artefacts as well as \textsc{Hi} detections. The bottom panel represents cases where there are no \textsc{Hi} detections but many imaging artefacts and RFI residuals.}\label{noisechar}
\end{figure}

\begin{figure*} 
  \centering
    \includegraphics[width = 180mm, height = 50mm]{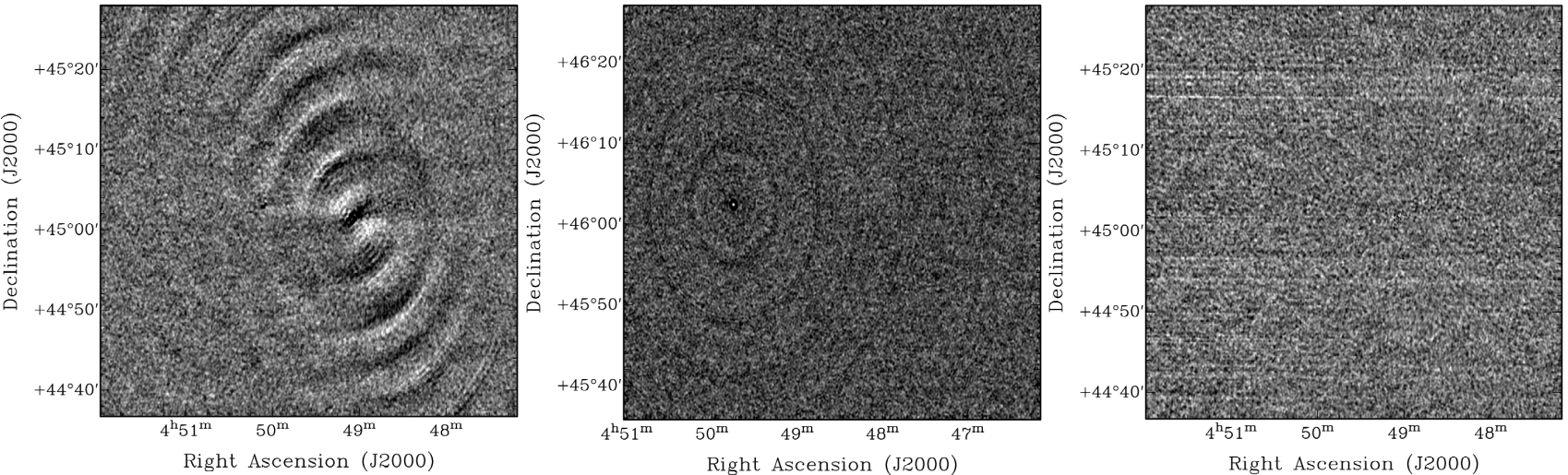}
  \caption{Examples of various types of imaging artefacts in the data cubes. The left panel shows artefacts caused by gain calibration errors over a small number of channels during half of the observation of the respective pointing. The middle panel shows artefacts due to bandpass imperfections after continuum subtraction. The right panel shows stripes caused by remaining RFI.}\label{egArtifacts}
\end{figure*}

Before compiling the list of sources detected, we investigated characteristics of the pixel value distribution in each of  the 35 individual cubes without primary beam corrections so as to avoid spatial noise variations in the mosaicked cube, as illustrated by the light grey histogram in Fig.~\ref{rms_histogram_spatial}. We found four distinct trends in the distributions as illustrated in Fig.~\ref{noisechar}. The top panel displays the behaviour of the noise when there are neither imaging artefacts nor \HI\ detections. Pixel values are then symmetrically distributed and near Gaussian. However, in cases where there are no imaging artefacts, but many \HI\ detections (second panel), the pixel distribution is heavily skewed to positive pixels. Most sources detected in cubes with these characteristics are expected to be real. The third panel shows a negative and a more extended positive wing, representing cases where there are artefacts in the cube as well as real sources. The bottom panel shows a case with RFI residuals and imaging artefacts but no \HI\ detections. In the latter two cases the wings are near symmetric and the positive peaks are unlikely to be due to real sources.
Examples of representative imaging artefacts in the data cubes are shown in Fig.~\ref{egArtifacts}. The left panel shows gain calibration errors over only about 10 channels during half of the observation. The middle panel shows the effect of an imperfect bandpass after continuum subtraction. The right panel shows horizontal stripes due to remaining RFI. There was no one dominant source that caused these artefacts.

To find \HI\ line emission from galaxies in the mosaicked cube, the spatial variation of the noise was largely removed by multiplying it by the weights cube given by (1/$\sigma^{2}$), where $\sigma$ is the weighted noise in each pointing. Source finding was then conducted in several steps using the GIPSY software. Firstly, the high resolution cube ($23\arcsec \times 16\arcsec$) was smoothed to a beam size of $30\arcsec \times 30\arcsec$ (bs30). Secondly, the high resolution and bs30 cubes were smoothed in velocity to four different resolutions, starting with Hanning smoothing (R2; 16.5 \kms) followed by a Gaussian smoothing kernel such that the FWHM of the resulting near-Gaussian spectral response function corresponds to four (R4; 33 \kms), six (R6; 49.5 \kms) and eight (R8; 66 \kms) channels, respectively. No down-sampling in spatial pixels or velocity channels was done after smoothing. Each of the 35 cubes was then searched at all eight different angular/velocity resolution combinations. Thirdly, for each resolution combination, channel maps were clipped at 8, 5, 4 and 3 times the rms noise level per channel. Lastly, a detection was accepted if it occurred at $8\sigma$ in a single velocity resolution element (i.e., in n-channels for Rn), at $5\sigma$ in 2 adjacent velocity resolution elements (2$\times$n channels for Rn), at $4\sigma$ in 3 adjacent velocity resolution elements and at $3\sigma$ in 4 adjacent velocity resolution elements. All channels within a velocity resolution element had to be above the specified noise clip level. All the pixels containing the \HI\ signal were combined into 3-dimensional masks. This automated source finding method resulted in 683 potential galaxy detection candidates.

A follow-up visual examination of all potential \HI\ detections was conducted independently by two persons. This resulted in the rejection of a total of 235 as imaging artefacts due to residual RFI and side lobes of continuum sources (see Fig.~\ref{egArtifacts} for examples). These artefacts occurred predominantly in the two pointings which had all short baselines removed due to RFI. As a result 90\% of the automatically detected sources in these two pointings were rejected. This is not too severe because the two pointings with short baselines removed represent only a negligible fraction (0.1\%) of all visibilities and (0.03\%) of our total survey volume, while the affected data are still sensitive to structures up to $\sim$7$'$ in size. 
The resulting list contained 448 remaining candidates. Visual inspection steps were taken to differentiate between real sources and those that resemble noise peaks in this list. We further removed 237 sources which were consistent with noise peaks or imaging artefacts based predominantly on their signal to local noise ratio. This led to a final detection list of 211 galaxies.

\section{Data Products of Detected HI Sources}\label{HIprofs}
In this section we describe the methods used to determine the \HI\ properties of galaxies detected. The structure of the \HI\ catalogue and the atlas of the detections is discussed in Sect.~\ref{sec4}.
 
\subsection{The Global HI Profiles}
The global \HI\ profiles were made from the primary-beam corrected flux within the channel-dependent  \HI\ emission masks of the detected galaxies at the 16.5 \kms\ velocity resolution. Each profile was then partitioned into three velocity bins of equal width, referred to as the receding ($r$), the middle ($m$) and approaching ($a$) bins. We then determined the peak flux densities ($F^{p}$) in the three velocity bins as $F^{p}_{r}$, $F^{p}_{m}$ and $F^{p}_{a}$, respectively.  The three values were used to categorise the shape of the global profile as  double-horned if $F^{p}_{r} > F^{p}_{m} < F^{p}_{a}$, Gaussian if $F^{r}_{r} < F^{p}_{m} > F^{p}_{a}$ and otherwise as asymmetric.  For the double-horned profiles the peak flux densities on both the receding (r) and approaching (a) sides were considered separately for calculating their 20\% and 50\% levels. In the Gaussian and asymmetric cases only the absolute peak flux density was used. We then linearly interpolated between the data points from the profile centre outward to determine the velocities ($v_{r,20}$, $v_{a,20}$, $v_{r,50}$, $v_{a,50}$) corresponding to levels at  20\% and 50\% of the peak flux densities. It should be noted however, that for \HI\ profiles with edges that are not consistently decreasing, this method tends to slightly underestimate the line widths.\\

\begin{flushleft}
The systemic velocity is determined according to;
\begin{equation}
 v_{sys} = 0.25(v_{r,20} + v_{a,20} + v_{r,50} +  v_{a,50})
\end{equation}
The profile line widths are measured using:
\end{flushleft}
\begin{equation}
w_{20} = v_{r,20} -  v_{a,20}
\end{equation}
\begin{equation}
w_{50} = v_{r,50} -  v_{a,50}
\end{equation}
The total \HI\ flux ($S_{\rm int}$) in Jy\,\kms\ was determined by integrating the global profiles according to;
\begin{equation}
S_{\rm int} = \Sigma S_{\rm \nu} \Delta v 
\end{equation}
where $S_{\rm \nu}$ is the \HI\ flux density in Jy at each channel and $\Delta v$ is the channel width in \kms.\\

For a particular galaxy, especially if it is spatially resolved, the size, shape and position of the \HI\ clean mask changes from channel to channel due to the rotating \HI\ disc. Thus the uncertainty in the flux also varies from channel to channel. To calculate the uncertainty in the flux of a global \HI\ profile at each channel, we replicated the clean mask that encloses the \HI\ emission and projected this at eight different line-free positions surrounding the detection. We then measured the signal in each of the eight projected masks and defined the uncertainty in the line flux from the galaxy as the rms scatter in the eight flux measurements in the projected masks. 

\subsection{Total HI maps}\label{totalHImaps}
The total \HI\ maps, showing the \HI\ column density distributions were made using the CLEAN masks that define the \HI\ emission from the CLEANed data cubes. These data cubes were corrected for primary beam attenuation. Pixels outside the CLEANed mask were set to zero and pixels inside the masks were summed up to build the \HI\ maps. The advantage of this method is an attained higher signal-to-noise ratio at some pixels of the \HI\ map. It does however, result in a non-uniform noise distribution across the map such that the $3\sigma$ column density level cannot be defined. We converted the pixel value of the \HI\ map to column densities (in atoms cm$^{-2}$) using;

\begin{flushleft}
\begin{equation}
N_{\rm HI} = 1.82 \times 10^{18} \int{T_{\rm b}} dv 
\end{equation}

where $T_{b}$ is the brightness temperature in Kelvin and $dv$ is the channel width in \kms\ of the integrated \HI\ emission line. For our survey, a typical $N_{\rm HI}$ sensitivity at the 3$\sigma$ level is $5.3\times10^{19}$ cm$^{-2}$ where $dv$ = 16.5 \kms.\\

The total \HI\ mass of a galaxy (in \MSUN) was determined by calculating: 

\begin{equation}
M_{\rm HI} = 2.36 \times 10^{5}D^{2}\int S_{\rm \nu} dv 
\end{equation}

where $D$ is the distance to the galaxy in Mpc measured according to:
\begin{equation}
D = \frac{v_{sys}}{H_{\rm 0}} 
\end{equation}

where H$_{0}$ is the Hubble constant and $ \int S_{\rm \nu} dv$ is the integrated global \HI\ profile flux in Jy\,\kms.
\end{flushleft}

We fitted a 2-dimensional Gaussian function to the \HI\ map of each of the 211 detections. We then estimated the \HI\ size ($D_{\rm HI}$) as defined by the fitted Gaussian's major ($\Theta_{x,g}$) and minor ($\Theta_{x,g}$) axis FWHM in arcseconds such that $D_{\rm HI} = \sqrt{\Theta_{x,g} \times \Theta_{y,g}}$. The \HI\ size ($D_{\rm HI}$) was then compared to that of the WSRT beam, $D_{\rm WSRT} = \sqrt{\Theta_{x} \times \Theta_{y}}$, where $\Theta_{x}$ and $\Theta_{y}$ are the beam FWHMs of 23\arcsec and 16\arcsec, respectively. An \HI\ detection was defined as spatially resolved if $D_{\rm HI}$ exceeded $1.5 \times D_{\rm WSRT}$, otherwise it is considered marginally resolved. The resulting position angles of the fitted Gaussians were not corrected for the minimal effect of the slightly elongated synthesised beam which is oriented exactly North-South. The distribution of fitted position angles, however, does not show an excess of North-South orientations.

\subsection{Radial Column Density Profiles}
For the spatially resolved galaxies, the radial \HI\ column density profiles were extracted from the total \HI\ maps in which the pixel column density units were converted to \MSUN pc$^{-2}$, by azimuthally averaging non-blank pixels in concentric elliptical annuli with radii along the semi-major axis of 5, 15, 25, 35,...,105\arcsec\ with a width of 10\arcsec\ and centred on the \HI-centroid position of the galaxy. This position was defined as the centroid of the 2-dimensional Gaussian function fitted to the total \HI\ map as described in Sect.~\ref{totalHImaps}.  No corrections for beam smearing were applied. We averaged over the receding and approaching side separately to identify possible asymmetries. From the radial profiles, the \HI\ radius (R$_{\text{\rm HI}}$) in kpc was measured as the radius where the mean surface density ($\Sigma_{\rm HI}$) drops to 1 \MSUN pc$^{-2}$.

\subsection{HI Velocity Fields}
The velocity fields of the spatially resolved \HI\ discs were produced by fitting a Gaussian to the R2.bs30 velocity profile at each pixel. The fitting algorithm used initial estimates for the amplitude, central velocity and velocity dispersion of the Gaussian function. The initial amplitude estimate simply is the peak flux density, whereas the initial estimates for the velocity centroid and dispersion are based on the flux-weighted first and second moment of the non-blank pixel profile, respectively.  The advantage of fitting Gaussians instead of measuring the standard flux-weighted moments is that we can specify ranges within which the fitted parameters and their uncertainties must lie in order for a fit to be accepted. Fits were accepted only if 1) the central velocity lies within the \HI\ mask, 2) the uncertainty in the velocity centroid is less than 10 \kms, and 3) the velocity dispersion lies within the range of $5 - 50$ \kms. This approach, however, does result in velocity fields that are patchy with holes where the signal-to-noise ratio is too low to obtain an acceptable fit.

\subsection{Position-Velocity Diagrams}
A position-velocity diagram (PVD) illustrates the shape of a projected rotation curve and the presence of a possible kinematic asymmetry of the rotating gas disc. Ideally, a PVD should be extracted along the kinematic major axis of an inclined rotating disc and centred on its dynamical centre. Usually, this geometry is derived from 2-dimensional velocity fields which are available only for the spatially resolved galaxies in our survey. Hence we followed different strategies for the resolved and marginally resolved galaxies.

For the PVDs of the resolved galaxies in our survey we estimated by eye the position angle of the receding side of the kinematic major axis by considering the direction perpendicular to the kinematic minor axis indicated by the green isovelocity contour of the velocity field depicted in the accompanying atlas. No detailed analysis of the velocity fields was carried out as this is planned for a forthcoming publication. The dynamical centres of the rotating gas discs were not derived from their 2-dimensional velocity fields but instead were assumed to coincide with the morphological centre of the UKIDSS image (see section ~\ref{SecOnCP}). In those cases where an infrared counterpart could not be identified, the dynamical centre was assumed to coincide with the centroid of the Gaussian function fitted to the \HI\ column density map.

For the PVDs of the marginally resolved galaxies we adopted the centroid and position angle of the 2-dimensional Gaussian fitted to the \HI\ column density maps. In those cases where an infrared counterpart could be identified, we adopted the morphological centre of the UKIDSS image as the dynamical centre of the PVD. In case the UKIDSS image showed a clear elongation, we estimated by eye the position angle of the infrared image and adopted this as the kinematic major axis. For the marginally resolved galaxies, uncertainties in the exact position angle and dynamical centre have little effect on the PVD.

We extracted the PVDs from the Hanning smoothed cubes (R2) as well as from the cubes smoothed to 66 \kms\ resolution (R8), both at the 23\arcsec $\times$ 16\arcsec\ angular resolution

\newpage
\subsection{Counterparts}\label{SecOnCP} 
The sky area covered by our survey has a Galactic foreground extinction in the $B$-band of up to A$_{B} \sim$ 4.5 mag. This makes optically identifying galaxies very difficult. However, the near-infrared wavelength opens a more transparent window since the extinction is less severe. We searched for NIR counterparts in the $K$-band images of the UKIRT Infrared Deep Sky Survey (UKIDSS; \citealp{Hewett2006} \citealp{Lawrence2007}, \citealp{Casali2007}, \citealp{Hambly2008}). "UKIDSS uses the UKIRT Wide Field Camera (WFCAM; \citealt{Casali2007} and a photometric system described in \citealt{Hewett2006}. The pipeline processing and science archive are described in Irwin et al (2008) and \citealp{Hambly2008}. We used data from the 10$^{th}$ Galactic Plane Survey (GPS) data release, which is described in detail in \citealt{Lucas2008}"\footnote{http://www.ukidss.org/}. 
The WSRT PP\,ZoA survey overlaps with the UKIDSS GPS \citep{Lucas2008} which has mapped 1800 deg$^2$ of the northern GP to a $K$-band depth of $K$ = 19.0 mag (AB system).

The UKIDSS images were searched in two steps. Firstly, the UKIDSS GPS catalogue was searched for galaxy counterpart candidates that were within 1\arcmin\ radius of the \HI\ positions. Only candidates closest to the \HI\ positions were considered as likely counterparts. These were then verified by a visual inspection of UKIDSS GPS images within the search radius in the $J, H$ and $K$ bands (pixel size of  0.2\arcsec). Secondly, to ensure that no potential counterparts were missed, the search radius was increased to 2\arcmin. For this step, only those \HI\ positions without a verified galaxy counterpart in the first step were cross-checked and verified by visual inspection as well. In some cases a UKIDSS image could not be used due to imaging artefacts. We then searched for 2MASS images instead, in the same manner as for UKIDSS.

Of all \HI\ detected galaxies, 62\% were found to have at least one near-infrared counterpart in the UKIDSS GPS catalogue, with an average offset of 2.6\arcsec\ from the \HI\ position. The other 38\% did not show any plausible counterpart close to the \HI\ position.

Additionally, we also searched the Wide-field Infrared Survey Explorer (WISE; Wright2010) images for counterparts, by inspecting the W1 ($3.4~\mu m$), W2 ($4.6~\mu m$) and W3 ($12~\mu m$) colour composite images. For this search we examined images 3\arcmin $\times$ 3\arcmin\ in size, specially constructed for the field using the method described in \citet{Jarrett2012}. Only those galaxies closest to the \HI\  position were considered and verified by eye as counterparts. In total 47\% of the galaxies detected in \HI\ were found to have a WISE counterpart.

\begin{figure}  
  \centering
    \includegraphics[width = 90mm, height = 100mm]{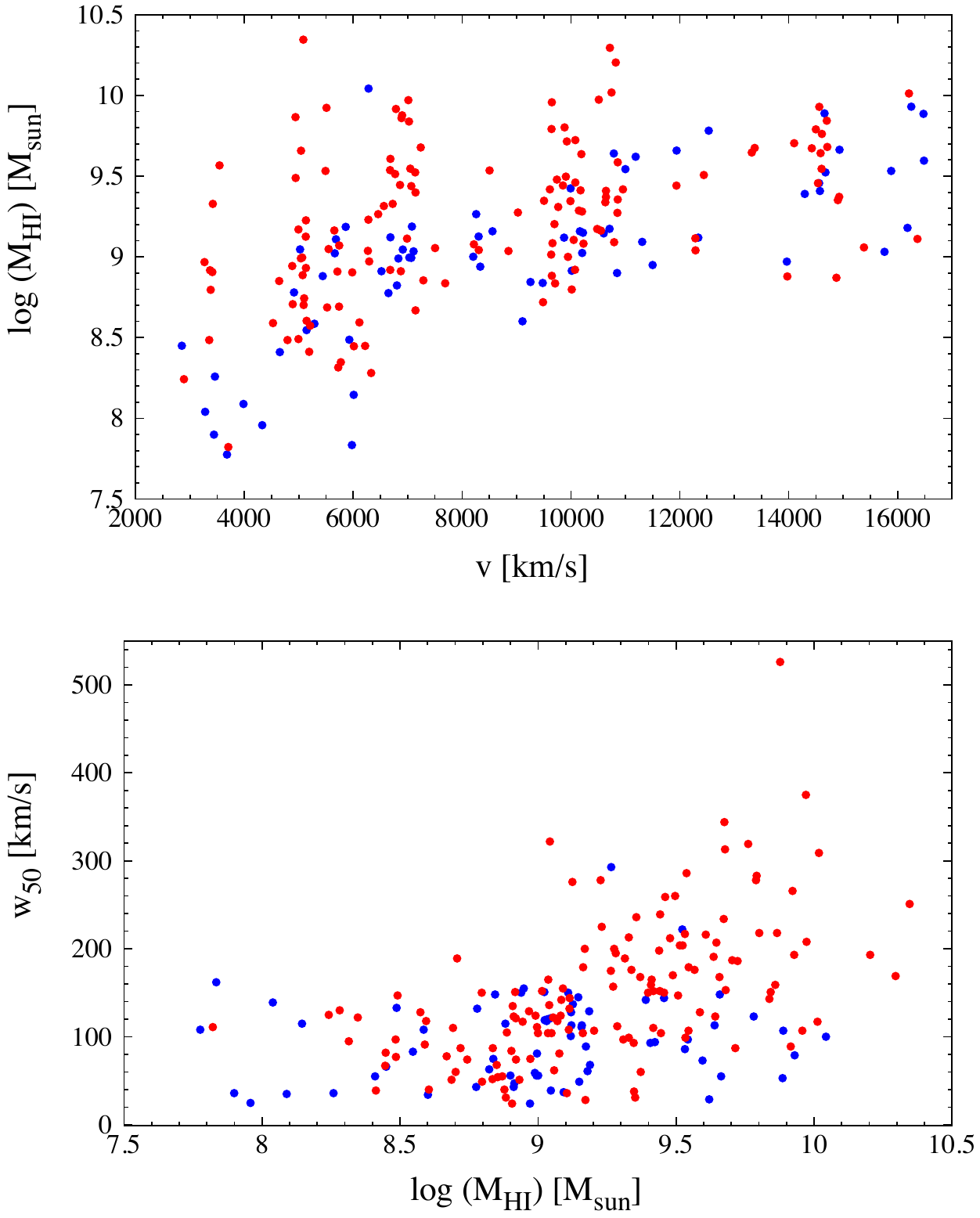}
   \caption{A comparison between global \textsc{Hi} properties of detected galaxies with near-infrared counterparts (red) and without (blue). The top panel shows the total \textsc{Hi} mass as a function of  radial velocity. The bottom panel shows the $w_{50}$ line width as a function of total \textsc{Hi} mass.}\label{counterpart}   
\end{figure}

In a further step, we investigated \HI\ properties that might have affected the recovery of a UKIDSS counterpart. They are displayed in Fig.~\ref{counterpart} where we plot the total \HI\ mass of all detected galaxies as a function of velocity and their $w_{50}$ line width as a function of \HI\ mass. Detections with near-infrared counterparts are shown in red and those without a counterpart in blue. We find relatively more counterparts for \HI-massive log($M_{\text{HI}}$/\MSUN) = 9.5 galaxies with large line widths ($w_{50} >150$ \kms). We also note that most galaxies below log($M_{\text{HI}}$/\MSUN) = 8.4 do not have identifiable near-infrared counterparts. These low \HI-mass detections are usually gas-rich, low surface brightness galaxies and are not easily detectable in the near-infrared.

\begin{figure}
  \centering
    \includegraphics[width = 90mm, height = 75mm]{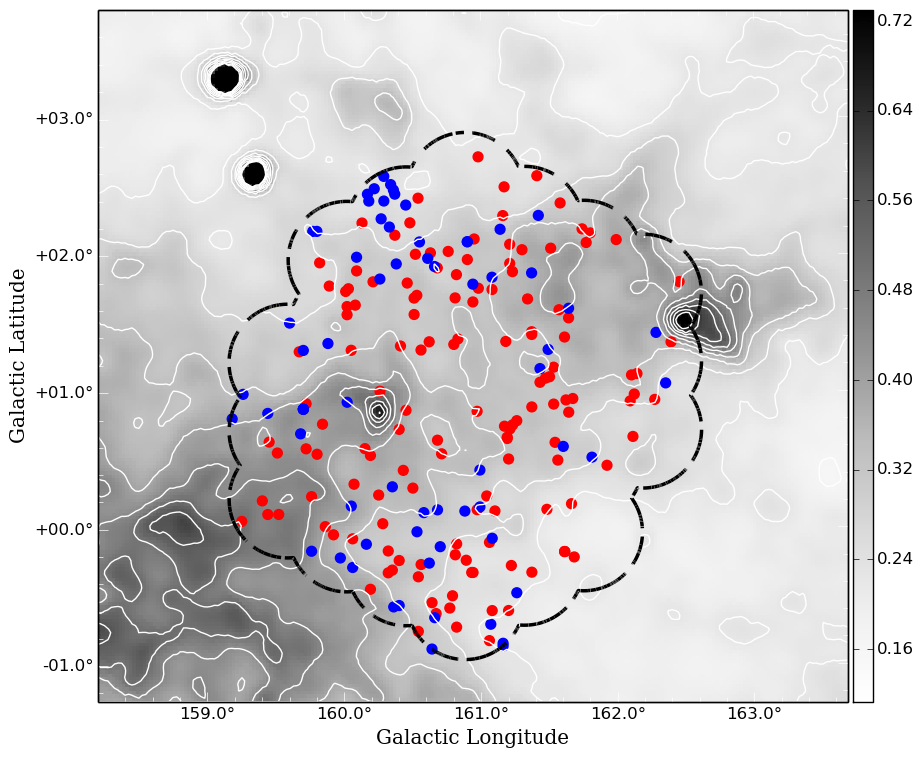}
   \caption{The spatial distribution of galaxies detected in \textsc{Hi}. Galaxies with a near-infrared counterpart are shown in red, those without a counterpart in blue. They are superimposed on a DIRBE/IRAS map of $A_{K}$ Galactic extinction values in the NIR $K$ band, ranging from 0.16 to 0.72 mag as shown on the greyscale bar. These correspond to $A_{B}$ = 1.8 to 8.0 mag, respectively in the $B$ band optical.}\label{dustmap}
\end{figure}

In Fig.~\ref{dustmap} we overlay the spatial distribution of the \HI\ detections on a $K$-band Galactic extinction ($A_{K}$) map from the DIRBE/IRAS\footnote{http://irsa.ipac.caltech.edu/applications/DUST/} data (\citealp{Schlegel1998}, \citealp{Schlafly2011}). Galaxies with near-infrared counterparts are indicated in red and those without in blue. The observed WSRT mosaic is outlined by the black dashed contour. The plot indicates that finding a galaxy counterpart at $\ell \approx 160\dg$ behind the GP does not seem to depend on Galactic extinction. The small group of \HI\ detections without a near-infrared counterpart in the top left of the plot is due to poor UKIDSS imaging in that area. 

\section{The HI catalogue and atlas}\label{sec4}
\subsection{The HI catalogue}
The derived \HI\ parameters of all detected galaxies are listed in tables presented. The first table (A1) lists the resolved galaxies and the second table (A2) the marginally resolved galaxies (see Sect.~\ref{totalHImaps} for details on how sources were divided into these two categories). The first 10 entries of each of these tables are shown in Tables~\ref{segHIprop} and \ref{segHIpropun} to illustrate their structure. The column entries of the 211 \HI\ detections are as follows:

\begin{description}
\item[\textbf{Column (1)}:] Galaxy designated identification based on the Right Ascension and Declination (J\,2000.0) of the \HI\ centroid.\\
\item[\textbf{Column (2)\&(3)}:] Galactic longitude and latitude in degrees.\\ 
\item[\textbf{Column (4)}:]  Systemic velocity in the barycentric-standard of rest (BSR) defined by the central velocity in \kms\ of the global \HI\ profile.\\
\item[\textbf{Column (5) \& (6)}:]  Observed \HI\ profile line widths and their uncertainties (in \kms) measured at 20\% and 50\% level of the \HI\ profile peak flux, respectively, determined at a velocity resolution of 16.5 \kms.\\
\item[\textbf{Column (7)}:] The integrated \HI\ flux ($S_{\rm int}$) corrected for primary beam attenuation, in Jy\,\kms.\\
\item[\textbf{Column (8)}:] Luminosity distance $D$ to the galaxy in Mpc derived from the systemic velocity measured in the BSR.\\
\item[\textbf{Column (9)}:] The total \HI\ mass in \MSUN.\\  
\item[\textbf{Column (10)}:] The \HI\ radius measured at the 1\,\MSUN\,pc$^{-2}$ surface density level ($R_{\text{\rm HI}}$) in kpc, given for resolved sources only.\\
\item[\textbf{Column (11)}:] Indication of an infrared counterpart, found in either UKIDSS (u), WISE (w) or 2MASS (m). 
\end{description}


\newpage
\onecolumn
\begin{centering}
\begin{landscape}
\begin{longtable}{lrrrllrrcrcc} 
\caption[]{{A sample table showing the structure of the catalogue for resolved galaxy detections. The full table is available as supplementary material.}}\label{segHIprop}\\

\hline \\[-1.8ex]
\thead{ZoA}              & \thead{$\ell$}   & \thead{$b$}   &\thead{$v_{rad}$}  & \thead{$w_{20}$}     &  \thead{$w_{50}$} & \thead{$S_{\rm int}$} & \thead{$D$}      &\thead{log\,($M_{\text{HI}}$)} &\thead{$R_{\text{HI}}$}  & \thead{counterpart} \\
\\
                         &  \thead{deg}     & \thead{deg}   & \thead{\kms}      &     \thead{\kms}     &   \thead{\kms}    &  \thead{Jy \kms}  &   \thead{Mpc}  & \thead{\MSUN}                 & \thead{kpc}             &             \\
\\
\thead{(1)}              &  \thead{(2)}     & \thead{(3)}   & \thead{(4)}       &     \thead{(5)}      &     \thead{(6)}   &   \thead{(7)}     &   \thead{(8)}  & \thead{(9)}                   & \thead{(10)}            & \thead{(11)}          \\

\hline \\[-1.8ex]
\endfirsthead

\hline \\[-1.8ex]
\endhead

\hline \\[-1.8ex]
\\
\endfoot

\hline
\endlastfoot
J044427.17+455116.7	&	159.25	&	0.06	&	5506	$\pm$	04	&		285	$\pm$	14	&		266	$\pm$	09	&	5.68	$\pm$	0.41	&	79	&	9.9	&	24.2	&	u~~$-$\\										
J044521.10+454432.8	&	159.44	&	0.11	&	5134	$\pm$	03	&		290	$\pm$	07	&		278	$\pm$	07	&	1.34	$\pm$	0.10	&	73	&	9.2	&	12.4	&	u~~$-$\\										
J044542.87+442101.0	&	160.54	&	-0.75	&	10717	$\pm$	01	&		198	$\pm$	04	&		169	$\pm$	03	&	3.57	$\pm$	0.10	&	153	&	10.3	&	33.5	&	u~~$-$\\										
J044602.33+443426.8	&	160.40	&	-0.56	&	6281	$\pm$	03	&		117	$\pm$	10	&		100	$\pm$	06	&	5.78	$\pm$	0.45	&	90	&	10.0	&	31.0	&	$-$~~$-$\\										
J044632.27+452152.2	&	159.86	&	0.02	&	9740	$\pm$	02	&		226	$\pm$	06	&		212	$\pm$	08	&	0.66	$\pm$	0.05	&	139	&	9.5	&	17.4	&	u~~w\\										
J044644.10+442004.0	&	160.67	&	-0.62	&	5654	$\pm$	05	&		194	$\pm$	13	&		179	$\pm$	13	&	0.94	$\pm$	0.09	&	81	&	9.2	&	12.7	&	u~~w\\										
J044644.74+444734.7	&	160.32	&	-0.32	&	5709	$\pm$	03	&		156	$\pm$	09	&		135	$\pm$	10	&	0.51	$\pm$	0.05	&	82	&	8.9	&	9.9	&	u~~$-$\\										
J044700.10+442439.7	&	160.64	&	-0.54	&	10747	$\pm$	02	&		330	$\pm$	07	&		309	$\pm$	05	&	1.86	$\pm$	0.09	&	154	&	10.0	&	35.8	&	u~~w\\										
J044706.91+453449.0	&	159.76	&	0.24	&	4994	$\pm$	03	&		216	$\pm$	08	&		200	$\pm$	06	&	1.24	$\pm$	0.06	&	71	&	9.2	&	14.0	&	u~~w\\										
J044727.30+445342.4	&	160.32	&	-0.16	&	5523	$\pm$	02	&	~\,69	$\pm$	06	&	~\,51	$\pm$   06	&	0.33	$\pm$	0.03	&	79	&	8.7	&	7.9	&	u~~$-$\\

\hline
\end{longtable}

\begin{longtable}{lrrrllrrcc} 
\caption[]{A sample table for marginally resolved galaxy detections. The full table is available as supplementary material.}\label{segHIpropun}\\

\hline \\[-1.8ex]
\thead{ZoA}              & \thead{$\ell$}   & \thead{$b$}   &\thead{$v_{rad}$} & \thead{$w_{20}$} & \thead{$w_{50}$} & \thead{$S_{\rm int}$} & \thead{$D$}      &\thead{log\,($M_{\text{HI}}$)} & \thead{counterpart} \\
\\
                         &  \thead{deg}     & \thead{deg}   & \thead{\kms}     &     \thead{\kms} &   \thead{\kms}   &  \thead{Jy \kms}  &   \thead{Mpc}  & \thead{\MSUN}                 &             \\
\\
\thead{(1)}              &  \thead{(2)}     & \thead{(3)}   & \thead{(4)}      &     \thead{(5)}  &     \thead{(6)}  &   \thead{(7)}     &   \thead{(8)}  & \thead{(9)}                   & \thead{(10)}          \\

\hline \\[-1.8ex]
\endfirsthead

\hline \\[-1.8ex]
\hline \\[-1.8ex]
\endhead

\hline \\[-1.8ex]
\\
\endfoot

\hline
\endlastfoot
J044524.14+451924.3	&	159.76	&	-0.16	&	6830	$\pm$	06	&	108	$\pm$	18	&	~\,59	$\pm$	16	&	0.43	$\pm$	0.07	&	98	&	9.0	&	$-$~~$-$\\
J044533.70+441127.9	&	160.64	&	-0.88	&	2852	$\pm$	01	&	~\,90	$\pm$	02	&	~\,66	$\pm$	03	&	0.71	$\pm$	0.05	&	41	&	8.4	&	$-$~~$-$\\
J044540.97+454045.8	&	159.52	&	0.11	&	9660	$\pm$	02	&	153	$\pm$	06	&	142	$\pm$	08	&	0.27	$\pm$	0.03	&	138	&	9.1	&	u~~$-$\\
J044541.51+455022.6	&	159.40	&	0.21	&	14584	$\pm$	03	&	217	$\pm$	09	&	123	$\pm$	07	&	0.43	$\pm$	0.05	&	208	&	9.6	&	u~~$-$\\
J044546.04+444901.5	&	160.19	&	-0.44	&	5134	$\pm$	05	&	~\,94	$\pm$	34	&	~\,51	$\pm$	10	&	0.68	$\pm$	0.09	&	73	&	8.9	&	u~~$-$\\
J044550.59+443552.5	&	160.36	&	-0.57	&	16249	$\pm$	05	&	~\,95	$\pm$	13	&	~\,79	$\pm$	16	&	0.67	$\pm$	0.10	&	232	&	9.9	&	$-$~~$-$\\
J044556.60+450741.4	&	159.97	&	-0.21	&	6803	$\pm$	03	&	~\,82	$\pm$	14	&	~\,63	$\pm$	08	&	0.30	$\pm$	0.04	&	97	&	8.8	&	$-$~~$-$\\
J044559.91+450104.1	&	160.06	&	-0.28	&	12528	$\pm$	05	&	171	$\pm$	10	&	123	$\pm$	42	&	0.80	$\pm$	0.13	&	179	&	9.8	&	$-$~~$-$\\
J044630.07+451655.0	&	159.92	&	-0.04	&	10229	$\pm$	05	&	170	$\pm$	12	&	124	$\pm$	18	&	0.24	$\pm$	0.04	&	146	&	9.1	&	u~~w\\

\hline
\end{longtable}

\end{landscape}
\end{centering}

\twocolumn
\subsection{The HI atlas}
\begin{figure} 
  \centering
    \includegraphics[width = 80mm, height = 40mm]{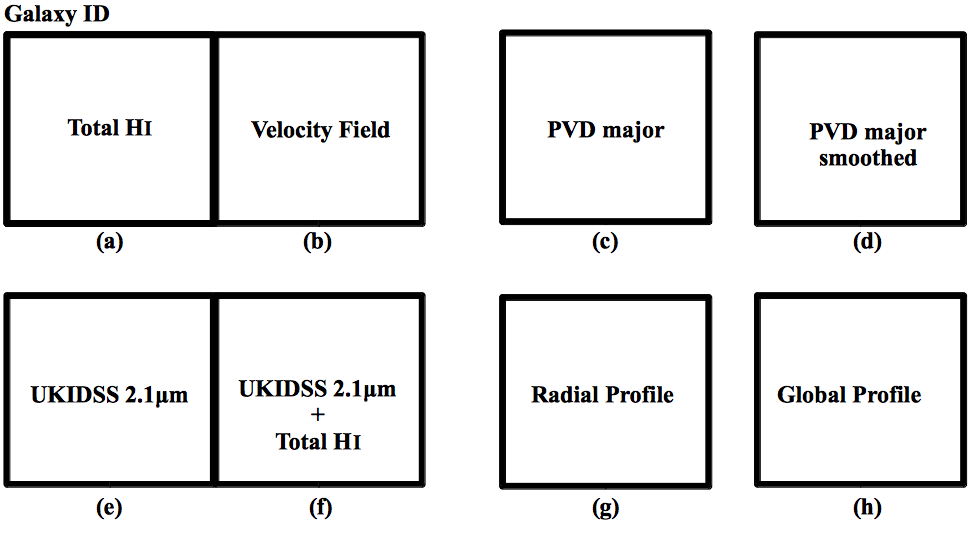}
   \caption{Schematics of the panel arrangement of the derived \textsc{Hi} data products for resolved galaxies.}\label{Atlas_Schematic_big}  
\end{figure}

\begin{figure} 
  \centering
    \includegraphics[width = 80mm, height = 15.5mm]{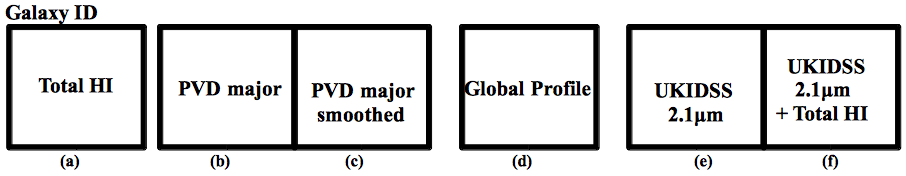}
   \caption{Schematics of the panel arrangement of the derived \textsc{Hi} data products for marginally resolved galaxies.}\label{Atlas_Schematic_small}
   \end{figure}

The \HI\ data products described in Sect. 4 are presented here for all detected galaxies. They are shown in an \HI\ atlas composed of two sections. The first section shows the resolved galaxies (one galaxy per page). The second provides maps of the marginally resolved galaxies (three per page). Schematics of the display panels are given in Figs.~\ref{Atlas_Schematic_big} and ~\ref{Atlas_Schematic_small}, respectively. An example page of the resolved and marginally resolved cases is presented in Figs.~\ref{example_big} and ~\ref{example_small}, respectively, the full atlas is available in supplementary material online ($http://www.ast.uct.ac.za/$$\sim$mpati/). The panels presented in the respective atlases are:\\

\large{\textbf{Total \HI\ map:}} The total \HI\ maps are presented at full angular resolution in panel (a) of Figs.~\ref{Atlas_Schematic_big} and ~\ref{Atlas_Schematic_small}. The \HI\ column density contour levels are at 1, 2, 4, 8, 16, 32... $\times$ 10$^{20}$ atoms/cm$^{2}$. The FWHM beam size is indicated by the hatched ellipse. The dashed line indicates the direction of the morphological major axis determined as the position angle of the fitted 2-dimensional Gaussian. The adopted \HI\ centre position is indicated by a small white cross.\\

\large{\textbf{Velocity fields:}} In panel (b) of Fig.~\ref{Atlas_Schematic_big} the \HI\ velocity fields are presented at an angular resolution of 30 arcseconds (not shown for marginally resolved galaxies in Fig.~\ref{Atlas_Schematic_small}). Pixels shown in grey scales are where the radial velocity was measured. Lighter greyscales and blue contours show the approaching side, darker and red contours the receding side of the rotating \HI\ disc. The systemic velocity as derived from the global \HI\ profile ($v_{sys}$) is shown by the green contour. Contours are drawn from $v_{sys}$ at intervals of 5, 10, 15, ..., 30, 35 \kms\ depending on the width of the global \HI\ profile. The dashed line indicates the adopted kinematic major axis and the white circle indicates the location of the adopted dynamical centre. The FWHM beam size is indicated by the hatched circle.\\ 

\large{\textbf{Position velocity diagrams:}} Shown in panels (c) and (d) of Fig.~\ref{Atlas_Schematic_big} and (b) and (c) of Fig.~\ref{Atlas_Schematic_small} are the major-axis position-velocity diagrams (PVDs) at full angular resolution (23\arcsec and 16\arcsec). The panels show the PVD at a velocity resolution of 16.5 \kms\ (left) and 66 \kms\ (right), respectively. The position angle is indicated in the top left corner of each panel. The horizontal dashed line indicates the systemic velocity and the vertical dashed line the adopted centre of rotation. The red contour outlines the \HI\ emission clean mask within which the integrated flux was determined. Contour levels  are -3 (dashed), -1.5 (dashed), 1.5, 3, 4.5, 6, 9, 12, 15, 20, 25, 30 times the local rms noise level.\\

\large{\textbf{Near-Infrared images:}} Panels (e) of Fig.~\ref{Atlas_Schematic_big} and Fig.~\ref{Atlas_Schematic_small} respectively show the $1\arcmin \times 1\arcmin$ and $0.5\arcmin \times 0.5\arcmin$, near-infrared UKIDSS GPS $K$-band ($2.1~\mu m$) images (\citealp{Lucas2008}, \citealp{Lawrence2007}). For cases where an UKIDSS image was unavailable due to imaging artefacts, a 2MASX $Ks$-band ($2.1~\mu m$) image is shown instead. The white cross indicates the \HI\ centre position. In panels (f) of Figs.~\ref{Atlas_Schematic_big} and Fig.~\ref{Atlas_Schematic_small}, we overlaid the contours of the total \HI\ map on a $2.5\arcmin \times 2.5\arcmin$ sized version of the near-infrared image, to show the extent of \HI\ in comparison to the NIR component of the galaxy. For the resolved galaxies, the white cross and circle indicate the \HI\ position and adopted dynamical centre respectively. For the marginally resolved galaxies, the dashed line, if present, indicates the position angle of the identified near-infrared counterpart.\\

\large{\textbf{Radial density profile:}} The radial \HI\ column density profiles, corrected for a face-on orientation, are shown only for the resolved galaxies in panel (g) of Fig.~\ref{Atlas_Schematic_big}. The red and blue points indicate the receding and approaching side respectively. The connected black points indicate the azimuthally averaged profile. The \HI\ radius $R_{\text{HI}}$ where the azimuthally averaged column density has dropped to 1 \MSUN pc$^{-2}$ is indicated by the vertical arrow. No correction for beam-smearing has been applied.\\

\large{\textbf{Global Profile:}} Figures.~\ref{Atlas_Schematic_big} and~\ref{Atlas_Schematic_small}, panels (d) and (h), show the global \HI\ profiles. The connected black dots give the primary-beam corrected integrated \HI\ flux density in each channel at a velocity resolution of 16.5 \kms. The grey thick line is the \HI\ profile derived from cubes that have been smoothed to 66 \kms. Indicated errors are derived as described in Sec.~\ref{HIprofs}. The vertical arrow points to the systemic velocity.\\


\begin{figure*}\begin{sideways}\begin{minipage}{24.4cm}\includegraphics[width = 250mm,height = 130mm]{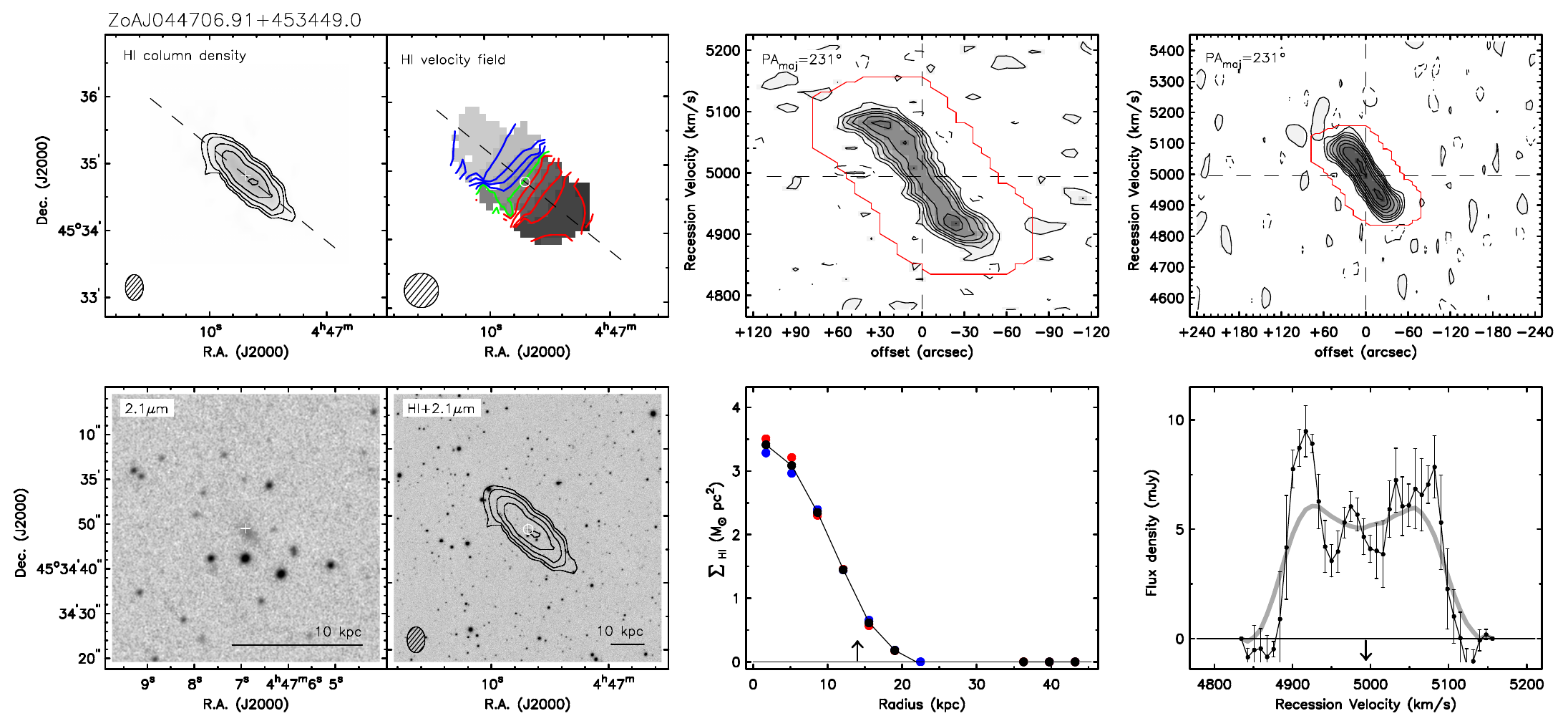}\caption{A sample atlas page for resolved sources.}\label{example_big}\end{minipage}\end{sideways}\centering\end{figure*} 

\begin{figure*}\begin{sideways}\begin{minipage}{24.4cm}\includegraphics[width = 250mm,height = 45mm]{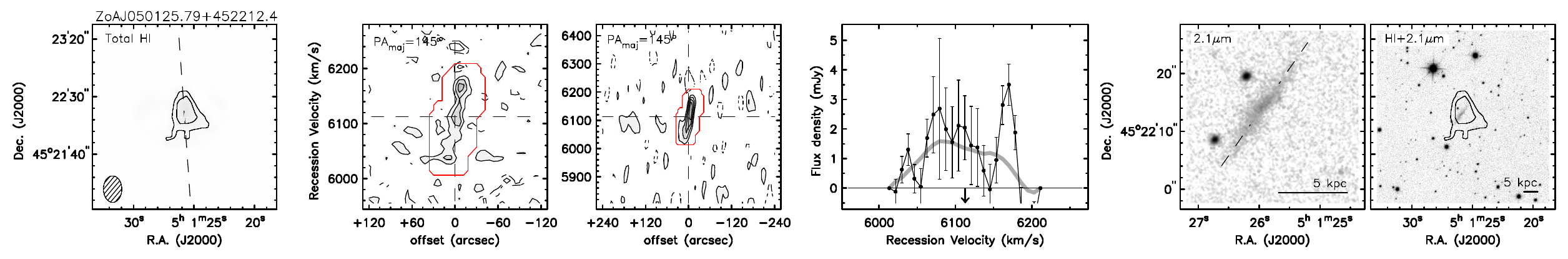}\\
\includegraphics[width = 250mm,height = 45mm]{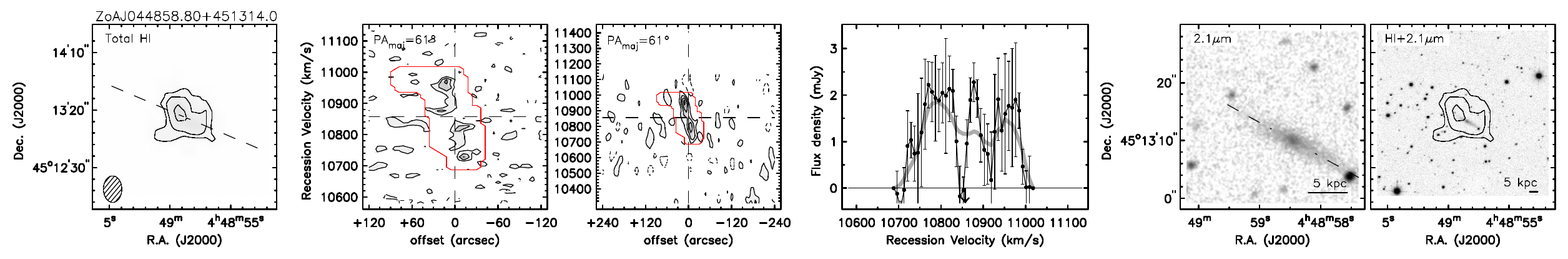}\\
\includegraphics[width = 250mm,height = 45mm]{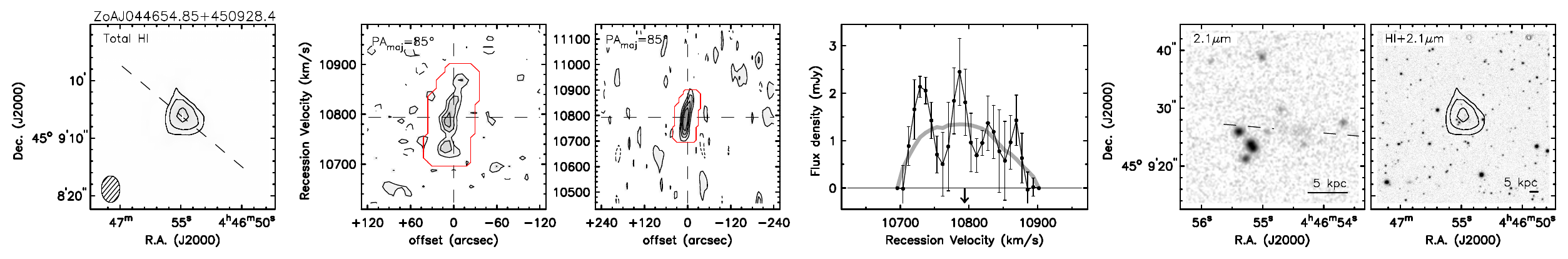}\\\caption{A sample atlas page for marginally resolved sources.}\label{example_small}\end{minipage}\end{sideways}\centering\end{figure*} 

\newpage
\subsection{Previous HI Detections}
We cross-correlated our detections with \HI\ detections reported in the literature. This was carried out using the NASA/IPAC Extragalactic Database (NED) and the Lyon/Meudon Extragalactic Database (HyperLeda). Using positional and velocity information, both databases were searched for galaxies previously detected in \HI\  over our entire volume. We also made a comparison with the unpublished results of the NRT 2MASX \HI\ follow-up survey (Kraan-Korteweg et al., in prep.).
Three galaxies were found, all of which had been observed with the NRT. All three were also detected in our WSRT survey. It is not surprising that only three galaxies were found in these databases, given the paucity of previous observations and detections over our WSRT survey area. The two galaxies with published \HI\ data were initially identified in optical searches by Weinberger (1980) and subsequently observed in \HI\  (Chamaraux et al. 1990, Paturel et al. 2003). The third galaxy was detected in the NRT 2MASX \HI\ follow-up survey.

\begin{figure}
  \hspace*{-0.7cm}
    \includegraphics[width = 90mm, height = 120mm]{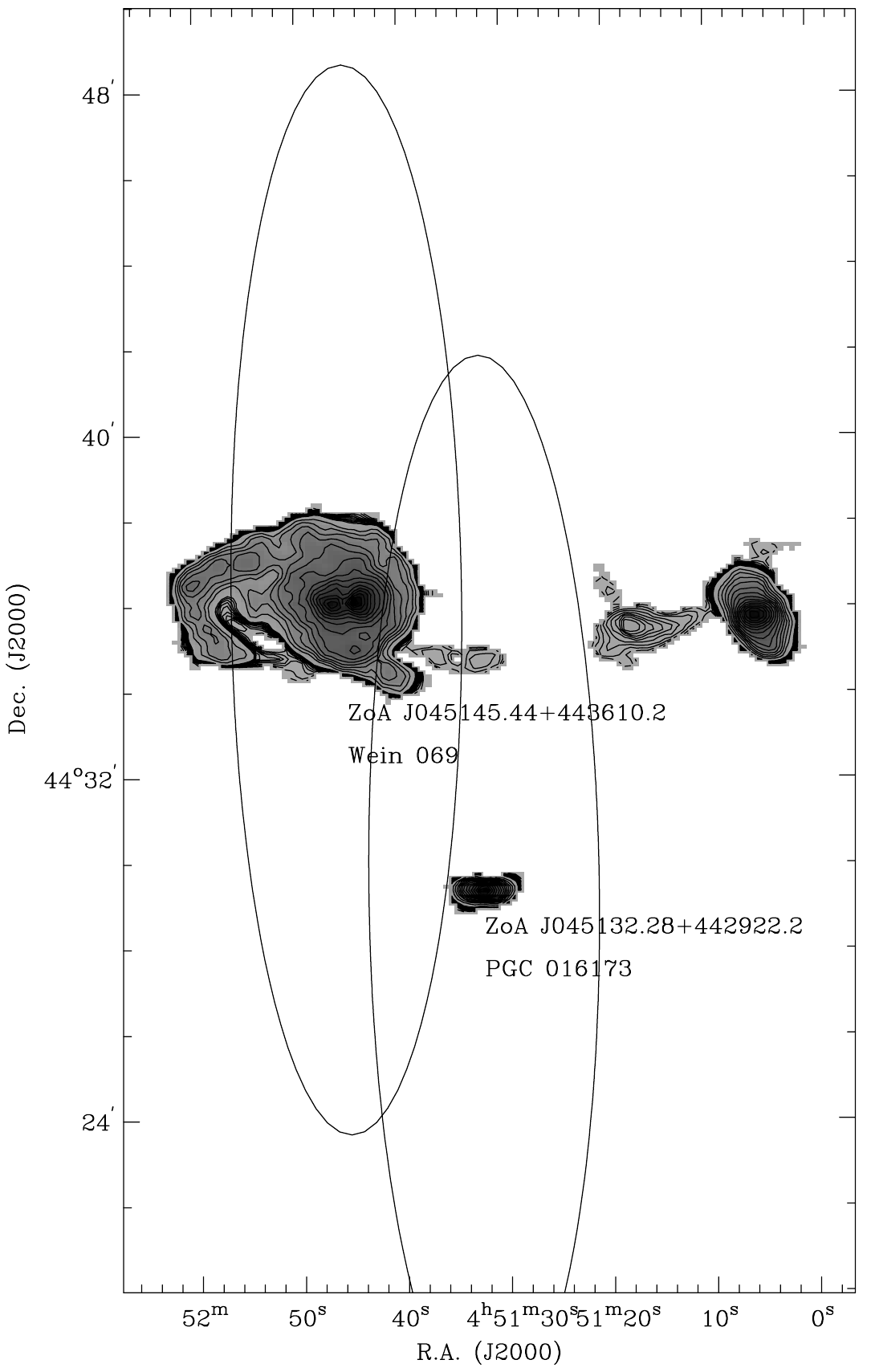}
   \caption{An illustration to show why our measured WSRT flux of ZoA\,J045132.28+442922.2 is six times lower than the NRT value reported by Paturel et al. (2003). The WSRT total \textsc{Hi} maps of ZoA\,J045145.44+443610.2 (Wein~069, on the left) and ZoA\,J045132.28+442922.2  (PGC~016173, at the bottom) with the NRT FWHM beam sizes of the reported literature detections overlaid. The NRT detection of the latter is confused by the former, which is much stronger.  In addition, we detected ZoA\,J045105.4+443544.0 at a similar recession velocity on the right.}\label{HIcounter}   
\end{figure}

The \HI\ parameters measured by them and those of our survey are given in Table~\ref{compHIparms}. The respective values of velocity and line widths of all three galaxies are in good agreement. The integrated fluxes ($S_{\rm int}$) of ZoA\,J045747.05+460717.5 and its literature counterpart are also in good agreement. However, our WSRT integrated flux measurement of ZoA\,J045132.28+442922.2 is six times lower than that of PGC~016173 as measured by \citet{Paturel2003}. To investigate this discrepancy, the relevant WSRT \HI\ data cube was carefully examined by eye. No indication of any missed \HI\ emission was found at that location. We then searched for a nearby, possibly confusing source within the large NRT beam (22\arcmin\ FWHM in Declination). Its presence is indeed confirmed and illustrated in Fig.~\ref{HIcounter} which shows the \HI\ column density map of the targeted galaxy in a larger field of view for the radial velocity range of $4752 - 5576$ \kms\ together with the respective NRT beams. In this total \HI\ map our WSRT \HI\ detection cross-identified with PGC~016173 corresponds to the southernmost galaxy in this group. It is obvious that the NRT beam picked up excess \HI\ flux from the nearby \HI-bright galaxy Wein~069 at top left, with an integrated flux as measured with the WSRT of 17.6 Jy\,\kms, explaining the higher NRT flux measurement.

For the third detection in common, ZoA\,J045145.44+443610.2 or Wein~069, we measured a slightly higher total integrated flux than Chamaraux et al (1990). This galaxy is in fact the large easternmost source in Fig.~\ref{HIcounter}. Its \HI\ diameter is large (72.8 kpc), resulting in an incomplete flux measurement for the narrow NRT beam. 

We cannot provide any further statistical comparison with the literature, given the fact that only three previous detections were cross-identified with our detections.

\begin{table}
\caption {A comparison of  \textsc{Hi} profile parameters with literature values.}\label{compHIparms}
\scriptsize
\begin{tabular}{llllll}
\hline
 ZoA& $v$  & $w_{50}$       & $w_{20}$       & $S_{\rm int}$ \\  
     &\kms\ &\kms\        &\kms\            & Jy\kms\        \\            
(1)&(2) & (3)            &(4)             &  (5)         \\         

\hline\hline
PGC~016173 (Wein~072) [1]& 5201              & 255                &323                &6.70                     \\
J045132.28+442922.2&5129$\pm$4 &276$\pm$11 &305$\pm$11& 1.06$\pm$0.07 \\
 &                       &                       &                   &                                 \\
Wein~069 [2]&5178&--                    &285             &14.60                        \\
J045145.44+443610.2&5086$\pm$1&251$\pm$2  &269$\pm$1& 17.63$\pm$0.32  \\
 &                       &                       &                   &                                  \\
04574731+4607167 [3]&   7199$\pm$9          &      370          &    408        & 4.27$\pm$0.56                         \\                   
J045747.05+460717.5& 7013$\pm$1 & 375$\pm$3       &    393$\pm$3   &     3.96$\pm$0.11\\  
\hline\hline\\
\end{tabular}
\small{[1] \citealp{Paturel2003}}\\
\small{[2] \citealp{Chamaraux1990}}\\
\small{[3] Kraan-Korteweg et al, in prep.}
\end{table}

\clearpage
\normalsize
\subsection{Completeness}
The completeness of our catalogue is defined as the fraction of sources detected by the survey from an underlying sample distribution. As a first estimate for this paper we use the integrated flux to estimate the completeness based on an empirical approach. Our method is based on inserting artificial sources throughout the surveyed volume and determining the rate at which they are recovered with our source finding scheme. 

The artificial galaxies were based on simulations by \citet{Obreschkow2009} who for about $3 \times 10^{7}$ galaxies evaluated the cosmic evolution of the atomic and molecular phases of the cold gas. Their simulations were based on the Semi-analytic Suite of SKA Simulated Skies (S$^{2}$-SAX; \citealp{DeLucia2007}) and are built on physical models applied to the semi-analytical model of evolving galaxies for the Millennium simulations \citep{Springel2005}. They cover a redshift range of $z = 0 - 1.2$, a sky area field of 10 deg$^2$ and comprise galaxies with \HI-peak flux $\geq$ 1$\mu$Jy \citep{Obreschkow2009b}.

For our survey volume the simulations predict \HI\ properties for 1183 galaxies spanning \HI\ masses and integrated fluxes ranging from log(M$_{\rm HI}$/\MSUN) = 6.05$ - $10.26 and $S_{\rm int}$ = 0.001$ - $4.149 Jy\,\kms, respectively. The set of evaluated \HI\ properties of each galaxy were used to build a detailed 3-dimensional model of the spatial and spectral distribution of the \HI\ line emission. All of the 3-dimensional models were inserted in a single large synthetic cube matching our survey volume and the distribution of large-scale structures. This cube was then spatially smoothed to match the 23\arcsec $\times$ 16\arcsec\ beam of our WSRT observations. This final synthetic cube served as our noise-free sky model which was subsequently added to our observed data cube.  

\begin{figure}  
  \centering
    \includegraphics[width = 88mm, height = 105mm]{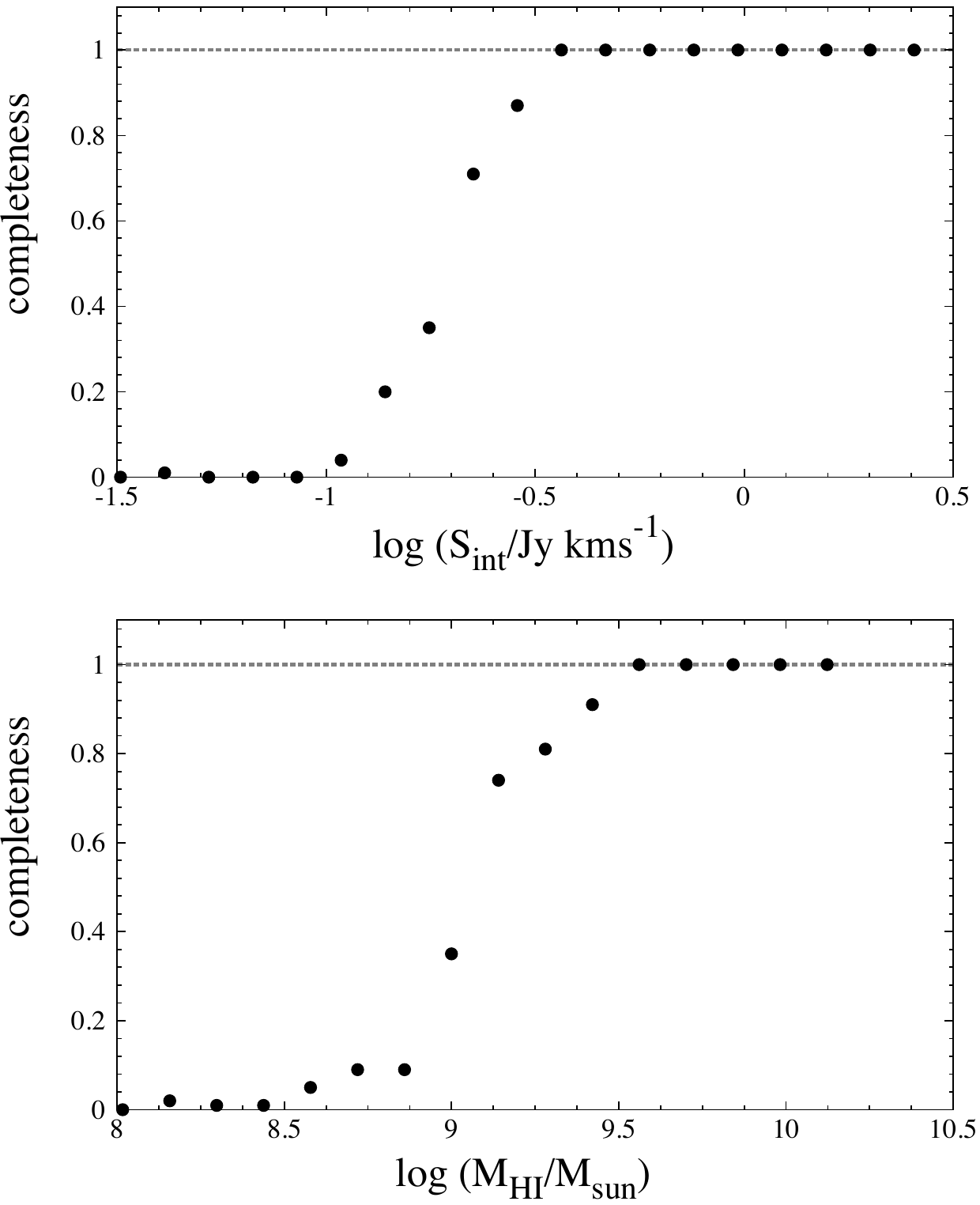}
   \caption{Completeness of the WSRT PP\,ZoA survey as function of the integrated \textsc{Hi} line flux (top panel) and the \textsc{Hi} mass (bottom panel). Points represent the fraction of artificial objects recovered in each bin. The grey dotted horizontal line represent where the survey is complete.}\label{CompJy}
\end{figure}

The data cube with the inserted artificial sources was then searched using the same procedure and detection criteria used to search for real galaxies as described in Sect.~3. No visual inspection of the artificial sources was done since their positions and redshifts were known a priori. The real sources were then removed from the list of detections such that only the detected artificial sources remained. A total of 101 artificial sources were recovered. The completeness for the whole WSRT survey was estimated by measuring the fraction of artificial sources recovered in given integrated flux and \HI\ mass bins as represented Fig.~\ref{CompJy}. 

From the detection rates of the simulated galaxies it follows that the WSRT PP\,ZoA survey is complete for objects with log($S_{\rm int}$/Jy\,\kms) $>$ $-0.45$ ($S_{\rm int}$ = 0.35 Jy\,\kms). The minimum integrated flux of an object to be detected by the WSRT PP\,ZoA survey is log($S_{\rm int}$/Jy\,\kms) $>$ $-0.98$ ($S_{\rm int}$ = 0.1 Jy\,\kms). From the bottom panel of Fig.~\ref{CompJy} we show that our survey is complete for galaxies with \HI\ masses of log($M_{\text{HI}}$/\MSUN) $>$ 9.5. 

We do however note that completeness is not just a measure of one parameter. A correction has to be made by integrating over another parameter such as the line width and peak flux as well as applying weighting to account for the relative source abundance in each parameter bin. In addition, given the depth of the survey, we expect the completeness to vary as a function of velocity. The completeness presented should therefore be regarded as a first approximation. In a forthcoming paper of this series we will provide a more robust verification and description. 

\section{The Distribution of HI Properties of Detected Galaxies}\label{sec6} 
\begin{figure*}
  \centering
    \includegraphics[width = 180mm, height = 100mm]{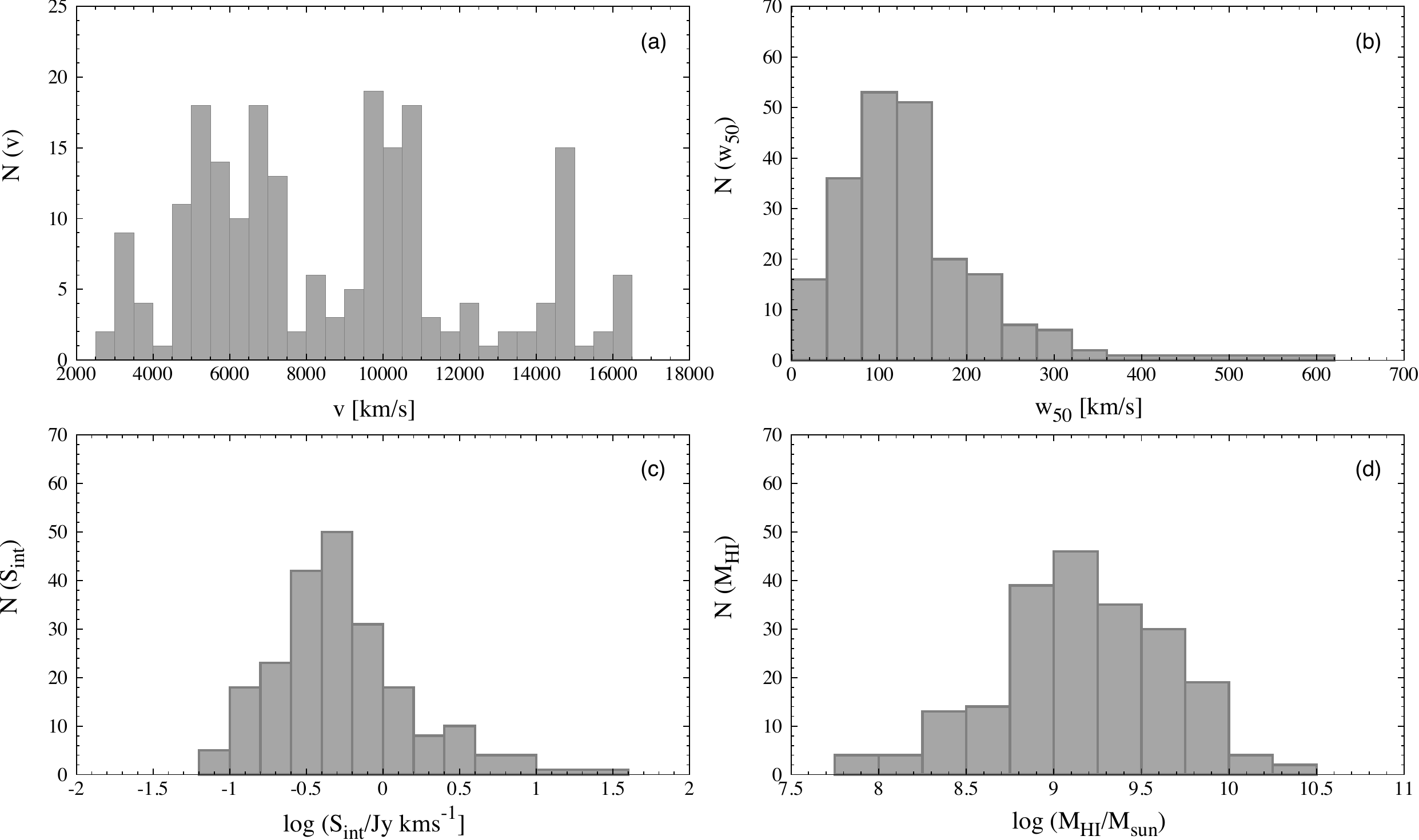}
   \caption{Distributions of selected \textsc{Hi} parameters of the WSRT ZoA\,PP detected galaxies. Top left: radial velocity ($v$). Top right: line width at the 50\% peak flux level ($w_{50}$). Bottom left:  logarithm of the integrated line flux (log $S_{\rm int}$/Jy\,\kms). Bottom right: logarithm of the total  \textsc{Hi} mass.}\label{vel_dist}
\end{figure*}

To investigate the \HI\ properties of the 211 detected galaxies, we plot the distributions of a number of their measured global \HI\ parameters in histograms (see Fig.~\ref{vel_dist}): their radial velocities, $w_{50}$ line widths, integrated line fluxes and total \HI\ masses.

We have visually identified four main overdensities along the line of sight by assessing their velocity distribution in combination with their spatial distribution. Since these features are located in the Auriga (Aur) constellation, we name them accordingly Aur~1 to Aur~4. The nearest peak in redshift space is at $~$3000 \kms\  (Aur~1) and may be a galaxy group, whereas the most prominent and broadest overdensity extends over $4500-7500$ \kms\ (Aur 2), i.e., at the approximate PPS redshift range. This is followed by a further distinctive peak at $9500-11000$ \kms\ (Aur~3) and a narrower peak close to the high-end redshift range of our survey, at 15000 \kms\ (Aur~4). Of all the detections, 41\% and 34\% are associated with the two most prominent peaks Aur~2 and Aur~3, respectively. In between these peaks, regions nearly completely devoid of galaxies are evident, indicating a wide range of cosmic environments probed with this interferometric survey over a wide velocity range. Further details on these structures are given in Sect.~\ref{sec7}.

\begin{figure}
  \centering
   \includegraphics[width = 85mm, height = 50mm]{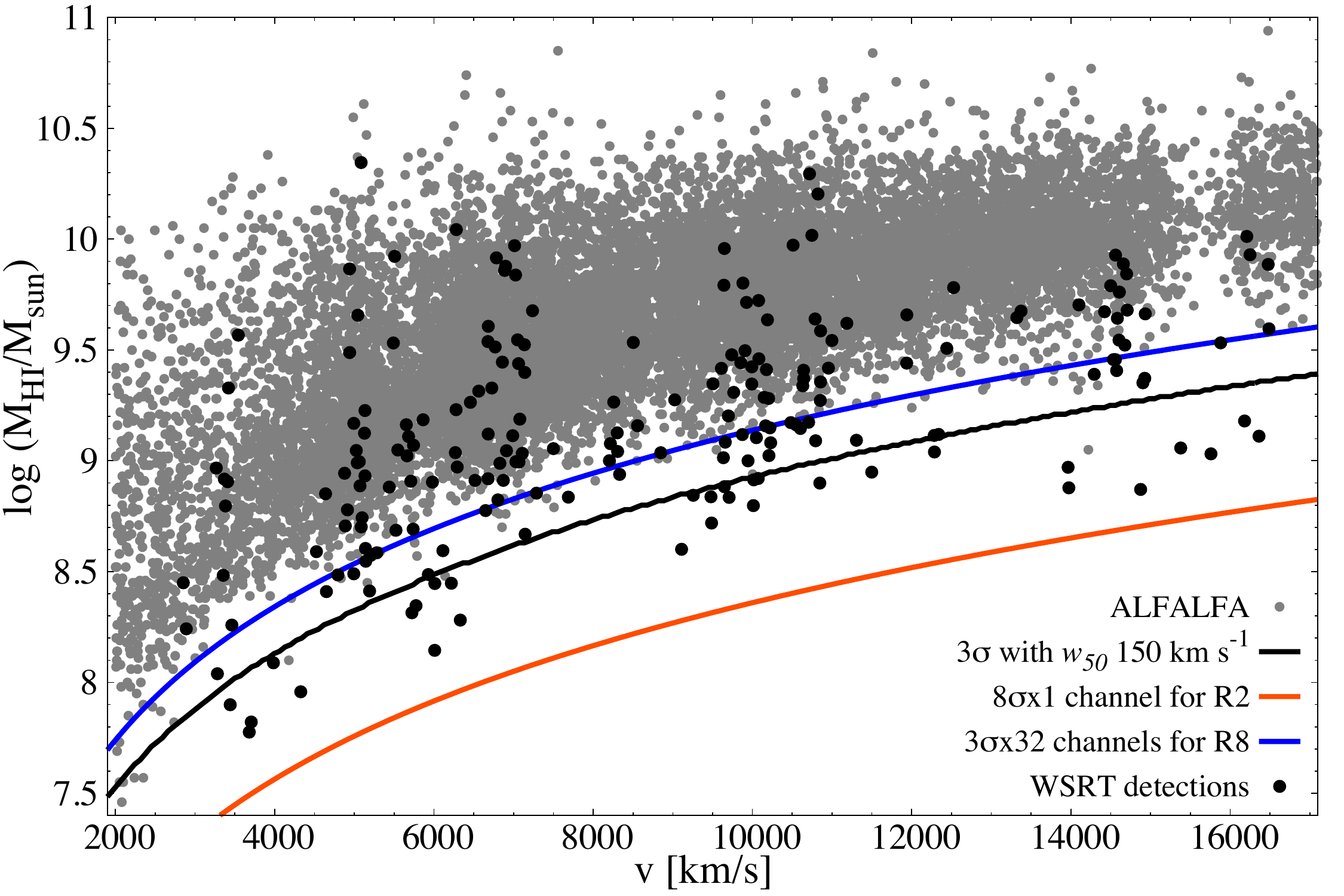}
   \caption{Logarithm of total \textsc{Hi} masses of galaxies detected in our survey (black dots) and the ALFALAFA survey (grey points) as a function of radial velocity. The black curve indicates the \textsc{Hi} mass limit for the WSRT ZoA\,PP survey assuming a 3$\sigma$ detection with a 150\kms\ line width. In orange we show the \textsc{Hi} mass limit of a detection at 8$\sigma$$\times$1 channel for a velocity resolution of 16.5 \kms\ (R2). The blue curve shows the \textsc{Hi} mass limit of a detection at 3$\sigma$$\times$32 channels for a velocity resolution of 66 \kms\ (R8)}\label{hi_mass_theo}   
\end{figure}

In the $w_{50}$ line widths histogram, the widths range from $24 - 526$ \kms, but most of the galaxies have $w_{50} < 225$ \kms\ and group around an average of 132 \kms. This is lower than the average $w_{50}$ of 186 \kms\ of the $\alpha$.40 catalogue of the Arecibo Legacy Fast ALFA survey (ALFALFA $\alpha$.40; \citealp{Haynes2011}). This difference can be explained by the relatively large coverage ($\sim$2800 deg$^2$) and lower sensitivity of 2.2 mJy of the ALFALFA survey, which leads to a higher average \HI\ mass of log($M_{\text{HI}}$ / \MSUN) = 9.5 (see Fig.~\ref{hi_mass_theo}) and therefore a larger average line width (c.f., bottom panel of Fig.~\ref{counterpart})
 
In the distribution of the integrated line fluxes, the average value is log($S_{\rm int}$/Jy\,\kms) = $-0.32$, and the fluxes span the range from log($S_{\rm int}$/Jy\,\kms) = $-1.40$ to 1.25, while the majority is restricted to a much tighter range of log($S_{\rm int}$/Jy\,\kms) = 0.6 to 1.0 ($0.35 < $$S_{\rm int}$$<$ 1.0 Jy\,\kms). 

The \HI\ mass distribution ranges from log($M_{\text{HI}}$/\MSUN) = 7.7  to  10.3. This is not excessively high compared to the most \HI-massive galaxies known, which can reach 10.8 (e.g.  \citealt{cluver2010}). This is due to the relatively small volume surveyed. The overall galaxy distribution has a mean \HI\ mass of log($M_{\text{HI}}$/\MSUN) = 9.1, which is below that of ALFALFA of log($M_{\text{HI}}$/\MSUN) = 9.5. There is a non-negligible fraction of lower \HI\ mass detections.

To assess where galaxies with different \HI\ masses are found in redshift, we plot in Fig.~\ref{hi_mass_theo} our  \HI\ masses as function of  radial velocity. The black curve shows the predicted \HI\ mass detection limit  of our  survey  at the 3$\sigma$ noise level for a galaxy with $w_{50}$ of 150 \kms. The two prominent peaks indicating clustering at about 6000 and 10000 \kms\  are once again evident. Galaxies found at these distances have  \HI\ masses ranging from log(M$_{\text{HI}}$/\MSUN) = $7.8 - 10.3$ and  $8.6 - 10.3$, respectively. To gauge how the \HI\ masses of our detections compare to those from other surveys, we also show \HI\ masses from the ALFALFA survey in grey. In general we find objects with lower \HI\ masses at all velocities which is consistent with the lower rms noise of $\sim$0.4 mJy of our WSRT survey compared to the 2.2 mJy of ALFALFA. 

\section{Large Scale Structures Crossing the ZoA}\label{sec7}
Our survey does not have a large enough areal coverage to explore the angular distribution of the large scale structures (LSS) at the different redshift peaks. However, it is deep enough to explore prominent LSS that cross the Galactic Plane in radial velocity space. We have found galaxies over the full radial velocity range of $cz$ = $2400 - 16600$ \kms. Their velocity distribution is not homogenous, which is not unexpected given that we targeted a cluster and its environment at about 6000\kms. Interestingly, behind the targeted cluster we also find evidence for either a clustering, a cross-section of a filament or a wall at larger distances. 

To explore the unveiled structures in the context of known LSS, we extracted galaxies with known redshifts from the literature (mainly the 2MRS) in the surroundings of our WSRT survey area. The 2MRS survey is homogenous and "all-sky" down to Galactic latitude $|b| = 5\dg$. However, it should be noted that it is NIR-selected and therefore favours early-type galaxies as opposed to the late-types typically detected in \HI\ samples. Given the lack of other deep all-sky galaxy imaging or \HI\ surveys, this remains the best  data set to compare our results with and to study the connectivity of known LSS across this area of the ZoA. It should be noted though that the peak sensitivity in redshift space of the 2MRS lies at about 12000\kms\ and that it is very sparsely sampled above $15000$ \kms.

\begin{figure*}  
  \centering
    \includegraphics[width = 180mm, height = 100mm]{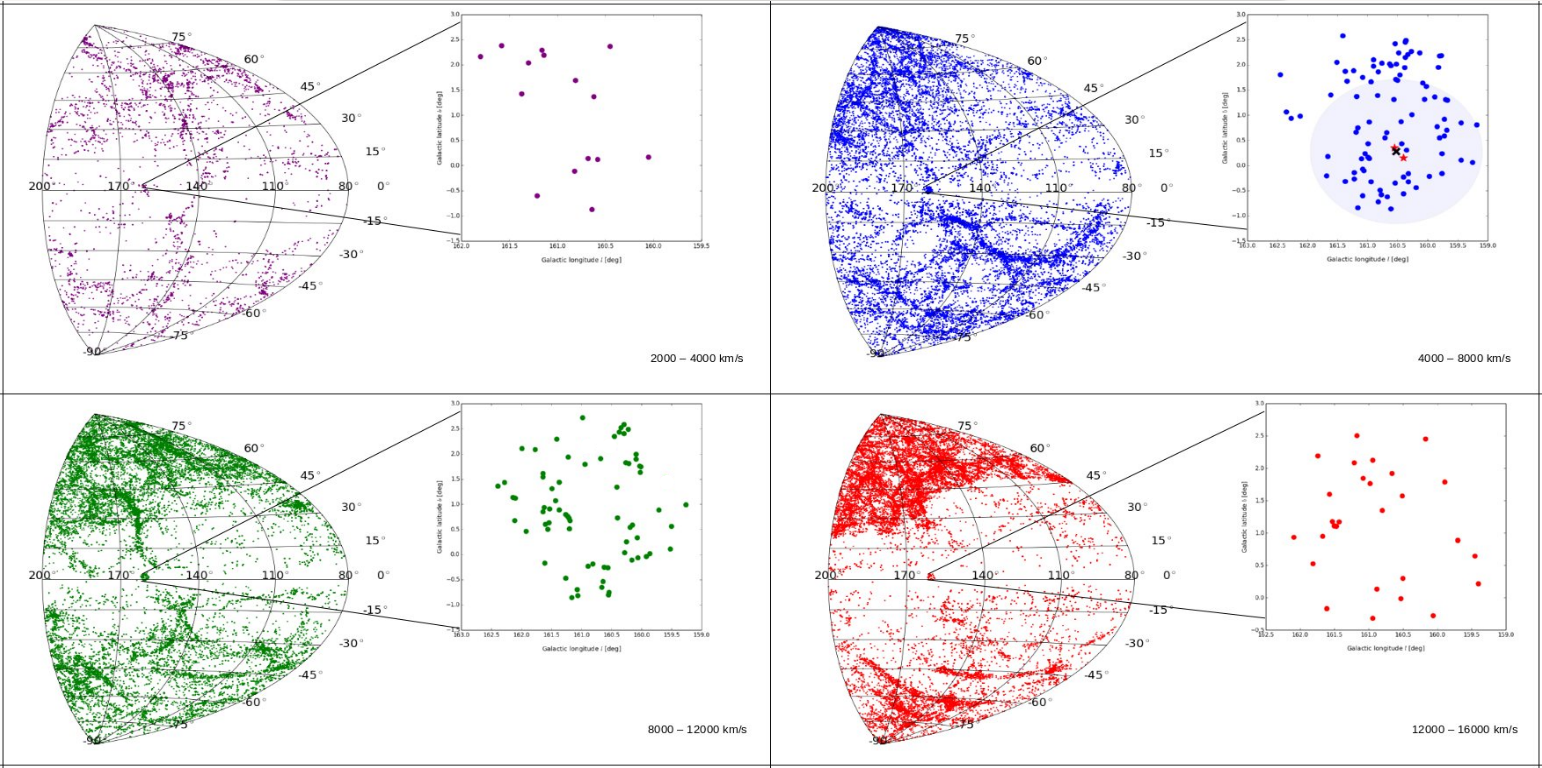}
   \caption{Aitoff projections of the sky-distribution in Galactic coordinates for $80\dg < \ell < 200\dg$ and $-90\dg < b < 90\dg$. Galaxies with known redshifts from the literature and the 2MRS are shown as dots. The images at the top right corner of each panel show a zoomed-in on the spatial distributions of galaxies detected by the WSRT ZoA\,PP survey and point to their location within the context of the large-scale structures they reside in. The galaxy sky distributions are separated into four radial velocity bins. Top left: $2000-4000$ \kms\ (Aur~1), top right: $4000 - 8000$ \kms\ (Aur~2), the light shaded area indicates 1 Abell radius of 85\arcmin\ at this distance, centred on the 3C\,129 cluster indicated by the black cross. Positions of the radio galaxies are indicated by red stars.  Bottom left: $8000 - 12000$ \kms\ (Aur~3) and bottom right: $12000 - 16000$ \kms\ (Aur~4).}\label{LSSdist}   
\end{figure*}

\begin{figure*}  
  \centering
    \includegraphics[width = 170mm, height = 80mm]{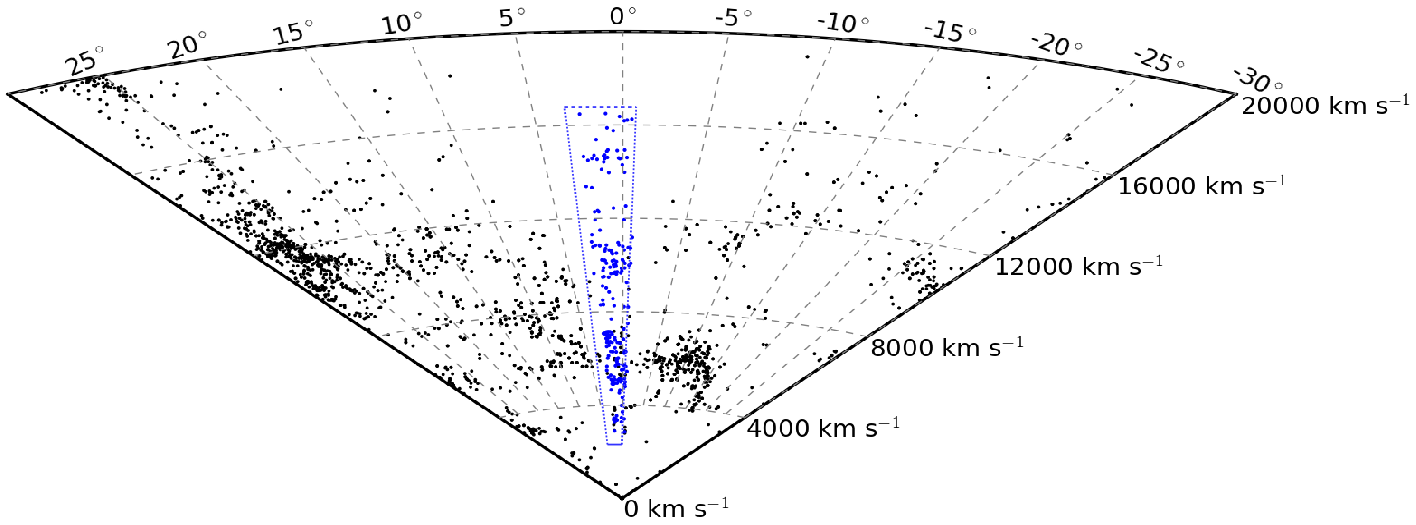}
   \caption{Redshift wedge diagram for $155\dg < \ell < 165\dg$ and $-30\dg < b < 30\dg$. Galaxies with known redshifts from the literature and the 2MRS out to $cz \approx 20000$ \kms\ are plotted in black. The WSRT PP\,ZoA survey region (at about -1$\dg < b < 4\dg$) is outlined by blue dashed lines and new detections  are represented by blue dots.}\label{distGPbottom} 
\end{figure*}

In Fig.~\ref{LSSdist}, the detected \HI\ galaxies are plotted together with adjacent galaxies with known redshifts  in an equal-area Aitoff sky projection for the Galactic coordinate range $80\dg < \ell < 200\dg$ and $-90\dg < b < 90\dg$, for four different radial velocity ranges. To clarify the effect of the new \HI\ detections in the small WSRT region (9.6 deg$^2$), we provide a zoomed-in plot next to the large area map for each of the redshift intervals. The different panels or colours represent the four overdensity peaks identified in Fig.~\ref{vel_dist}, namely in purple (Aur~1; $2000 - 4000$ \kms), blue (Aur~2; $4000 - 8000$ \kms), green (Aur~ 3; $8000 - 12000$ \kms) and red (Aur~ 4; $12000 - 16000$ \kms). The picture of the new ZoA crossing features becomes even clearer when assessing their positions in a redshift wedge (Fig.~\ref{distGPbottom}) which displays the Galactic longitude range of $155\dg - 165\dg$ for the latitude range $-30\dg < b < 30\dg$. The blue dashed lines outline our surveyed volume.

New features include a small number of nearby galaxies (Aur~1). The most prominent feature is Aur~2, around 6000 \kms, an overdensity associated with the 3C\,129 X-ray cluster and the two strong radio sources, 3C\,129 and 3C\,129.1 (see Sect.~\ref{intro}). Further out in the $8000 - 12000$ \kms\ (green) redshift range a new striking feature peaking at 10000 \kms\ dominates. Additionally, we are seeing hints of the edge of another structure, C31, identified by \citet{erdogdu2006}, at  $14000 - 16000$ \kms. However, the number of detections at these higher radial velocities is relatively low because it lies close to the high-velocity limit of our survey where the \HI\ mass sensitivity is reduced. In addition, the 2MRS has lost most of its sensitivity at  $ \geq 15000$ \kms. Nevertheless, it clearly marks a high-density region, as demonstrated in Fig.~\ref{hi_mass_theo} and \ref{distGPbottom}. For these reasons, concrete conclusions regarding this most distant galaxy density peak, cannot be drawn at present.

The newly identified structures already pointed out in Sec. 6 are discussed in detail in the following subsections, starting with the two most prominent structures, Aur 2 and Aur 3, followed by a short overview of the minor structures Aur 1 and Aur 4.

\newpage
\subsubsection*{The Perseus-Pisces Supercluster Connection}
The Perseus-Pisces Supercluster (PPS) is the most noticeable linear feature in the northern sky  around $cz \approx 6000$ \kms\ (e.g. \citealp{Giovanelli1985}, \citealp{Giovanelli1986b}, \citealp{Haynes1988}, \citealp{Wegner1993}). The main ridge of the PPS is a linear filamentary structure of about 45 Mpc in length \citep{Seeberger1994}, bounded in $\ell$ between  $80\dg$ and  180$\dg$. It comprises a rich chain of Abell clusters  such as ACO\,262, ACO\,347 and ACO\,426 (\citealp{Gregory1981}, \citealp{Wegner1993}).

In the top right panel of Fig.~\ref{LSSdist} the PPS can be traced from $\ell,b \approx 80\dg,15\dg$, entering the ZoA at $\ell \approx 90\dg$ and emerging  on the other side at $\ell,b \approx 90\dg,-5\dg$, the connection of which was confirmed to extend across the Milky Way by the NRT ZoA survey (Kraan-Korteweg et al. in prep). The filament bends all the way to the south of the ZoA at $\ell,b \approx 120\dg,-30\dg$, where it connects to ACO\,2634 ($\ell, b \approx 103\dg,-33\dg$) and ACO\,2666 ($\ell,b \approx 107\dg,-33\dg$) at  $cz \approx$ 8000\kms\ \citep{Merighi1986}. It then bends at a slight angle, vanishing into the ZoA at $\ell \approx 160\dg$ and reappears  aligned with the main ridge at $cz \approx$ 6000\kms.

It is based on this observation that \citet{Focardi1984} proposed an extension of the PPS across the ZoA towards a clump of galaxies associated with ACO\,569 (see also \citealp{Focardi1986}). This suggestion stemmed from a "conglomerate of galaxies" hinted at in \citet{Weinbeger1980} at $\ell \approx 160\dg$, where \citet{Spinrad1975} had measured radial velocities of the two radio galaxies mentioned in Sect.~1. The lack of data in the obscured ZoA prompted further extensive studies (\citealp{Focardi1984}, \citealp{Focardi1986}, \citealp{Chamaraux1990}, \citealp{Lu1995}, \citealp{Saurer1997}, \citealp{Pantoja1997}) ultimately confirming this link. It is clear now that our new observations not only substantiate this continuity, but  also prove that this connection incorporates yet another major cluster that forms part of the PPS.

The \HI\ detections in our survey are associated with the 3C\,129 X-ray cluster with a mass of M$_{\rm X} = 5 \times 10^{14}$\MSUN, centred at $\ell,b \approx 160.52\dg,0.28\dg$ \citep{Leahy2000} which when added to the overall Perseus-Pisces chain, makes the PPS one of the largest known structures in the local Universe. In fact, \citet{Chamaraux1990} showed that its dimension could be as large as $150 - 300$ Mpc, with the addition of other clusters in the south of the ZoA. Moreover, it has been postulated in the past that the PPS could be as densely populated as, or even more, than the Great Attractor (\citealp{Saunders1991}, \citealp{Strauss1995}, \citealp{Hudson1997}). Adding this new mass overdensity to the overall mass of the PPS might lead to significant infall motions of galaxies in its immediate surroundings. A complete and detailed dynamical examination of this galaxy cluster, including flows associated with the structure, is one of the questions that will be addressed in a forthcoming paper of this series.

\subsubsection*{Behind the Perseus-Pisces Supercluster}
The background of the PPS (see the bottom left panel of Fig.~\ref{LSSdist}) comprises galaxies in the velocity range $8000 - 12000$ \kms. In this range, the most dominant filamentary structure is evident. The Aur~3 overdensity of galaxies is part of this structure, which can be traced from  $\ell,b \approx 170\dg,45\dg$ to the North of the ZoA, traversing the ZoA at $\ell \approx 160\dg$, emerging from the ZoA at $\ell \approx 150\dg$ to the South. It continues towards a cluster dominated by early type galaxies at $\ell, b, cz \approx 143\dg,-22\dg, 10720$ \kms\ (J7; \citealp{Wegner1996}). It then bends at a slight angle and diffuses into a sheet of galaxies at $\ell \gtrsim  143\dg$, $b \lesssim 40\dg$. The new \HI\ detections in the ZoA at $\ell \approx 160\dg$ are not associated with any known cluster. However, they coincide with a structure, CID\,15 predicted by \citet{erdogdu2006} in the 2MRS reconstructed density and velocity maps.

\subsubsection*{Additional ZoA Structures}
Both the top left ($cz \approx 2000 - 4000$ \kms) and bottom right ($cz \approx 12000 - 16000$ \kms) panels of Fig.~\ref{LSSdist} indicate hints of galaxy overdensities. The nearer one, Aur~1, is too widely spread over the survey area to associate the detections with any large-scale structure in the vicinity. The more distant one, Aur~4 at the far edge of the survey volume does coincide with another structure, C31, predicted  by \citet{erdogdu2006} (see their Fig. 10) at $cz \approx 14000 - 16000$ \kms, although the actual galaxy map (Fig.~\ref{LSSdist}) does not give a clear hint of a connection to this feature. It seems to emerge from below the Galactic plane and ends in the ZoA without continuing into the northern Galactic hemisphere. There is not enough data for a quantitative analysis.

\section{Summary} 
We have conducted a blind 21cm \HI-line imaging survey with the WSRT of the area where the Perseus-Pisces Supercluster (PPS) crosses the Zone of Avoidance, centred at $\ell,b  \approx 160\dg,0.5\dg$. The survey comprises an effective area of $\sim$9.6 deg$^2$ observed at an angular resolution of 23\arcsec $\times$ 16\arcsec\ and a velocity resolution of 16.5 \kms\ with an average rms noise level of 0.36 mJy/beam over the radial velocity range of $cz = 2400-16600$ \kms. We detected 211 \HI\ sources, of which only three had previously been detected in \HI.  We have presented the results as a data catalogue with derived \HI-parameters and as an \HI\ atlas with total \HI\ maps and velocity fields, major axis position-velocity diagrams, global line profiles and overlays with near-infrared images. A total of 80 sources were resolved by the WSRT. We found near-infrared counterparts for 130 (62\%) of our detections from UKIDSS and 100 (47\%) mid-infrared counterparts from WISE. In cases where a UKIDSS image could not be found due to imaging artefacts we searched a 2MASS image at comparable wavelength to the UKIDSS $K$-band. The radial velocity distribution of our \HI\ detections shows four overdensities crossing the Galactic Plane. The most prominent is a connection between the PPS above and below the ZoA and associated with the 3C\,129 X-ray cluster. The new detections support previous claims that the PPS filament extends to the cluster A\,569. Of particular interest is another ZoA structure behind the PPS at a radial velocity of 10000 \kms. While predicted previously, it has now been detected for the first time. It seems to belong to a prominent large scale filament at 10000 \kms\ which comprises one other known cluster, J7, residing just below the ZoA. The data presented here provide input for an upcoming detailed analysis of the 3C\,129 cluster and the other overdensities detected, to be presented in forthcoming papers in this series.

\section*{Acknowledgments}
\small
MR acknowledges financial support from the Ubbo Emmius Fund of the University of Groningen. MR, RKK, ACS, TJ and EE acknowledge the research support provided by the South African National Research Foundation. The SA and NL authors of this collaboration all benefitted tremendously from collaborative exchanges support by the NRF/NWO bilateral agreement for Astronomy and Astronomy Enabling Technologies. MV acknowledges financial support from the DAGAL network for the People Programme (Marie Curie Actions) of the European UnionÕs Seventh Framework Programme FP7/2007-2013/ under REA grant agreement number PITNGA-2011-289313. MV acknowledges support by a Vici grant from the Innovational Research Incentives Scheme of the Netherlands Organisation for Scientific Research (NWO). MV also acknowledges the Leids Kerkhoven Bosscha Fonds (LKBF) for travel support. We would also like to thank K. Said, B. Frank, R. Juraszek, B. Koribalski, P. Serra, L. Fl\''{o}er, L. Staveley-Smith and I. Wong for their various contributions to the project. The Westerbork Synthesis Radio Telescope is operated by the ASTRON (Netherlands Institute for Radio Astronomy) with support from the Netherlands Foundation for Scientific Research (NWO). 

This publication makes use of data products from the Two Micron All Sky Survey, which is a joint project of the University of Massachusetts and the Infrared Processing and Analysis Center/California Institute of Technology, funded by the National Aeronautics and Space Administration and the National Science Foundation. 

This research has also made use of the UKIRT Infrared Deep Sky Survey database. 

This publication also makes use of data products from the Wide-field Infrared Survey Explorer, which is a joint project of the University of California, Los Angeles, and the Jet Propulsion Laboratory/California Institute of Technology, funded by the National Aeronautics and Space Administration. 

This research also makes use of the HyperLEDA database (http://leda.univ-lyon1.fr) and the NASA/IPAC Extragalactic Database (NED) which is operated by the Jet Propulsion Laboratory, California Institute of Technology, under contract with the National Aeronautics and Space Administration


\normalsize
\bibliographystyle{mnras}
\bibliography{References} 





section*{The HI Catalogue}\label{Atlas}

\onecolumn
\begin{centering}
\begin{landscape}
\begin{longtable}{lrrrrrrrrcrc} 
\caption*{\textbf{Table A1.} A catalogue of resolved WSRT \textsc{Hi} detections.}\label{resolved}\\

\hline \\[-1.8ex]
\thead{ZoA}            &  \thead{$\ell$}    &  \thead{$b$}    & \thead{$v_{rad}$}  &  \thead{$w_{20}$}  &  \thead{$w_{50}$}  &  \thead{$S_{\rm int}$}  &  \thead{$D$}  & \thead{log\,($M_{\text{HI}}$)}  & \thead{$R_{\text{HI}}$}   &  \thead{counterpart} \\
\\
                          &   \thead{deg}      &  \thead{deg}    &  \thead{\kms}       &      \thead{\kms}      &    \thead{\kms}      &   \thead{Jy \kms}     &    \thead{Mpc}   &  \thead{\MSUN}      &  \thead{kpc}      &              \\
\\
\thead{(1)}               &   \thead{(2)}      &  \thead{(3)}    &  \thead{(4)}         &      \thead{(5)}        &      \thead{(6)}      &    \thead{(7)}        &    \thead{(8)}   &  \thead{(9)}         &  \thead{(10)}       &  \thead{(11)}          \\

\hline \\[-1.8ex]
\endfirsthead

\multicolumn{9}{c}{{\textbf{Table A1}} -- Continued} \\[0.5ex]
\hline \\[-1.8ex]
\thead{ZoA}            &  \thead{$\ell$}    &  \thead{$b$}    & \thead{$v_{rad}$}  &  \thead{$w_{20}$}  &  \thead{$w_{50}$}  &  \thead{$S_{\rm int}$}  &  \thead{$D$}  & \thead{log\,($M_{\text{HI}}$)}  & \thead{$R_{\text{HI}}$}   &  \thead{counterpart} \\
\\
                          &   \thead{deg}      &  \thead{deg}    &  \thead{\kms}       &      \thead{\kms}      &    \thead{\kms}      &   \thead{Jy \kms}     &    \thead{Mpc}   &  \thead{\MSUN}      &  \thead{kpc}      &              \\
\\
\thead{(1)}               &   \thead{(2)}      &  \thead{(3)}    &  \thead{(4)}         &      \thead{(5)}        &      \thead{(6)}      &    \thead{(7)}        &    \thead{(8)}   &  \thead{(9)}         &  \thead{(10)}       &  \thead{(11)}          \\

\hline \\[-1.8ex]
\endhead

\hline \\[-1.8ex]
\\
\endfoot

\hline
\endlastfoot

J044427.17+455116.7		&		159.25		&		0.06		&		5506	$\pm$	04		&			285	$\pm$	14		&			266	$\pm$	09		&		5.68	$\pm$	0.41		&		79		&		9.9		&		24.2	&			u~~$-$\\										
J044521.10+454432.8		&		159.44		&		0.11		&		5134	$\pm$	03		&			290	$\pm$	07		&			278	$\pm$	07		&		1.34	$\pm$	0.10		&		73		&		9.2		&		12.4		&		u~~$-$\\										
J044542.87+442101.0		&		160.54		&		-0.75		&		10717	$\pm$	01		&			198	$\pm$	04		&			169	$\pm$	03		&		3.57	$\pm$	0.10		&		153		&		10.3		&		33.5		&		u~~$-$\\										
J044602.33+443426.8		&		160.40		&		-0.56		&		6281	$\pm$	03		&			117	$\pm$	10		&			100	$\pm$	06		&		5.78	$\pm$	0.45		&		90		&		10.0		&		31.0		&	 u~~$-$\\          
J044632.27+452152.2		&		159.86		&		0.02		&		9740	$\pm$	02		&			226	$\pm$	06		&			212	$\pm$	08		&		0.66	$\pm$	0.05		&		139		&		9.5		&		17.4		&		u~~w\\										
J044644.10+442004.0		&		160.67		&		-0.62		&		5654	$\pm$	05		&			194	$\pm$	13		&			179	$\pm$	13		&		0.94	$\pm$	0.09		&		81		&		9.2		&		12.7		&		u~~w\\										
J044644.74+444734.7		&		160.32		&		-0.32		&		5709	$\pm$	03		&			156	$\pm$	09		&			135	$\pm$	10		&		0.51	$\pm$	0.05		&		82		&		8.9		&		9.9		&		u~~$-$\\										
J044700.10+442439.7		&		160.64		&		-0.54		&		10747	$\pm$	02		&			330	$\pm$	07		&			309	$\pm$	05		&		1.86	$\pm$	0.09		&		154		&		10.0		&		35.8		&		u~~w\\										
J044706.91+453449.0		&		159.76		&		0.24		&		4994	$\pm$	03		&			216	$\pm$	08		&			200	$\pm$	06		&		1.24	$\pm$	0.06		&		71		&		9.2		&		14.0		&		u~~w\\										
J044727.30+445342.4		&		160.32		&		-0.16		&		5523	$\pm$	02		&		~\,69	$\pm$	06		&		~\,51	$\pm$   06		&		0.33	$\pm$	0.03		&		79		&		8.7		&		7.9		&	 $-$~~$-$\\
J044743.99+441946.5		&		160.79		&		-0.49		&		4878	$\pm$	04		&			146	$\pm$	10		&			117	$\pm$	10		&		0.76	$\pm$	0.05		&		70		&		8.9		&		9.9		&		u~~w\\										
J044837.97+454508.0		&		159.80		&		0.55		&		5745	$\pm$	02		&			136	$\pm$	04		&			118	$\pm$	10		&		0.74	$\pm$	0.06		&		82		&		9.1		&		10.8		&		u~~w\\										
J044840.48+452428.9		&		160.07		&		0.33		&		10956	$\pm$	02		&			164	$\pm$	07		&			152	$\pm$	06		&		0.45	$\pm$	0.05		&		157		&		9.4		&		17.6		&		u~~w\\										
J044907.68+443018.3		&		160.81		&		-0.19		&		10644	$\pm$	04		&			195	$\pm$	08		&			168	$\pm$	22		&		0.43	$\pm$	0.04		&		152		&		9.4		&		16.4		&		u~~w\\										
J044957.27+460315.1		&		159.72		&		0.92		&		4942	$\pm$	01		&			185	$\pm$	03		&			170	$\pm$	02		&		2.59	$\pm$	0.08		&		71		&		9.5		&		15.3		&		u~~w\\										
J045002.21+452645.5		&		160.19		&		0.54		&		9611	$\pm$	03		&			125	$\pm$	07		&			110	$\pm$	08		&		0.59	$\pm$	0.06		&		137		&		9.4		&		16.9		&	 $-$~~$-$\\          
J045005.46+453021.9		&		160.15		&		0.59		&		9642	$\pm$	03		&			307	$\pm$	10		&			283	$\pm$	08		&		1.38	$\pm$	0.08		&		138		&		9.8		&		25.8		&		u~~w\\										
J045013.10+440833.5		&		161.22		&		-0.27		&		5547	$\pm$	02		&			124	$\pm$	08		&			104	$\pm$	06		&		0.76	$\pm$	0.05		&		79		&		9.1		&		11.2		&		u~~w\\										
J045022.07+442207.7		&		161.06		&		-0.10		&		5739	$\pm$	02		&			125	$\pm$	04		&			110	$\pm$	14		&		0.31	$\pm$	0.03		&		82		&		8.7		&		6.7		&	 $-$~~$-$\\          
J045026.43+451135.1		&		160.43		&		0.43		&		4639	$\pm$	03		&			130	$\pm$	06		&		~\,68	$\pm$	35		&		0.69	$\pm$	0.05		&		66		&		8.9		&		7.9		&		u~~w\\					
J045105.43+443544.0		&		160.97		&		0.14		&		5044	$\pm$	01		&			187	$\pm$	04		&			168	$\pm$	03		&		3.71	$\pm$	0.11		&		72		&		9.7		&		20.5		&		u~~$-$\\										
J045132.28+442922.2		&		161.10		&		0.13		&		5129	$\pm$	04		&			305	$\pm$	11		&			276	$\pm$	11		&		1.06	$\pm$	0.07		&		73		&		9.1		&		11.1		&		u~~w\\										
J045137.03+461912.8		&		159.70		&		1.31		&		7030	$\pm$	04		&			134	$\pm$	18		&			~\,81	$\pm$	08		&		0.42	$\pm$	0.04		&		100		&		9.0		&		9.5		&		$-$~~$-$\\					
J045145.44+443610.2		&		161.04		&		0.24		&		5086	$\pm$	00		&			270	$\pm$	01		&			251	$\pm$	02		&		17.63	$\pm$	0.32		&		73		&		10.3		&		36.4		&		u~~w\\										
J045156.49+450320.0		&		160.71		&		0.55		&		4525	$\pm$	03		&			188	$\pm$	06		&			~\,91	$\pm$	20		&		0.39	$\pm$	0.06		&		65		&		8.6		&		7.4		&		u~~w\\					
J045208.67+434927.2		&		161.68		&		-0.21		&		7239	$\pm$	01		&			330	$\pm$	03		&			313	$\pm$	04		&		1.90	$\pm$	0.09		&		103		&		9.7		&		20.3		&		u~~w\\										
J045220.74+454136.5		&		160.26		&		1.01		&		6458	$\pm$	01		&			190	$\pm$	03		&			175	$\pm$	04		&		0.92	$\pm$	0.06		&		92		&		9.3		&		14.5		&		u~~w\\										
J045223.95+452744.1		&		160.45		&		0.87		&		5982	$\pm$	02		&			~\,97	$\pm$	04		&			~\,84	$\pm$	06		&		0.47	$\pm$	0.05		&		85		&		8.9		&		9.7		&	 $-$~~$-$\\          
J045227.46+444542.0		&		160.99		&		0.43		&		5144	$\pm$	02		&			102	$\pm$	04		&			~\,83	$\pm$	13		&		0.28	$\pm$	0.03		&		73		&		8.6		&		6.0		&		$-$~~$-$\\					
J045331.19+443916.6		&		161.20		&		0.51		&		10148	$\pm$	02		&			125	$\pm$	07		&			112	$\pm$	06		&		0.39	$\pm$	0.04		&		145		&		9.3		&		14.4		&		u~~w\\										
J045408.61+444522.2		&		161.19		&		0.66		&		5063	$\pm$	02		&			129	$\pm$	06		&			111	$\pm$	05		&		0.81	$\pm$	0.05		&		72		&		9.0		&		9.9		&	 $-$~~$-$\\          
J045414.22+461633.3		&		160.02		&		1.63		&		9024	$\pm$	02		&			246	$\pm$	09		&			200	$\pm$	04		&		0.48	$\pm$	0.05		&		129		&		9.3		&		15.4		&		u~~w\\										
J045414.90+450315.7		&		160.97		&		0.86		&		6269	$\pm$	02		&			208	$\pm$	05		&			165	$\pm$	05		&		0.57	$\pm$	0.05		&		90		&		9.0		&		8.7		&		u~~w\\										
J045420.49+454723.7		&		160.41		&		1.34		&		10856	$\pm$	02		&			161	$\pm$	06		&			128	$\pm$	09		&		0.68	$\pm$	0.05		&		155		&		9.6		&		19.7		&		u~~w\\										
J045426.94+444941.2		&		161.17		&		0.75		&		5072	$\pm$	03		&			132	$\pm$	08		&			105	$\pm$	08		&		0.63	$\pm$	0.04		&		72		&		8.9		&		8.9		&		u~~w\\										
J045430.27+461345.1		&		160.08		&		1.64		&		6726	$\pm$	02		&			229	$\pm$	07		&			213	$\pm$	05		&		0.98	$\pm$	0.06		&		96		&		9.3		&		14.5		&		u~~w\\										
J045442.75+462112.8		&		160.01		&		1.74		&		9881	$\pm$	06		&			293	$\pm$	11		&			218	$\pm$	33		&		1.35	$\pm$	0.09		&		141		&		9.8		&		25.6		&		u~~w\\										
J045443.51+444709.1		&		161.23		&		0.76		&		9926	$\pm$	01		&			106	$\pm$	03		&			~\,87	$\pm$	04		&		1.09	$\pm$	0.05		&		142		&		9.7		&		23.5		&		u~~w\\					
J045445.29+453922.2		&		160.56		&		1.31		&		4941	$\pm$	02		&			292	$\pm$	06		&			218	$\pm$	07		&		6.17	$\pm$	0.10		&		71		&		9.9		&		22.9		&		u~~w\\										
J045446.73+442143.1		&		161.56		&		0.50		&		10510	$\pm$	02		&			224	$\pm$	04		&			208	$\pm$	06		&		1.77	$\pm$	0.07		&		150		&		10.0		&		27.2		&		u~~w\\										
J045514.42+453815.1		&		160.62		&		1.37		&		3355	$\pm$	03		&			116	$\pm$	07		&			~\,97	$\pm$	08		&		0.56	$\pm$	0.05		&		48		&		8.5		&		5.9		&		u~~w\\					
J045546.73+444512.9		&		161.37		&		0.89		&		10177	$\pm$	02		&			179	$\pm$	03		&			165	$\pm$	33		&		0.52	$\pm$	0.05		&		145		&		9.4		&		16.0		&	 $-$~~$-$\\          
J045604.55+452940.6		&		160.83		&		1.39		&		6684	$\pm$	02		&			242	$\pm$	08		&			216	$\pm$	06		&		1.90	$\pm$	0.06		&		95		&		9.6		&		16.9		&		u~~w\\										
J045616.24+455602.6		&		160.51		&		1.69		&		6990	$\pm$	03		&			153	$\pm$	12		&			132	$\pm$	07		&		0.55	$\pm$	0.04		&		100		&		9.1		&		11.4		&		u~~w\\										
J045626.80+455541.2		&		160.53		&		1.71		&		6901	$\pm$	03		&			546	$\pm$	06		&			526	$\pm$	11		&		3.26	$\pm$	0.12		&		99		&		9.9		&		25.6		&		u~~w\\										
J045635.00+460222.4		&		160.46		&		1.80		&		6679	$\pm$	02		&			304	$\pm$	07		&			286	$\pm$	07		&		1.62	$\pm$	0.08		&		95		&		9.5		&		20.5		&		u~~w\\										
J045714.03+451227.7		&		161.18		&		1.37		&		6279	$\pm$	02		&			241	$\pm$	04		&			225	$\pm$	06		&		0.89	$\pm$	0.06		&		90		&		9.2		&		13.7		&		u~~w\\										
J045721.12+454135.5		&		160.81		&		1.69		&		3420	$\pm$	01		&			117	$\pm$	02		&			~\,99	$\pm$	02		&		3.76	$\pm$ 0.10		&		49		&		9.3		&		13.6		&		u~~w\\					
J045723.71+463354.4		&		160.13		&		2.24		&		7064	$\pm$	02		&			214	$\pm$	07		&			198	$\pm$	07		&		1.14	$\pm$	0.09		&		101		&		9.4		&		16.3		&		u~~w\\										
J045726.29+440242.9		&		162.11		&		0.67		&		9698	$\pm$	06		&			148	$\pm$	18		&			107	$\pm$	14		&		0.35	$\pm$	0.04		&		139		&		9.2		&		12.5		&		u~~w\\										
J045735.72+444850.6		&		161.53		&		1.18		&		14422	$\pm$	02		&			252	$\pm$	04		&			234	$\pm$	08		&		0.47	$\pm$	0.05		&		206		&		9.7		&		21.2		&		m~~w\\										
J045742.65+453422.1		&		160.94		&		1.66		&		6293	$\pm$	03		&			108	$\pm$	07		&			~\,75	$\pm$	08		&		0.49	$\pm$	0.03		&		90		&		9.0		&		10.1		&		u~~w\\					
J045747.05+460717.5		&		160.52		&		2.01		&		7013	$\pm$	01		&			393	$\pm$	03		&			375	$\pm$	03		&		3.96	$\pm$	0.11		&		100		&		10.0		&		30.7		&		u~~w\\										
J045750.73+461921.4		&		160.37		&		2.15		&		6861	$\pm$	01		&			117	$\pm$	05		&			104	$\pm$	04		&		1.23	$\pm$	0.07		&		98		&		9.4		&		17.3		&		u~~w\\										
J045752.30+455539.3		&		160.68		&		1.91		&		10822	$\pm$	01		&			231	$\pm$	09		&			193	$\pm$	03		&		2.82	$\pm$	0.10		&		155		&		10.2		&		33.7		&		u~~w\\										
J045800.33+462839.4		&		160.27		&		2.27		&		7079	$\pm$	03		&			102	$\pm$	07		&			~\,68	$\pm$	10		&		0.64	$\pm$	0.06		&		101		&		9.2		&		12.1		&		$-$~~$-$\\					
J045808.63+450525.8		&		161.37		&		1.42		&		3543	$\pm$	00		&			191	$\pm$	01		&			176	$\pm$	01		&		6.01	$\pm$	0.10		&		51		&		9.6		&		16.9		&		u~~w\\										
J045809.05+454743.1		&		160.82		&		1.86		&		6562	$\pm$	04		&			240	$\pm$	17		&			189	$\pm$	08		&		0.99	$\pm$	0.06		&		94		&		9.3		&		13.4		&		u~~$-$\\										
J045810.84+460229.4		&		160.63		&		2.02		&		7137	$\pm$	01		&			219	$\pm$	03		&			204	$\pm$	05		&		1.36	$\pm$	0.07		&		102		&		9.5		&		18.3		&		u~~w\\										
J045828.65+441328.4		&		162.09		&		0.93		&		14608	$\pm$	03		&			338	$\pm$	07		&			319	$\pm$	29		&		0.56	$\pm$	0.05		&		209		&		9.8		&		24.6		&		u~~w\\										
J045830.71+463953.7		&		160.17		&		2.45		&		14658	$\pm$	03		&			134	$\pm$	14		&			107	$\pm$	08		&		0.75	$\pm$	0.13		&		209		&		9.9		&		31.8		&		$-$~~$-$\\										
J045835.03+453110.8		&		161.08		&		1.75		&		5037	$\pm$	05		&			154	$\pm$	23		&			124	$\pm$	10		&		0.80	$\pm$	0.06		&		72		&		9.0		&		9.0		&		u~~$-$\\										
J045838.09+461747.7		&		160.48		&		2.24		&		7051	$\pm$	02		&			200	$\pm$	05		&			179	$\pm$	07		&		1.46	$\pm$	0.06		&		101		&		9.5		&		17.3		&		u~~w\\										
J045843.03+455647.1		&		160.76		&		2.03		&		6770	$\pm$	03		&			244	$\pm$	06		&			204	$\pm$	28		&		1.47	$\pm$	0.07		&		97		&		9.5		&		16.9		&		u~~w\\										
J045847.49+441350.2		&		162.12		&		0.98		&		6888	$\pm$	01		&			190	$\pm$	03		&			159	$\pm$	06		&		3.19	$\pm$	0.08		&		98		&		9.9		&		24.1		&	 $-$~~$-$\\          
J045855.81+454808.9		&		160.90		&		1.97		&		5098	$\pm$	02		&			135	$\pm$	05		&			~\,74	$\pm$	14		&		0.44	$\pm$	0.03		&		73		&		8.7		&		7.5		&		u~~w\\					
J045908.04+440500.0		&		162.27		&		0.94		&		6785	$\pm$	01		&			116	$\pm$	03		&			~\,89	$\pm$	02		&		3.71	$\pm$	0.07		&		97		&		9.9		&		23.9		&		u~~w\\					
J045909.87+451619.0		&		161.34		&		1.68		&		4888	$\pm$	03		&			202	$\pm$	08		&			189	$\pm$	09		&		0.44	$\pm$	0.04		&		70		&		8.7		&		7.5		&		u~~w\\										
J045920.95+441945.4		&		162.10		&		1.12		&		10641	$\pm$	03		&			194	$\pm$	08		&			159	$\pm$	07		&		0.47	$\pm$	0.05		&		152		&		9.4		&		17.3		&	 $-$~~$-$\\          
J045936.64+445709.6		&		161.64		&		1.54		&		10190	$\pm$	02		&			211	$\pm$	04		&			191	$\pm$	05		&		0.86	$\pm$	0.06		&		146		&		9.6		&		21.5		&		u~~w\\										
J045941.86+452913.7		&		161.23		&		1.88		&		7025	$\pm$	01		&			156	$\pm$	03		&			143	$\pm$	02		&		2.92	$\pm$	0.10		&		100		&		9.8		&		29.4		&		u~~w\\										
J045946.93+455125.0		&		160.95		&		2.12		&		13320	$\pm$	03		&			222	$\pm$	09		&			207	$\pm$	09		&		0.52	$\pm$	0.05		&		190		&		9.6		&		21.0		&		u~~w\\										
J045958.09+440554.1		&		162.35		&		1.06		&		5681	$\pm$	05		&			188	$\pm$	27		&			150	$\pm$	10		&		0.83	$\pm$	0.06		&		81		&		9.1		&		11.5		&	 u~~w\\          
J050007.95+452211.1		&		161.37		&		1.87		&		4914	$\pm$	02		&			146	$\pm$	05		&			132	$\pm$	06		&		0.52	$\pm$	0.04		&		70		&		8.8		&		7.6		&	 u~~w\\          
J050039.87+453120.2		&		161.30		&		2.04		&		3382	$\pm$	03		&			216	$\pm$	16		&			150	$\pm$	06		&		1.15	$\pm$	0.05		&		48		&		8.8		&		7.4		&		u~~w\\										
J050117.24+454732.3		&		161.16		&		2.29		&		3368	$\pm$	02		&			179	$\pm$	06		&			151	$\pm$	04		&		1.52	$\pm$	0.07		&		48		&		8.9		&		8.7		&		u~~w\\										
J050237.59+461133.0		&		160.98		&		2.72		&		9648	$\pm$	01		&			122	$\pm$	03		&			107	$\pm$	02		&		2.02	$\pm$	0.10		&		138		&		10.0		&		28.4		&		u~~w\\										
J050251.93+451609.7		&		161.74		&		2.19		&		14499	$\pm$	02		&			324	$\pm$	06		&			278	$\pm$	08		&		0.61	$\pm$	0.05		&		207		&		9.8		&		23.9		&		u~~w\\										
J050256.86+451230.4		&		161.80		&		2.16		&		3267	$\pm$	02		&			146	$\pm$	11		&			129	$\pm$	05		&		1.78	$\pm$	0.08		&		47		&		9.0		&		10.2		&	 m~~$-$\\          
J050320.82+450122.0		&		161.99		&		2.11		&		8504	$\pm$	01		&			120	$\pm$	03		&			~\,99	$\pm$	05		&		0.99	$\pm$	0.05		&		121		&		9.5		&		17.9		&		u~~$-$\\

\hline
\end{longtable}
\end{landscape}
\end{centering}

\begin{centering}
\begin{landscape}
\begin{longtable}{lrrrrrrrrcc} 
\caption*{\textbf{Table A2.} A catalogue of the marginally resolved WSRT \textsc{Hi} detections.}\label{unresolved}\\

\hline \\[-1.8ex]
\thead{ZoA}           & \thead{$\ell$}   & \thead{$b$}   &\thead{$v_{rad}$} & \thead{$w_{20}$} & \thead{$w_{50}$} & \thead{$S_{\rm int}$} & \thead{$D$} &\thead{log\,($M_{\text{HI}}$)} & \thead{counterpart} \\
\\
                         &  \thead{deg}     & \thead{deg}   & \thead{\kms}      &     \thead{\kms}     &   \thead{\kms}     &  \thead{Jy \kms}    &   \thead{Mpc}  & \thead{\MSUN}      &             \\
\\
\thead{(1)}              &  \thead{(2)}     & \thead{(3)}   & \thead{(4)}        &     \thead{(5)}       &     \thead{(6)}     &   \thead{(7)}       &   \thead{(8)}  & \thead{(9)}     & \thead{(10)}          \\

\hline \\[-1.8ex]
\endfirsthead

\multicolumn{9}{c}{{\textbf{Table A2.}} -- Continued} \\[0.5ex]
\hline \\[-1.8ex]
\thead{ZoA}           & \thead{$\ell$}   & \thead{$b$}   &\thead{$v_{rad}$} & \thead{$w_{20}$} & \thead{$w_{50}$} & \thead{$S_{\rm int}$} & \thead{$D$} &\thead{log\,($M_{\text{HI}}$)} & \thead{counterpart} \\
\\
                         &  \thead{deg}     & \thead{deg}   & \thead{\kms}      &     \thead{\kms}     &   \thead{\kms}     &  \thead{Jy \kms}    &   \thead{Mpc}  & \thead{\MSUN}      &             \\
\\
\thead{(1)}              &  \thead{(2)}     & \thead{(3)}   & \thead{(4)}        &     \thead{(5)}       &     \thead{(6)}     &   \thead{(7)}       &   \thead{(8)}  & \thead{(9)}     & \thead{(10)}          \\

\hline \\[-1.8ex]
\endhead

\hline \\[-1.8ex]
\\
\endfoot

\hline
\endlastfoot

J044524.14+451924.3		&		159.76		&		-0.16		&		6830	$\pm$	06		&		108	$\pm$	18		&		~\,59	$\pm$	16		&		0.43	$\pm$	0.07		&		98		&		9.0		&	 u~~$-$\\
J044533.70+441127.9		&		160.64		&		-0.88		&		2852	$\pm$	01		&		~\,90	$\pm$	02		&		~\,66	$\pm$	03		&		0.71	$\pm$	0.05		&		41		&		8.4		&		$-$~~$-$\\
J044540.97+454045.8		&		159.52		&		0.11		&		9660	$\pm$	02		&		153	$\pm$	06		&		142	$\pm$	08		&		0.27	$\pm$	0.03		&		138		&		9.1		&		u~~$-$\\
J044541.51+455022.6		&		159.40		&		0.21		&		14584	$\pm$	03		&		217	$\pm$	09		&		123	$\pm$	07		&		0.43	$\pm$	0.05		&		208		&		9.6		&		u~~$-$\\
J044546.04+444901.5		&		160.19		&		-0.44		&		5134	$\pm$	05		&		~\,94	$\pm$	34		&		~\,51	$\pm$	10		&		0.68	$\pm$	0.09		&		73		&		8.9		&		u~~$-$\\
J044550.59+443552.5		&		160.36		&		-0.57		&		16249	$\pm$	05		&		~\,95	$\pm$	13		&		~\,79	$\pm$	16		&		0.67	$\pm$	0.10		&		232		&		9.9		&		$-$~~$-$\\
J044556.60+450741.4		&		159.97		&		-0.21		&		6803	$\pm$	03		&		~\,82	$\pm$	14		&		~\,63	$\pm$	08		&		0.30	$\pm$	0.04		&		97		&		8.8		&	 u~~$-$\\
J044559.91+450104.1		&		160.06		&		-0.28		&		12528	$\pm$	05		&		171	$\pm$	10		&		123	$\pm$	42		&		0.80	$\pm$	0.13		&		179		&		9.8		&		$-$~~$-$\\
J044630.07+451655.0		&		159.92		&		-0.04		&		10229	$\pm$	05		&		170	$\pm$	12		&		124	$\pm$	18		&		0.24	$\pm$	0.04		&		146		&		9.1		&		u~~w\\
J044635.75+441909.5		&		160.66		&		-0.65		&		9992	$\pm$	03		&		128	$\pm$	11		&		~\,94	$\pm$	07		&		0.55	$\pm$	0.07		&		143		&		9.4		&	 u~~w\\
J044652.86+440859.1		&		160.82		&		-0.72		&		7501	$\pm$	08		&		151	$\pm$	24		&		122	$\pm$	21		&		0.42	$\pm$	0.07		&		107		&		9.1		&	  $-$~~$-$\\
J044654.85+450928.4		&		160.06		&		-0.07		&		10794	$\pm$	02		&		170	$\pm$	06		&		155	$\pm$	08		&		0.22	$\pm$	0.02		&		154		&		9.1		&		u~~w\\
J044657.71+444649.5		&		160.35		&		-0.30		&		16209	$\pm$	06		&		163	$\pm$	29		&		117	$\pm$	14		&		0.81	$\pm$	0.08		&		232		&		10.0		&	 $-$~~$-$\\
J044705.89+450301.9		&		160.16		&		-0.11		&		10165	$\pm$	04		&		158	$\pm$	09		&		111	$\pm$	20		&		0.29	$\pm$	0.03		&		145		&		9.2		&	 u~~$-$\\
J044717.85+441654.2		&		160.77		&		-0.58		&		4993	$\pm$	02		&		157	$\pm$	05		&		147	$\pm$	06		&		0.26	$\pm$	0.03		&		71		&		8.5		&		u~~w\\
J044719.35+435436.2		&		161.06		&		-0.82		&		10077	$\pm$	04		&		197	$\pm$	13		&		186	$\pm$	11		&		1.08	$\pm$	0.17		&		144		&		9.7		&		u~~w\\
J044725.76+443638.5		&		160.54		&		-0.35		&		6219	$\pm$	02		&		~\,92	$\pm$	06		&		~\,82	$\pm$	05		&		0.15	$\pm$	0.02		&		89		&		8.5		&	 $-$~~$-$\\
J044726.92+462337.0		&		159.18		&		0.81		&		5864	$\pm$	02		&		149	$\pm$	04		&		129	$\pm$	09		&		0.92	$\pm$	0.10		&		84		&		9.2		&		$-$~~$-$\\
J044727.96+444723.6		&		160.40		&		-0.23		&		7288	$\pm$	04		&		~\,90	$\pm$	09		&		~\,54	$\pm$	12		&		0.28	$\pm$	0.03		&		104		&		8.9		&		u~~w\\
J044731.28+434829.0		&		161.16		&		-0.86		&		11186	$\pm$	02		&		~\,54	$\pm$	08		&		~\,29	$\pm$	06		&		0.69	$\pm$	0.06		&		160		&		9.6		&		$-$~~$-$\\
J044735.05+455908.2		&		159.51		&		0.56		&		9945	$\pm$	03		&		129	$\pm$	14		&		104	$\pm$	07		&		0.21	$\pm$	0.03		&		142		&		9.0		&	 $-$~~$-$\\
J044736.43+434854.0		&		161.16		&		-0.84		&		5440	$\pm$	01		&		126	$\pm$	05		&		115	$\pm$	03		&		0.53	$\pm$	0.05		&		78		&		8.9		&		$-$~~$-$\\
J044744.05+460456.5		&		159.45		&		0.64		&		12441	$\pm$	02		&		159	$\pm$	05		&		147	$\pm$	06		&		0.43	$\pm$	0.05		&		178		&		9.5		&	 $-$~~$-$\\
J044752.21+451900.3		&		160.05		&		0.17		&		3280	$\pm$	02		&		158	$\pm$	05		&		139	$\pm$	11		&		0.21	$\pm$	0.02		&		47		&		8.0		&		$-$~~$-$\\
J044752.94+435840.2		&		161.07		&		-0.70		&		8204	$\pm$	03		&		~\,81	$\pm$	10		&		~\,56	$\pm$	08		&		0.31	$\pm$	0.03		&		117		&		9.0		&	 u~~w\\
J044754.01+443847.3		&		160.56		&		-0.26		&		10074	$\pm$	04		&		100	$\pm$	12		&		~\,74	$\pm$	09		&		0.17	$\pm$	0.02		&		144		&		8.9		&	 $-$~~$-$\\
J044808.96+450326.3		&		160.28		&		0.04		&		9853	$\pm$	04		&		259	$\pm$	07		&		239	$\pm$	30		&		0.59	$\pm$	0.07		&		141		&		9.4		&		u~~w\\
J044809.24+443634.5		&		160.62		&		-0.25		&		9873	$\pm$	04		&		120	$\pm$	08		&		101	$\pm$	15		&		0.28	$\pm$	0.04		&		141		&		9.1		&		$-$~~$-$\\
J044819.38+440155.8		&		161.08		&		-0.60		&		4794	$\pm$	03		&		~\,92	$\pm$	07		&		~\,77	$\pm$	09		&		0.28	$\pm$	0.04		&		68		&		8.5		&		$-$~~$-$\\
J044830.74+455010.9		&		159.72		&		0.59		&		6012	$\pm$	06		&		133	$\pm$	15		&		~\,67	$\pm$	23		&		0.16	$\pm$	0.04		&		86		&		8.4		&		$-$~~$-$\\
J044832.96+462704.2		&		159.26		&		0.99		&		8557	$\pm$	06		&		128	$\pm$	21		&		113	$\pm$	14		&		0.41	$\pm$	0.07		&		122		&		9.2		&		$-$~~$-$\\
J044837.47+461304.6		&		159.44		&		0.85		&		5930	$\pm$	04		&		147	$\pm$	12		&		133	$\pm$	10		&		0.18	$\pm$	0.04		&		85		&		8.5		&		$-$~~$-$\\
J044845.01+435610.4		&		161.20		&		-0.60		&		3410	$\pm$	01		&		~\,37	$\pm$	04		&		~\,24	$\pm$	02		&		1.42	$\pm$	0.09		&		49		&		8.9		&		$-$~~$-$\\
J044850.72+455649.4		&		159.68		&		0.70		&		5976	$\pm$	03		&		168	$\pm$	06		&		162	$\pm$	11		&		0.04	$\pm$	0.02		&		85		&		7.9		&		$-$~~$-$\\
J044850.94+444943.4		&		160.53		&		-0.02		&		14294	$\pm$	02		&		163	$\pm$	03		&		142	$\pm$	12		&		0.25	$\pm$	0.02		&		204		&		9.4		&		$-$~~$-$\\
J044857.72+443728.8		&		160.70		&		-0.13		&		16180	$\pm$	03		&		~\,77	$\pm$	08		&		~\,61	$\pm$	10		&		0.12	$\pm$	0.02		&		231		&		9.2		&		$-$~~$-$\\
J044858.80+451314.0		&		160.25		&		0.25		&		10857	$\pm$	04		&		276	$\pm$	08		&		236	$\pm$	20		&		0.40	$\pm$	0.04		&		155		&		9.4		&		u~~w\\
J044900.08+441936.1		&		160.93		&		-0.32		&		14604	$\pm$	03		&		212	$\pm$	07		&		107	$\pm$	25		&		0.34	$\pm$	0.04		&		209		&		9.5		&		u~~w\\
J044900.78+441902.8		&		160.94		&		-0.32		&		6332	$\pm$	05		&		146	$\pm$	21		&		130	$\pm$	11		&		0.10	$\pm$	0.03		&		90		&		8.3		&		$-$~~$-$\\
J044912.74+442451.3		&		160.89		&		-0.23		&		8216	$\pm$	02		&		156	$\pm$	05		&		~\,81	$\pm$	04		&		0.37	$\pm$	0.04		&		117		&		9.1		&		$-$~~$-$\\
J044927.63+443253.7		&		160.82		&		-0.11		&		3707	$\pm$	04		&		156	$\pm$	11		&		111	$\pm$	11		&		0.10	$\pm$	0.02		&		53		&		7.8		&		u~~$-$\\
J044931.06+435844.3		&		161.26		&		-0.47		&		9257	$\pm$	04		&		220	$\pm$	13		&		148	$\pm$	10		&		0.17	$\pm$	0.03		&		132		&		8.8		&		$-$~~$-$\\
J044934.80+451030.3		&		160.35		&		0.31		&		6649	$\pm$	02		&		~\,84	$\pm$	06		&		~\,43	$\pm$	07		&		0.28	$\pm$	0.03		&		95		&		8.8		&		$-$~~$-$\\
J044935.92+445252.1		&		160.58		&		0.12		&		3985	$\pm$	02		&		~\,65	$\pm$	05		&		~\,35	$\pm$	07		&		0.16	$\pm$	0.02		&		57		&		8.1		&		$-$~~$-$\\
J044944.31+460225.7		&		159.70		&		0.88		&		14933	$\pm$	02		&		~\,77	$\pm$	04		&		~\,55	$\pm$	12		&		0.43	$\pm$	0.04		&		213		&		9.7		&		$-$~~$-$\\
J044945.15+460232.2		&		159.70		&		0.88		&		13964	$\pm$	01		&		~\,60	$\pm$	03		&		~\,24	$\pm$	04		&		0.10	$\pm$	0.02		&		199		&		9.0		&	 u~~$-$\\
J044945.20+460234.8		&		159.70		&		0.88		&		15881	$\pm$	03		&		106	$\pm$	09		&		~\,86	$\pm$	10		&		0.28	$\pm$	0.04		&		227		&		9.5		&	 u~~w\\
J044945.25+460229.7		&		159.70		&		0.88		&		8258	$\pm$	03		&		400	$\pm$	07		&		293	$\pm$	12		&		0.56	$\pm$	0.05		&		118		&		9.3		&	 u~~w\\
J044945.98+455144.8		&		159.84		&		0.77		&		5774	$\pm$	02		&		129	$\pm$	04		&		122	$\pm$	05		&		0.14	$\pm$	0.02		&		82		&		8.3		&		$-$~~$-$\\
J045002.27+444907.5		&		160.68		&		0.14		&		3461	$\pm$	01		&		~\,52	$\pm$	04		&		~\,36	$\pm$	03		&		0.32	$\pm$	0.02		&		49		&		8.3		&		$-$~~$-$\\
J045006.21+450306.5		&		160.50		&		0.30		&		14904	$\pm$	02		&		~\,69	$\pm$	04		&		~\,31	$\pm$	07		&		0.21	$\pm$	0.03		&		213		&		9.4		&		u~~w\\
J045034.24+435929.0		&		161.37		&		-0.32		&		5724	$\pm$	06		&		137	$\pm$	16		&		~\,95	$\pm$	16		&		0.13	$\pm$	0.02		&		82		&		8.3		&		u~~w\\
J045035.09+442218.2		&		161.08		&		-0.07		&		4329	$\pm$	01		&		~\,43	$\pm$	02		&		~\,25	$\pm$	04		&		0.10	$\pm$	0.01		&		62		&		8.0		&		$-$~~$-$\\
J045045.12+443915.9		&		160.88		&		0.13		&		14558	$\pm$	01		&		176	$\pm$	03		&		144	$\pm$	05		&		0.28	$\pm$	0.03		&		208		&		9.5		&		$-$~~$-$\\
J045104.83+454944.1		&		160.02		&		0.93		&		16487	$\pm$	07		&		~\,93	$\pm$	26		&		~\,73	$\pm$	15		&		0.30	$\pm$	0.05		&		236		&		9.6		&		$-$~~$-$\\
J045117.24+443525.9		&		160.99		&		0.16		&		5026	$\pm$	01		&		~\,61	$\pm$	04		&		~\,39	$\pm$	03		&		0.91	$\pm$	0.05		&		72		&		9.0		&		$-$~~$-$\\
J045127.67+461953.7		&		159.67		&		1.30		&		5087	$\pm$	05		&		118	$\pm$	21		&		~\,60	$\pm$	10		&		0.40	$\pm$	0.04		&		73		&		8.7		&		u~~w\\
J045136.02+452435.7		&		160.40		&		0.73		&		9648	$\pm$	01		&		~\,46	$\pm$	03		&		~\,31	$\pm$	04		&		0.17	$\pm$	0.02		&		138		&		8.9		&		u~~w\\
J045202.73+435400.5		&		161.61		&		-0.17		&		14928	$\pm$	09		&		~\,86	$\pm$	33		&		~\,60	$\pm$	20		&		0.22	$\pm$	0.04		&		213		&		9.4		&		u~~w\\
J045202.83+435359.9		&		161.61		&		-0.17		&		11938	$\pm$	02		&		200	$\pm$	04		&		152	$\pm$	12		&		0.40	$\pm$	0.05		&		171		&		9.4		&		u~~$-$\\
J045202.99+435400.5		&		161.61		&		-0.17		&		13974	$\pm$	03		&		~\,49	$\pm$	06		&		~\,40	$\pm$	09		&		0.08	$\pm$	0.02		&		200		&		8.9		&		u~~$-$\\
J045210.82+463120.9		&		159.60		&		1.51		&		10603	$\pm$	03		&		158	$\pm$	09		&		145	$\pm$	10		&		0.26	$\pm$	0.03		&		151		&		9.2		&		$-$~~$-$\\
J045217.56+450844.6		&		160.68		&		0.65		&		6680	$\pm$	04		&		147	$\pm$	09		&		121	$\pm$	33		&		0.39	$\pm$	0.04		&		95		&		8.9		&		u~~w\\
J045231.94+461230.3		&		159.88		&		1.36		&		7109	$\pm$	03		&		133	$\pm$	09		&		120	$\pm$	07		&		0.44	$\pm$	0.05		&		102		&		9.0		&		$-$~~$-$\\
J045254.16+441202.2		&		161.48		&		0.14		&		16363	$\pm$	04		&		162	$\pm$	13		&		108	$\pm$	12		&		0.10	$\pm$	0.02		&		234		&		9.1		&		u~~w\\
J045255.89+460258.8		&		160.05		&		1.31		&		7689	$\pm$	02		&		~\,99	$\pm$	07		&		~\,87	$\pm$	08		&		0.24	$\pm$	0.03		&		110		&		8.8		&		$-$~~$-$\\
J045343.57+440459.8		&		161.66		&		0.18		&		5142	$\pm$	02		&		~\,92	$\pm$	04		&		~\,40	$\pm$	08		&		0.32	$\pm$	0.03		&		73		&		8.6		&		u~~w\\
J045358.21+461405.5		&		160.02		&		1.57		&		7142	$\pm$	02		&		165	$\pm$	07		&		150	$\pm$	05		&		1.02	$\pm$	0.06		&		102		&		9.4		&		u~~$-$\\
J045411.23+444554.1		&		161.19		&		0.67		&		9991	$\pm$	03		&		112	$\pm$	07		&		~\,93	$\pm$	14		&		0.46	$\pm$	0.04		&		143		&		9.3		&		$-$~~$-$\\
J045428.52+462803.0		&		159.89		&		1.78		&		14698	$\pm$	03		&		186	$\pm$	06		&		151	$\pm$	16		&		0.67	$\pm$	0.07		&		210		&		9.8		&		u~~w\\
J045430.92+444645.3		&		161.21		&		0.73		&		10202	$\pm$	04		&		221	$\pm$	12		&		195	$\pm$	11		&		0.38	$\pm$	0.04		&		146		&		9.3		&		u~~w\\
J045449.54+462052.1		&		160.03		&		1.76		&		9505	$\pm$	01		&		~\,78	$\pm$	03		&		~\,38	$\pm$	06		&		0.51	$\pm$	0.04		&		136		&		9.3		&		$-$~~$-$\\
J045457.48+463735.6		&		159.82		&		1.95		&		6872	$\pm$	05		&		169	$\pm$	10		&		123	$\pm$	33		&		0.36	$\pm$	0.05		&		98		&		8.9		&		$-$~~$-$\\
J045459.27+444700.0		&		161.26		&		0.79		&		10012	$\pm$	02		&		~\,62	$\pm$	05		&		~\,49	$\pm$	06		&		0.13	$\pm$	0.01		&		143		&		8.8		&		u~~w\\
J045516.70+442734.8		&		161.54		&		0.63		&		10050	$\pm$	02		&		105	$\pm$	05		&		~\,36	$\pm$	04		&		0.26	$\pm$	0.02		&		144		&		9.1		&		u~~w\\
J045519.33+442356.6		&		161.60		&		0.60		&		10016	$\pm$	02		&		~\,63	$\pm$	05		&		~\,47	$\pm$	07		&		0.17	$\pm$	0.01		&		143		&		8.9		&		$-$~~$-$\\
J045541.13+462255.5		&		160.09		&		1.89		&		9488	$\pm$	02		&		~\,98	$\pm$	05		&		~\,87	$\pm$	17		&		0.12	$\pm$	0.02		&		136		&		8.7		&		u~~w\\
J045543.03+461423.6		&		160.21		&		1.81		&		9765	$\pm$	02		&		201	$\pm$	04		&		~\,97	$\pm$	14		&		0.44	$\pm$	0.04		&		140		&		9.3		&		u~~w\\
J045543.43+441116.9		&		161.81		&		0.52		&		15759	$\pm$	04		&		135	$\pm$	12		&		118	$\pm$	14		&		0.09	$\pm$	0.02		&		225		&		9.0		&		$-$~~$-$\\
J045545.27+455110.8		&		160.51		&		1.57		&		15380	$\pm$	02		&		~\,73	$\pm$	05		&		~\,62	$\pm$	08		&		0.10	$\pm$	0.02		&		220		&		9.0		&		$-$~~$-$\\
J045547.62+452909.9		&		160.80		&		1.35		&		14101	$\pm$	04		&		213	$\pm$	85		&		187	$\pm$	08		&		0.53	$\pm$	0.06		&		201		&		9.7		&		u~~w\\
J045548.78+464858.3		&		159.77		&		2.18		&		4651	$\pm$	04		&		~\,77	$\pm$	08		&		~\,55	$\pm$	20		&		0.25	$\pm$	0.06		&		66		&		8.4		&		$-$~~$-$\\
J045551.34+440335.6		&		161.92		&		0.46		&		9708	$\pm$	02		&		~\,67	$\pm$	06		&		~\,52	$\pm$	09		&		0.15	$\pm$	0.01		&		139		&		8.8		&		$-$~~$-$\\
J045555.08+464704.8		&		159.80		&		2.18		&		6517	$\pm$	04		&		~\,75	$\pm$	09		&		~\,43	$\pm$	16		&		0.40	$\pm$	0.05		&		93		&		8.9		&	 u~~$-$\\
J045558.82+461234.1		&		160.26		&		1.83		&		10221	$\pm$	02		&		119	$\pm$	11		&		~\,49	$\pm$	06		&		0.28	$\pm$	0.04		&		146		&		9.1		&		$-$~~$-$\\
J045607.91+462638.4		&		160.09		&		1.99		&		9110	$\pm$	04		&		~\,91	$\pm$	11		&		~\,34	$\pm$	14		&		0.10	$\pm$	0.01		&		130		&		8.6		&		$-$~~$-$\\
J045625.85+443832.0		&		161.53		&		0.91		&		10550	$\pm$	03		&		165	$\pm$	13		&		104	$\pm$	07		&		0.27	$\pm$	0.03		&		151		&		9.2		&		u~~w\\
J045633.37+443124.4		&		161.64		&		0.85		&		10483	$\pm$	01		&		~\,61	$\pm$	09		&		~\,28	$\pm$	02		&		0.28	$\pm$	0.02		&		150		&		9.2		&		u~~w\\
J045647.57+444925.8		&		161.43		&		1.07		&		9635	$\pm$	02		&		164	$\pm$	06		&		152	$\pm$	06		&		0.23	$\pm$	0.03		&		138		&		9.0		&		u~~w\\
J045653.00+443516.6		&		161.62		&		0.94		&		10633	$\pm$	03		&		296	$\pm$	06		&		176	$\pm$	474		&		0.40	$\pm$	0.05		&		152		&		9.3		&		u~~w\\
J045656.49+461134.0		&		160.38		&		1.94		&		6911	$\pm$	02		&		138	$\pm$	04		&		121	$\pm$	06		&		0.48	$\pm$	0.05		&		99		&		9.0		&	 u~~w\\
J045705.33+443341.0		&		161.67		&		0.95		&		14875	$\pm$	02		&		~\,67	$\pm$	04		&		~\,55	$\pm$	06		&		0.07	$\pm$	0.01		&		212		&		8.9		&		u~~w\\
J045705.96+444834.3		&		161.47		&		1.10		&		14708	$\pm$	04		&		189	$\pm$	13		&		153	$\pm$	10		&		0.46	$\pm$	0.04		&		210		&		9.7		&		u~~$-$\\
J045711.02+444742.6		&		161.50		&		1.11		&		14561	$\pm$	01		&		206	$\pm$	03		&		193	$\pm$	15		&		0.83	$\pm$	0.04		&		208		&		9.9		&		u~~$-$\\
J045715.06+445255.8		&		161.43		&		1.17		&		14677	$\pm$	03		&		246	$\pm$	08		&		222	$\pm$	11		&		0.32	$\pm$	0.04		&		210		&		9.5		&	 u~~w\\
J045751.11+455706.4		&		160.66		&		1.92		&		14574	$\pm$	02		&		153	$\pm$	05		&		~\,93	$\pm$	06		&		0.25	$\pm$	0.03		&		208		&		9.4		&		$-$~~$-$\\
J045757.17+460156.9		&		160.61		&		1.98		&		5284	$\pm$	02		&		149	$\pm$	06		&		108	$\pm$	08		&		0.29	$\pm$	0.02		&		75		&		8.6		&		$-$~~$-$\\
J045757.30+462330.0		&		160.33		&		2.21		&		6009	$\pm$	02		&		123	$\pm$	06		&		115	$\pm$	07		&		0.08	$\pm$	0.03		&		86		&		8.1		&		$-$~~$-$\\
J045804.57+445518.5		&		161.49		&		1.31		&		9480	$\pm$	06		&		152	$\pm$	30		&		~\,75	$\pm$	13		&		0.16	$\pm$	0.03		&		135		&		8.9		&		$-$~~$-$\\
J045811.77+450602.6		&		161.37		&		1.44		&		10081	$\pm$	02		&		268	$\pm$	03		&		259	$\pm$	05		&		0.59	$\pm$	0.06		&		144		&		9.5		&		u~~w\\
J045815.72+453936.1		&		160.94		&		1.79		&		11498	$\pm$	04		&		167	$\pm$	09		&		155	$\pm$	34		&		0.14	$\pm$	0.03		&		164		&		9.0		&		$-$~~$-$\\
J045816.80+463749.5		&		160.18		&		2.40		&		16477	$\pm$	03		&		~\,81	$\pm$	06		&		~\,53	$\pm$	09		&		0.59	$\pm$	0.09		&		235		&		9.9		&		$-$~~$-$\\
J045816.94+460925.8		&		160.55		&		2.10		&		8332	$\pm$	04		&		195	$\pm$	09		&		150	$\pm$	25		&		0.26	$\pm$	0.03		&		119		&		8.9		&		$-$~~$-$\\
J045816.96+453621.8		&		160.98		&		1.76		&		12287	$\pm$	05		&		147	$\pm$	18		&		136	$\pm$	12		&		0.15	$\pm$	0.03		&		176		&		9.0		&		$-$~~$-$\\
J045842.68+463243.4		&		160.29		&		2.40		&		8303	$\pm$	02		&		150	$\pm$	06		&		137	$\pm$	07		&		0.40	$\pm$	0.05		&		119		&		9.1		&		$-$~~$-$\\
J045852.18+463855.9		&		160.22		&		2.49		&		11306	$\pm$	04		&		~\,56	$\pm$	10		&		~\,37	$\pm$	18		&		0.20	$\pm$	0.04		&		162		&		9.1		&	 u~~w\\
J045852.98+445329.6		&		161.61		&		1.40		&		5208	$\pm$	03		&		147	$\pm$	07		&		128	$\pm$	20		&		0.29	$\pm$	0.03		&		74		&		8.6		&		u~~w\\
J045900.14+453437.7		&		161.08		&		1.84		&		12338	$\pm$	02		&		144	$\pm$	07		&		128	$\pm$	06		&		0.18	$\pm$	0.02		&		176		&		9.1		&		$-$~~$-$\\
J045906.31+462358.9		&		160.45		&		2.37		&		3681	$\pm$	03		&		117	$\pm$	12		&		108	$\pm$	07		&		0.09	$\pm$	0.03		&		53		&		7.7		&		$-$~~$-$\\
J045913.88+463039.1		&		160.37		&		2.45		&		7068	$\pm$	02		&		104	$\pm$	05		&		~\,56	$\pm$	14		&		0.41	$\pm$	0.05		&		101		&		9.0		&		$-$~~$-$\\
J045920.35+463222.8		&		160.36		&		2.48		&		5662	$\pm$	04		&		224	$\pm$	14		&		151	$\pm$	10		&		0.68	$\pm$	0.07		&		81		&		9.0		&		$-$~~$-$\\
J045926.99+463431.4		&		160.34		&		2.52		&		10788	$\pm$	03		&		147	$\pm$	08		&		113	$\pm$	12		&		0.78	$\pm$	0.10		&		154		&		9.6		&		$-$~~$-$\\
J045930.60+455248.7		&		160.90		&		2.10		&		6682	$\pm$	04		&		270	$\pm$	19		&		113	$\pm$	08		&		0.62	$\pm$	0.05		&		95		&		9.1		&		$-$~~$-$\\
J045930.82+455247.2		&		160.90		&		2.10		&		11940	$\pm$	03		&		270	$\pm$	24		&		148	$\pm$	06		&		0.66	$\pm$	0.04		&		171		&		9.7		&	 u~~w\\
J045932.30+463928.6		&		160.29		&		2.58		&		10998	$\pm$	03		&		122	$\pm$	11		&		~\,97	$\pm$	07		&		0.60	$\pm$	0.08		&		157		&		9.5		&		$-$~~$-$\\
J045932.66+441837.9		&		162.14		&		1.13		&		9908	$\pm$	01		&		272	$\pm$	04		&		260	$\pm$	03		&		0.66	$\pm$	0.05		&		142		&		9.5		&		u~~w\\
J045937.49+450218.9		&		161.57		&		1.60		&		14535	$\pm$	05		&		183	$\pm$	14		&		150	$\pm$	14		&		0.28	$\pm$	0.03		&		208		&		9.5		&		$-$~~$-$\\
J045940.69+462146.7		&		160.54		&		2.42		&		5193	$\pm$	01		&		~\,59	$\pm$	03		&		~\,39	$\pm$	04		&		0.20	$\pm$	0.02		&		74		&		8.4		&		u~~$-$\\
J045953.38+453145.6		&		161.21		&		1.94		&		10852	$\pm$	03		&		172	$\pm$	07		&		157	$\pm$	09		&		0.33	$\pm$	0.03		&		155		&		9.3		&		u~~w\\
J045955.96+445945.0		&		161.64		&		1.61		&		10208	$\pm$	02		&		133	$\pm$	06		&		119	$\pm$	08		&		0.21	$\pm$	0.03		&		146		&		9.0		&		$-$~~$-$\\
J050031.81+453733.6		&		161.21		&		2.08		&		12284	$\pm$	02		&		153	$\pm$	05		&		144	$\pm$	05		&		0.18	$\pm$	0.02		&		175		&		9.1		&		u~~w\\
J050045.76+454454.4		&		161.14		&		2.19		&		3442	$\pm$	01		&		~\,46	$\pm$	02		&		~\,36	$\pm$	04		&		0.14	$\pm$	0.01		&		49		&		7.9		&	 u~~$-$\\
J050120.46+442255.1		&		162.28		&		1.43		&		10706	$\pm$	02		&		133	$\pm$	05		&		~\,89	$\pm$	15		&		0.27	$\pm$	0.03		&		153		&		9.2		&		$-$~~$-$\\
J050124.98+441446.0		&		162.39		&		1.36		&		8849	$\pm$	01		&		113	$\pm$	03		&		104	$\pm$	05		&		0.29	$\pm$	0.03		&		126		&		9.0		&		$-$~~$-$\\
J050125.79+452212.4		&		161.51		&		2.05		&		6113	$\pm$	02		&		153	$\pm$	07		&		118	$\pm$	07		&		0.22	$\pm$	0.04		&		87		&		8.6		&		u~~w\\
J050212.15+453526.1		&		161.42		&		2.29		&		10847	$\pm$	03		&		~\,87	$\pm$	06		&		~\,56	$\pm$	15		&		0.14	$\pm$	0.02		&		155		&		8.9		&		$-$~~$-$\\
J050217.54+455441.1		&		161.17		&		2.50		&		13374	$\pm$	01		&		371	$\pm$	04		&		344	$\pm$	04		&		0.55	$\pm$	0.05		&		191		&		9.7		&		$-$~~$-$\\
J050231.16+451050.9		&		161.77		&		2.09		&		8306	$\pm$	05		&		340	$\pm$	18		&		322	$\pm$	13		&		0.33	$\pm$	0.04		&		119		&		9.0		&		m~~w\\
J050309.74+453102.6		&		161.58		&		2.38		&		2891	$\pm$	02		&		165	$\pm$	04		&		125	$\pm$	12		&		0.44	$\pm$	0.06		&		41		&		8.3		&		m~~$-$\\
J050329.24+454555.0		&		161.41		&		2.58		&		5490	$\pm$	04		&		271	$\pm$	09		&		217	$\pm$	12		&		2.37	$\pm$	0.19		&		78		&		9.5		&		u~~$-$\\
J050335.40+442816.5		&		162.45		&		1.80		&		7146	$\pm$	04		&		~\,88	$\pm$	09		&		~\,78	$\pm$	12		&		0.19	$\pm$	0.03		&		102		&		8.7		&		u~~$-$\\
\hline
\end{longtable}
\end{landscape}
\end{centering}

\twocolumn


\bsp	
\label{lastpage}
\end{document}